\documentclass[12pt,preprint]{aastex}

\usepackage{verbatim}
\usepackage{paralist}

\shorttitle{Structure of the neutral ISM in nearby Galaxies}
\shortauthors{Bagetakos et al.}

\begin{document}

\title{The Fine--Scale Structure of the neutral Interstellar Medium in nearby Galaxies}
\author{I. Bagetakos\altaffilmark{1},
E. Brinks\altaffilmark{1},
F. Walter\altaffilmark{2},
W.J.G. de Blok\altaffilmark{3},
A. Usero\altaffilmark{4},
A.K. Leroy\altaffilmark{2,5,*},
J.W. Rich\altaffilmark{6}
and R.C. Kennicutt, Jr.\altaffilmark{7}}
\altaffiltext{1}{Centre for Astrophysics Research, Science \& Technology Research Institute, University of Hertfordshire, Hatfield AL10 9AB, United Kingdom.}
\altaffiltext{2}{Max--Planck--Institut f\"ur Astronomie, K\"onigstuhl 17, 69117 Heidelberg, Germany.}
\altaffiltext{3}{Department of Astronomy, University of Cape Town, Private Bag X3, Rondebosch 7701, South Africa.}
\altaffiltext{4}{Observatorio Astron\'omico Nacional, C/ Alfonso XII 3, Madrid 28014, Spain}
\altaffiltext{5}{National Radio Astronomy Observatory, 520 Edgemont Road, Charlottesville, VA 22903, USA}
\altaffiltext{6}{Research School of Astronomy \& Astrophysics, ANU, Mount Stromlo Observatory, Cotter Road, Weston Creek, ACT 2611, Australia.}
\altaffiltext{7}{Institute of Astronomy, University of Cambridge, Madingley Road, Cambridge CB3 0HA, UK.}
\altaffiltext{*}{Hubble Fellow}


\begin{abstract}
We present an analysis of the properties of \ion{H}{1} holes detected in 20 galaxies that are part of ``The \ion{H}{1} Nearby Galaxy Survey" (THINGS). We detected more than 1000 holes in total in the sampled galaxies. Where they can be measured, their sizes range from about 100 pc (our resolution limit) to about 2\,kpc, their expansion velocities range from 4 to 36 $\mathrm{km\,s^{-1}}$, and their ages are estimated to range between 3 and 150\,Myr. The holes are found throughout the disks of the galaxies, out to the edge of the \ion{H}{1} disk; 23\% of the holes fall outside $R_{25}$. We find that shear limits the age of holes in spirals (shear is less important in dwarf galaxies) which explains why \ion{H}{1} holes in dwarfs are rounder, on average than in spirals. Shear, which is particularly strong in the inner part of spiral galaxies, also explains why we find that holes outside $R_{25}$ are larger and older. We derive the scale height of the \ion{H}{1} disk as a function of galactocentric radius and find that the disk flares up in all galaxies. We proceed to derive the surface and volume porosity ($Q_\mathrm{2D}$ and $Q_\mathrm{3D}$) and find that this correlates with the type of the host galaxy: later Hubble types tend to be more porous.

The size distribution of the holes in our sample follows a power law with a slope of $a_{\mathrm{\nu}}\sim -2.9$. Assuming that the holes are the result of massive star formation, we derive values for the supernova rate (SNR) and star formation rate (SFR) which scales with the SFR derived based on other tracers. If we extrapolate the observed number of holes to include those that fall below our resolution limit, down to holes created by a single supernova, we find that our results are compatible with the hypothesis that \ion{H}{1} holes result from star formation. 

\end{abstract}
\keywords{galaxies: ISM --- galaxies: structure --- ISM: bubbles --- ISM: structure}

\maketitle

\section{Introduction}
\label{sec:intro}
Studies of neutral atomic hydrogen (\ion{H}{1}) in nearby galaxies using synthesis instruments such as the Westerbork Synthesis Radio Telescope (WSRT), the Very Large Array (VLA), Australia Telescope Compact Array (ATCA) and more recently the Giant Metre--Wave Radio Telescope (GMRT) have shown the frothy structure of the interstellar medium (ISM). Filamentary structures characterize the gas distribution of the Galaxy and nearby galaxies and have been referred to as the ``cosmic bubble bath'' \citep{bra75} or the ``violent interstellar medium'' \citep{mcc79}. Particularly striking are elliptical features representing a deficiency of interstellar matter, usually bordered by neutral or ionized regions of higher density. These have been described as bubbles and shells, based on their appearance at different wavelengths.

These structures were first recognised as such in the Large Magellanic Cloud (LMC) by \citet{wes66} and in the Small Magellanic Cloud (SMC) by \citet{hin67}. The explanation put forward at the time was that these shells were the result of ``super supernovae''. In our own galaxy, there has long been evidence for shell-like structures. \citet{hei79,hei84} lists a number of early observations of expanding \ion{H}{1} supershells in the Milky Way. As he emphasized ``Identification of shells is a somewhat subjective process due to the fact that some shells are detected at only one velocity and for those shells that do change as a function of velocity usually only one hemisphere of the shell is observed". In addition, distances to these shells are very uncertain, which shifted the focus to nearby galaxies. So far, of order a dozen galaxies have been observed at spatial and velocity resolutions adequate to map and analyse \ion{H}{1} holes. They are listed in Table~\ref{tbl-stu}. 

Following \citet{chu04} and references therein, bubbles and shells are assumed to be created by single stars, as a result of the combined effects of a stellar wind followed by the star exploding as a supernova. Superbubbles and supergiant shells require multiple stars or some other mechanism for their formation. Since bubbles and shells appear as holes in maps of neutral hydrogen, we will be using this term throughout this paper. As mentioned, stellar winds and supernovae from young massive stars in OB associations and Super Star Clusters (SSCs) are considered the main driving forces behind the creation of holes and shells whose radii can reach up to hundreds of parsecs (e.g., \citealt{wea77}; \citealt{mcc87}). Recent observations and analysis of four nearby disk galaxies by \citet{rel07} has shown that H$\alpha$ shells in giant \ion{H}{2} regions are the likely progenitors of neutral hydrogen shells. In addition, \citet{wei09b} found that \ion{H}{1} holes in Holmberg\,II could in principle be the result of stellar feedback from multiple SF episodes, even though similar stellar populations were also found in regions where no \ion{H}{1} holes were detected. However, a one to one correlation between \ion{H}{1} holes and active star forming regions such as OB associations has not been established as yet. Moreover numerical models by \citet{sil08} suggest that the H$\alpha$ shells studied by \citet{rel07} cannot evolve into the largest \ion{H}{1} shells observed unless a physical mechanism other than supernovae (SNe) is invoked.  

The kinetic energies inferred on the basis of the typical expansion velocities measured (of order a few tens of km\,s$^{-1}$) range from 10$^{50}$ to 10$^{54}$ erg. One of the problems of this ``standard" theory of mechanical input is the existence of a growth-rate discrepancy. Observations show that shells are smaller than what is predicted by theory. \citet{oey06} suggests that this discrepancy can be attributed to one, or a combination of, the following reasons: an overestimate of the input power, an underestimate of the ambient density and/or pressure, enhanced radiative cooling in the interior of the bubbles, or the transfer of mechanical energy from the holes to cosmic rays through the strong MHD turbulence and magnetic fields that govern their interior. 

Because of a failing consensus on the hypothesis that star clusters power supergiant shells, several alternative explanations have been put forward to explain structures which are found in environments where star formation is suppressed or for those cases for which the energy requirement to create the structure is believed to far surpass what can be delivered by even a super star cluster (SSC). Among the mechanisms proposed are infall of high velocity clouds (e.g., \citealt{ten88}; \citealt{ran96}; \citealt{san99}; \citealt{mur04}), ram pressure enlarging embryonic shells \citep{bur02}, and turbulence, coupled with cooling and gravitational instability \citep{wad00,dib05}. At one time even gamma-ray bursts have been invoked (\citealt{loe98}; \citealt{efr99}; \citealt{per00}). 

The study of shells and supershells is relevant to many areas of galaxy research. Supergiant shells are linked to superbubble blow out and might lie at the origin of a galactic fountain \citep{bre80}. If shells do indeed break out of the disk, the halo or in extreme cases even the intergalactic medium can be enriched by metals produced in the massive stars before they exploded as SNe \citep{ten00}. Numerical simulations and modeling suggest that actively star forming dwarf irregular galaxies are most prone to mass loss \citep{mac99,sil01}. Given that in the currently accepted $\Lambda$CDM cosmology dwarf galaxies dominated at large look--back times, this could explain the rapid enrichment and mixing of heavy elements shortly after the first galaxies were formed. On more local scales, \ion{H}{1} holes play an important role in the evolution of the ISM and are ultimately linked to the process of star formation. Expanding holes compress their surrounding medium leading to conditions conducive to star formation (e.g., \citealt{ten05}). Examples for this can be found in the LMC where CO observations point to the presence of molecular clouds on the rims of several supergiant shells \citep{yam01}. The Magellanic Clouds are not the only example. \citet{ler06} show similar such accumulations of molecular gas on the interface between two neighbouring shells in IC\,10 which seemingly are running into each other. In other words, shells could be driving self--regulating (propagating or stochastic) star formation \citep{elm02}. On yet smaller scales, the processes leading to shells and supergiant shells provide energy input on the largest turbulent scales which then cascade down to ever smaller scales. And finally, the energy input from SNe in the form of expanding shells provides positive feedback as it raises the velocity dispersion in the gas, shutting off further star formation until such time as the gas has cooled down, in a dynamical sense, allowing SF to recommence. On a global scale, the star formation rate in disk galaxies is then controlled by the porosity of the ISM \citep{sil97}. This implies a feedback loop which likely causes the ISM to have a one dimensional velocity dispersion of 6--10 km\,s$^{-1}$ irrespective of galaxy type \citep{dib06,ler08,tam09}.

\begin{deluxetable}{lccrr}
\tabletypesize{\scriptsize}
\tablecaption{Previously published studies of \ion{H}{1} holes in galaxies \label{tbl-stu}}
\tablewidth{0pt}
\tablehead{
\colhead{Galaxy} & \colhead{FWHM beam} & \colhead{FWHM beam} & \colhead{Velocity Resolution} & \colhead{Reference}\\
\colhead{} & \colhead{(\arcsec)} & \colhead{(pc)} & \colhead{(km\,s$^{-1}$)} & \colhead{}}
\startdata
M\,31&$24\times36$ &$80\times120$&8.2&1, 2\\
M\,33&$12\times24$&$40\times40$&8.2&3\\
M\,101&$13\times16$&500&8.2&4\\
Holmberg\,II&$4.0\times4.5$&60&2.6&5\\
SMC&98&28&1.6&6, 7, 8\\
NGC\,2403&9&144&6.2&9, 10\\
IC\,10&$4.7\times5.0$&$44\times20$&12.2&11\\
IC\,2574&$6.4\times5.9$&$100\times91$&2.6&12, 13\\
LMC&60&15&1.6&14\\
NGC\,6822&$42.4\times12.0$&$100\times29$&1.6&15\\
DDO\,47&$7.8\times7.2$&$150\times140$&2.6&16\\
Holmberg\,I&$8.2\times7.0$&$140\times120$&2.6&17\\
M\,82&$1.4\times1.2$&$22\times19$&10.3&18\\
NGC\,6946&$12\times14$&$350\times400$&4.1&19\\
DDO\,43&$14\times11$&$375\times290$&2.6&20\\
DDO\,88&$14.5\times13.1$&$520\times470$&2.6&21\\
NGC\,1569&$12.1\times10.8$&$114\times95$&2.6&22\\
IC\,1613&$7.4\times7.0$&$26\times24$&2.6&23\\
\enddata
\tablecomments{Where multiple studies exist for a galaxy, only the one with the best resolution and/or sensitivity is listed.}
\tablerefs{
(1) \citealt{bri81};
(2) \citealt{bri86};
(3) \citealt{deu90};
(4) \citealt{kam91};
(5) \citealt{puc92};
(6) \citealt{sta97};
(7) \citealt{sta99};
(8) \citealt{hat05};
(9) \citealt{thi98};
(10) \citealt{mas99};
(11) \citealt{wil98};
(12) \citealt{wal99};
(13) \citealt{ric08};
(14) \citealt{kim99};
(15) \citealt{deb06};
(16) \citealt{wal01};
(17) \citealt{ott01};
(18) \citealt{wil02};
(19) \citealt{boo04};
(20) \citealt{sim05b};
(21) \citealt{sim05a};
(22) \citealt{muh06};
(23) \citealt{sil06}.}
\end{deluxetable}

The resolution and sensitivity levels down to which the nearby galaxies listed in Table~\ref{tbl-stu} have been observed vary greatly, making any statistical comparison fraught with difficulty. The purpose of this paper is to improve upon the status quo by adding 18 new galaxies and revisiting two (Holmberg\,II and IC\,2574). All 20 galaxies were observed at similar angular and velocity resolution and down to similar detection thresholds as part of ``The \ion{H}{1} Nearby Galaxy Survey" (THINGS). The identification of \ion{H}{1} holes was carried out in a consistent manner which enables us for the first time to study these features in a systematic way across a variety of Hubble types.

A description of the observations is given in Sec.~\ref{sec:Observations}. In Sec.~\ref{sec:Char} we present the global characteristics of the 20 galaxies under study. The complete catalogue of \ion{H}{1} holes and an outline of our results follows in Sec.~\ref{sec:Holes} and the statistical analysis of the holes is presented in Sec.~\ref{sec:Stats}. In Sec.~\ref{sec:Summary} we present a summary of our results.

\section{Observations}
\label{sec:Observations}
The present study is based on observations carried out within the framework of THINGS project. THINGS is based on a sample of 34 nearby systems (2-15 Mpc) observed at high angular ($\approx 6 \arcsec$) and velocity (2.6-5.2 km\,s$^{-1}$) resolution \citep{wal08}. The observations were carried out with the NRAO\footnote{The National Radio Astronomy Observatory is a facility of the National Science Foundation operated under cooperative agreement by Associated Universities, Inc.} VLA in B, C, and D configurations totaling $\sim 500$ hours (including archival data). A summary of the observations is given in \citet{wal08}. THINGS was designed to complement the Spitzer Infrared Nearby Galaxies Survey \citep[SINGS;][]{sings} and the \textit{GALEX} Nearby Galaxies Survey \citep[NGS;][]{ngs} therefore providing us with multi--wavelength coverage of these galaxies. 

For our analysis we selected 20 galaxies from the THINGS sample based on certain criteria. The basic criterion was to include at least one pair of galaxies from each Hubble type in order to be able to compare \ion{H}{1} holes across all Hubble types. The galaxies in each pair were also selected to vary substantially in their star formation rates and \ion{H}{1} masses and care was taken to select galaxies with sufficiently high signal-to-noise ratios. The list of selected galaxies is given in Table~\ref{tbl-glob}. In what follows we split our sample in dwarf and spiral galaxies based on their morphological type. All Irr and magellanic type galaxies were classified as dwarfs.

\begin{deluxetable}{lllllrrrrrcc}
\tabletypesize{\scriptsize}
\rotate
\tablecaption{Characteristics of the galaxies in the sample \label{tbl-glob}}
\tablewidth{0pt}
\tablehead{
\colhead{Name} & \colhead{Alternative} & \colhead{$\alpha$(J2000)} & \colhead{$\delta$(J2000)} &  \colhead{Type} & \colhead{D} &  \colhead{Incl.\tablenotemark{a}} & \colhead{PA\tablenotemark{b}} &  \colhead{H\,{\sc i} Mass} & \colhead{log(SFR)} & \colhead{log($D_\mathrm{25})$} & \colhead{Resolution}\\
\colhead{} & \colhead{Name} & \colhead{({\em h~~~m~~~s})} & \colhead{$(\arcdeg\ \ \ \arcmin\ \ \ \arcsec)$} & \colhead{} & \colhead{(Mpc)} & \colhead{(deg)} & \colhead{(deg)} & \colhead{$(10^8\,\mathrm{M_{\sun}})$} & \colhead{$(\mathrm{M_{\sun}\,yr^{-1}})$} & \colhead{$(\mathrm{0.1\arcmin})$} & \colhead{(pc)}
}
\startdata
NGC\,628&M\,74&01 36 41.8&+15 47 00.0&SA(s)c&7.3&7&20&$38.0$& 0.08&1.99&218\\
NGC\,2366&DDO\,42&07 28 53.4&+69 12 51.1&IB(s)m&3.4&64&40&$6.5$&$-1.02$\tablenotemark{c}&1.64&107\\
NGC\,2403& \nodata &07 36 51.1&+65 36 02.9&SAB(s)cd&3.2&63&124&$25.8$&$-0.07$&2.20&87\\
Holmberg\,II&DDO\,50&08 19 05.0&+70 43 12.0&Im&3.4&41&177&$5.9$&$-1.13$&1.82&107\\
DDO\,53& \nodata &08 34 07.2&+66 10 54.0&Im&3.6&31&132&$0.6$&$-2.10$&0.89&103\\
NGC\,2841& \nodata &09 22 02.6&+50 58 35.4&SA(r)b&14.1&74&153&$85.6$&$-0.70$&1.80&405\\
Holmberg\,I&DDO\,63&09 40 32.3&+71 10 56.0&IAB(s)m&3.8&12&50&$1.4$&$-2.23$&1.52&128\\
NGC\,2976& \nodata &09 47 15.3&+67 55 00.0&SAc&3.6&65&335&$1.4$&$-0.98$&1.90&87\\
NGC\,3031&M\,81&09 55 33.1&+69 03 54.7&SA(s)ab&3.6&59&330&$36.3$&$0.03$&2.33&132\\
NGC\,3184& \nodata &10 18 17.0&+41 25 28.0&SAB(rs)cd&11.1&16&179&$30.6$&$0.16$&1.87&281\\
IC\,2574&DDO\,81&10 28 27.7&+68 24 59.4&SAB(s)m&4.0&53&55&$14.7$&$-0.93$&2.11&111\\
NGC\,3521& \nodata &11 05 48.6&$-00$ 02 09.2&SAB(rs)bc&10.7&73&340&$80.2$&$0.52$&1.92&376\\
NGC\,3627&M\,66&11 20 15.0&+12 59 29.6&SAB(s)b&9.3&62&173&$8.2$&$0.39$&2.01&249\\
NGC\,4214& \nodata &12 15 39.2&+36 19 37.0&IAB(s)m&2.9&44&65&$4.1$&$-1.28$&1.83&98\\
NGC\,4449& \nodata &12 28 11.9&+44 05 40.0&IBm&4.2&60&230&$11.0$&$-0.33$\tablenotemark{d}&1.67&267\\
NGC\,4736&M\,94&12 50 53.0&+41 07 13.2&(R)SA(r)ab&4.7&41&296&$4.0$&$-0.37$&1.89&130\\
DDO\,154&NGC\,4789A&12 54 05.9&+27 09 09.9&IB(s)m&4.3&66&375&$3.6$&$-2.44$&1.29&147\\
NGC\,5194&M\,51&13 29 52.7&+47 11 43.0&SA(s)bc&8.0&42&172&$25.4$&$0.78$&1.89&221\\
NGC\,6946& \nodata &20 34 52.2&+60 09 14.4&SAB(rs)cd&5.9&33&243&$41.5$&$0.68$&2.06&135\\
NGC\,7793& \nodata &23 57 49.7&$-32$ 35 27.9&SA(s)d&3.9&50&290&$8.9$&$-0.29$&2.02&142\\
\enddata
\tablecomments{All values taken from \citet{wal08} unless stated otherwise.}
\tablenotetext{a}{Average inclination}
\tablenotetext{b}{Average position angle}
\tablenotetext{c}{\citet{hun04}}
\tablenotetext{d}{\citet{hun99}}

\end{deluxetable}

The data reduction process is described in detail in \citet{wal08}. The AIPS package was used to reduce the data for every galaxy and to create the datacubes. Two different weighting schemes were applied, \textit{natural weighted} (NA, resulting in 'low' resolution, high sensitivity maps) and \textit{robust weighted} (RO, resulting in higher resolution maps at the cost of increased noise), and a residual-flux rescaling method was performed on both versions of the cubes. For this study we used both the NA and RO cubes for the detection of the \ion{H}{1} holes as is described in Sec.~\ref{sec:HoleDet}

\section{Global Characteristics}
\label{sec:Char}
In order to derive some of the properties of the holes the scale height of the \ion{H}{1} layer needs to be known. In an isothermal disk the scale height z is given by:
\begin{equation}
z=\frac{v^2}{\pi G \mu} 
\end{equation}
\citep{kel70} where $v$ is the velocity dispersion and $\mu$ the total mass surface density. However, the disk of a galaxy consists of different components, stars and gas, which have different velocity dispersions i.e matter is not isothermal. In such a case where the gas is affected by the gravitational potential of the stars and its own self-gravity, its velocity dispersion is given by:
\begin{equation}
v_{g}^2=\frac{2\pi G \mu z_{g}^{2}}{z_g+z_\ast}
\end{equation}
\citep{dop94} where $z_g$ and $z_\ast$ are the scale heights of gas and stars respectively. The 1$\sigma$ line of sight velocity dispersion of the gas was calculated by taking the FWHM of spectra in quiescent regions of the galaxy and was found to be of order $6-10$ km\,s$^{-1}$ consistent with what was found by \citet{ler08} and \citet{tam09}. We adopted a constant velocity dispersion for the gas for each of our targets. For $z_\ast$ we used $z_\ast=l_\ast/7.3$ following \citet{ler08}, where $l_\ast$ is the disk scale length. This average flattening ratio was measured by \citet{kre02} assuming $z_\ast$ does not vary with radius, which is a reasonable assumption. The total mass surface density was taken as the sum of the stars and gas (\ion{H}{1} and H$_2$) mass surface densities \citep{ler08}. The gas surface density is corrected to account for Helium and metals. For the stars we used the $3.6\mu m$ surface brightness profile which traces the older stars and consequently the bulk of the star mass. In regular spirals, the effect of the dark matter halo on the scale height of the gas can be ignored to first order, because its effective contribution to the mass surface density within the typical radii dealt with is insignificant compared to that of baryonic matter. It should be noted that in the case of dwarf galaxies the influence on scale height of the dark matter distribution becomes significant. In that case the approach by \citet{dop94} would need to be replaced by a more sophisticated approach, such as the one proposed by \citet{nar02}.

\begin{figure}
\begin{center}
\includegraphics[angle=0,scale=.69]{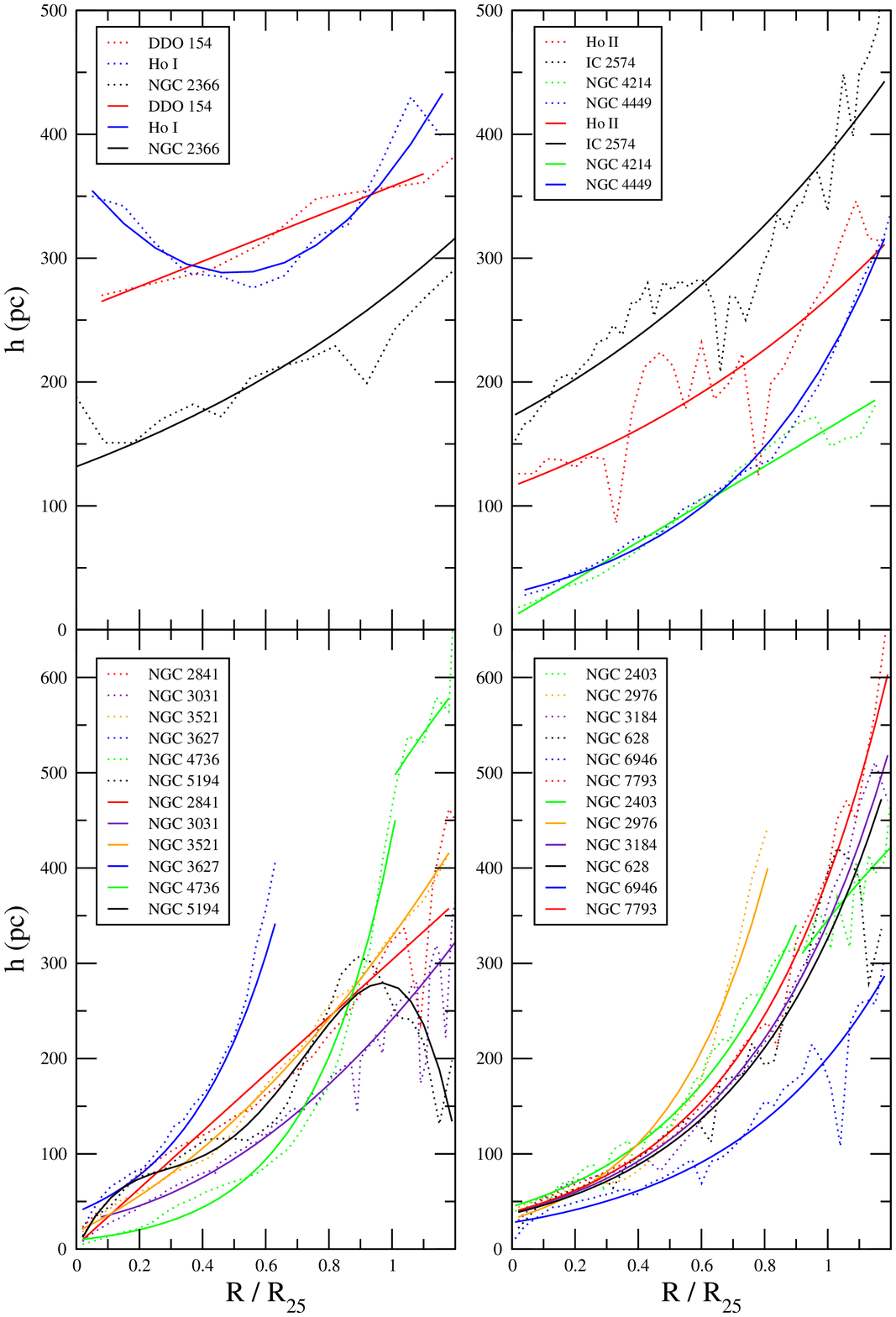}
\end{center}
\caption{Scale heights as a function of galactocentric (R) radius normalised to R$_{25}$ for dwarf galaxies (top panels), early-type (bottom left) and late-type (bottom right) spirals. The dotted lines correspond to the scale height calculated using Eq. 2 and the solid lines to a best fit function. \label{fig:SH}}
\end{figure}

The mass surface density decreases with radius, initially exponentially as it follows the light and gradually flattening well beyond R$_{25}$ where HI dominates. The observed almost constant velocity dispersion of the gas combined with a radial decline in mass surface density implies that the gas layer flares, i.e., the scale height increases with radius. This is confirmed by observations of the Milky Way \citep{kal09} and nearby galaxies \citep{bri84,san08}. We therefore decided to use a radially dependent mass surface density to explicitly give radially dependent values for the scale height; an average value would introduce large uncertainties in the inner and outer parts of a galaxy. Fig.~\ref{fig:SH} illustrates the scale heights for all galaxies in our sample out to $1.2\,R_{25}$. The horizontal axis represents galactocentric radius normalised to $R_{25}$. The dotted lines correspond to the scale height calculated using Eq. 2 and the solid lines to a best fit exponential function used to parameterise its behaviour. For galaxies for which the \ion{H}{1} disk extends to more than $1.2\,R_{25}$ we extrapolated out to the edge of the \ion{H}{1} disk at radius $R_\mathrm{max}$. It is clear that in the majority of objects the scale height increases exponentially with radius and the best fit function has the form:
\begin{equation}
h(r)\sim e^{(a\frac{R}{R_{25}})}
\end{equation}
In dwarf galaxies the average exponent \textit{a} is $\sim0.8$, in early-type spirals is $\sim3.6$ and in late-type spirals it is $\sim2.3$. At $R>R_{25}$ the scale height tends to flatten to some asymptotic value as the gas and dark matter dominate the matter distribution and therefore we decided to extrapolate using a logarithmic function. This extrapolation only affects 3 galaxies (NGC\,2403, NGC\,4449 and NGC\,4736) with a small number of holes.

The method described appears to underestimate the scale height in the inner part of spiral galaxies by up to a factor of 2. This can be understood as the velocity dispersion is dependent on radius as well (see e.g., \citealt{tam09}). The velocity dispersion tends to be higher in the inner disk due to the higher star formation and supernova rate. There is still considerable uncertainty surrounding this finding, a full analysis of which is beyond the scope of this paper. The scale height is merely used to determine the volume density of the \ion{H}{1} gas in the disk which in turn is used to determine the energy requirements of the holes. As is discussed later, the uncertainty in the volume density has the smallest effect on the calculation of these energies compared to other uncertainties. We therefore decided to adopt a constant gas velocity dispersion throughout each galaxy.  Table~\ref{tbl-char} summarises the values adopted for all the galaxies in our sample. The scale height listed is an average over the entire \ion{H}{1} disk. When deriving the properties for a an \ion{H}{1} hole, though, the scale height at the location of the hole was calculated. 

\begin{deluxetable}{lrrcc}
\tabletypesize{\scriptsize}
\tablecaption{Global derived characteristics of the galaxies in the sample \label{tbl-char}}
\tablewidth{0pt}
\tablehead{
\colhead{Galaxy} & \colhead{$R_\mathrm{max}$} & \colhead{$V_\mathrm{rot}$} & \colhead{$\sigma_\mathrm{gas}$} &
 \colhead{\textit{$<$h$>$}}\\
\colhead{} & \colhead{(kpc)} & \colhead{(km\,s$^{-1}$)} & \colhead{(km\,s$^{-1}$)} &  \colhead{($\times 100$ pc)}
\\
\colhead{(1)} & \colhead{(2)} & \colhead{(3)} & \colhead{(4)} & \colhead{(5)}
}
\startdata
NGC\,628&20.0&213&7&4.9\\
NGC\,2366&6.8&49&9&5.1\\
NGC\,2403&16.3&138&8&3.6\\
Holmberg\,II&6.8&38&7&3.4\\
DDO\,53&2.0&23&8&2.9\\
NGC\,2841&46.0&282&9&4.8\\
Holmberg\,I&4.5&50&9&6.4\\
NGC\,2976&2.5&74&9&1.2\\
NGC\,3031&14.0&199&8&1.4\\
NGC\,3184&18.1&207&7&3.1\\
IC\,2574&9.3&70&7&2.9\\
NGC\,3521&23.8&220&10&3.3\\
NGC\,3627&7.8&203&13&1.4\\
NGC\,4214&4.8&48&6&1.4\\
NGC\,4449&5.1&42&10&4.0\\
NGC\,4736&9.5&174&9&3.6\\
DDO\,154&6.8&44&8&5.4\\
NGC\,5194&11.9&137&14&1.6\\
NGC\,6946&17.6&204&8&2.8\\
NGC\,7793&7.8&119&8&2.6\\
\enddata
\tablecomments{(2) The maximum radius of the H\,I disk. (3) The rotation velocity of the galaxy at $R_\mathrm{max}$. (4) The velocity dispersion of the H\,I gas. (5) Average scale height over the H\,I disk.}
\end{deluxetable}

\section{The \ion{H}{1} hole catalogues}
\label{sec:Holes}
\subsection{Hole detection}
\label{sec:HoleDet}

Nearly all of the \ion{H}{1} hole studies reported above are based on identifications made ``by eye''. This method carries selection biases which are difficult to quantify and is therefore subjective to a certain degree. To address this, several attempts have been made to create automated detection algorithms. The first such attempt was made by \citet{thi98} who used models of expanding structures to identify shells in observed data cubes; they applied their code to NGC\,2403. \citet{mas99} refined the code to include more realistic three-dimensional hydrodynamical simulations of shells and revisited NGC\,2403 where they detected 601 expanding shells. The same technique was also applied to the Canadian Galactic Plane Survey (CGPS) by \citet{mas02}. These codes, although they are quite robust in discovering regular, expanding shells around young stars fail to detect non--spherical or incomplete shells. These are the type of features we are predominantly finding in our data. A different approach to an automated procedure was proposed by \citet{dai03} who used a neural networks algorithm to detect expanding shells based on velocity spectra. This method was successfully tested on data from the CGPS. More recently \citet{ehl05} developed a model independent algorithm which identifies \ion{H}{1} holes as depressions in the channel maps without any constraints on the shape or expansion velocity of the hole, but requiring the hole is ``complete". They applied this code on the Leiden--Dwingeloo Survey data and identified about 1000 structures in the Milky Way. 

A cursory inspection of our data on galaxies in the Local Volume, with varying orientation parameters, sensitivity and linear resolution, quickly convinced us that none of the automated methods can be applied successfully to our entire sample. The human brain is superior when it comes to recognising irregular and open structures. We therefore decided to identify the \ion{H}{1} holes ``by eye'', despite the fact that it is less objective than an algorithm and quite a time--consuming process.

The search for \ion{H}{1} holes in the 20 THINGS galaxies was conducted with the KARMA\footnote{The KARMA visualization software package was developed by Richard Gooch, formerly of the Australia Telescope National Facility (ATNF).} visualization software package which contains a large number of modules (tasks) several of which were used for this project. In particular, the task {\sc kvis} was used to display and investigate the integrated \ion{H}{1} maps as well as digital movies of the data cube. The task {\sc kpvslice} was also extensively used as it offers many ways to look at a data cube by producing interactive position-velocity (\textit{pV}) cuts in any orientation through a data cube. 

The detection process involved several steps which were followed in a consistent way for all 20 galaxies. First, the task {\sc kvis} was used to inspect the natural weighted integrated map for the most obvious, more extended structures. The second step was to search the higher resolution robust weighted integrated \ion{H}{1} map for smaller holes. From these two steps we generated a list of candidate holes. The task {\sc kvis} was then used to inspect the natural weighted and robust weighted data cubes by displaying the channel maps sequentially, much like a movie. This was done for two reasons: although some of the holes are clearly visible in the integrated maps, many holes do not show up at all due, for example, to the inclination of the galaxy which means that holes along the line of sight tend to blend, or are ``filled in'' with emission at other velocities. Also, holes which are still fully contained within the \ion{H}{1} disk can only be reliably detected in channel maps or \textit{pV} diagrams. The second reason to display the channel maps as a movie was to discard holes that are chance superpositions of structures, such as parts of spiral arms, rather than genuine holes in the ISM. In order to set standard criteria to classify a feature as a genuine hole, we decided that a hole should be visible over at least 3 or 5 channels (depending on whether the velocity resolution is 5.2 or 2.6 km\,s$^{-1}$, respectively), which is of order the full width at half maximum corresponding to the typical \ion{H}{1} velocity dispersion encountered. This further eliminates spurious features.

In order to eliminate as much as possible any subjectivity, each hole in the revised list was examined individually using the task {\sc kpvslice}. The signature of a hole in a \textit{pV} diagram depends on the type of the hole as is discussed later in this paper, but overall this method is better suited for the detection of the smallest features. In addition, the inspection of \textit{pV} diagrams is used to determine some of the basic properties of the holes, such as position, type, heliocentric and expansion velocities.
For every feature detected, two sets of properties were determined. The first set involves the observed properties and the second set the derived properties. In addition, a quality ranking \textit{Q} is assigned, which represents a (subjective) judgement regarding the probability of the feature being actually a shell in the \ion{H}{1} distribution; its values range from 1 (low quality) to 9 (high quality). Even though this property is not directly quantifiable it depends on many aspects of the appearance of a hole and it follows certain criteria: 

\begin{enumerate}
\item the hole must be present in at least 3 or 5 consecutive channel maps depending on whether the velocity resolution is 5.2 or 2.6 km\,s$^{-1}$, respectively,
\item the center of the hole must remain stationary across the channel maps in which it is observed,
\item there must be sufficient contrast (at least a 50\% drop in surface brightness) between the hole and its immediate surroundings,
\item the shape of the hole in position-velocity space must be clearly defined and adequately described by an ellipse.
\end{enumerate}

Overall we detected more than 4000 candidate \ion{H}{1} holes across the 20 galaxies. We decided to stay on the conservative side and only keep holes detected at high significance, i.e. above a certain threshold. This threshold was determined by comparing the holes we found in IC\,2574 with those found in a parallel study by \citet{ric08}. Both authors looked for holes independently assigning a threshold parameter \textit{Q} value to each hole. We then compared our catalogs based on the position and size of each hole. A hole was considered a match when its position was within half a beam size and its diameter agreed to better than 75\% . From this comparison we found that over 75\% of our $Q>5$ holes had been identified by both authors and we therefore decided to use this value as our quality threshold. Once all the low \textit{Q} holes were removed from our lists we were left with $\sim 1000$ \ion{H}{1} holes. 

\subsection{Observed properties}
In what follows we describe the observed characteristics of each hole.

\begin{figure}
\begin{center}
\includegraphics[angle=-90,scale=.60]{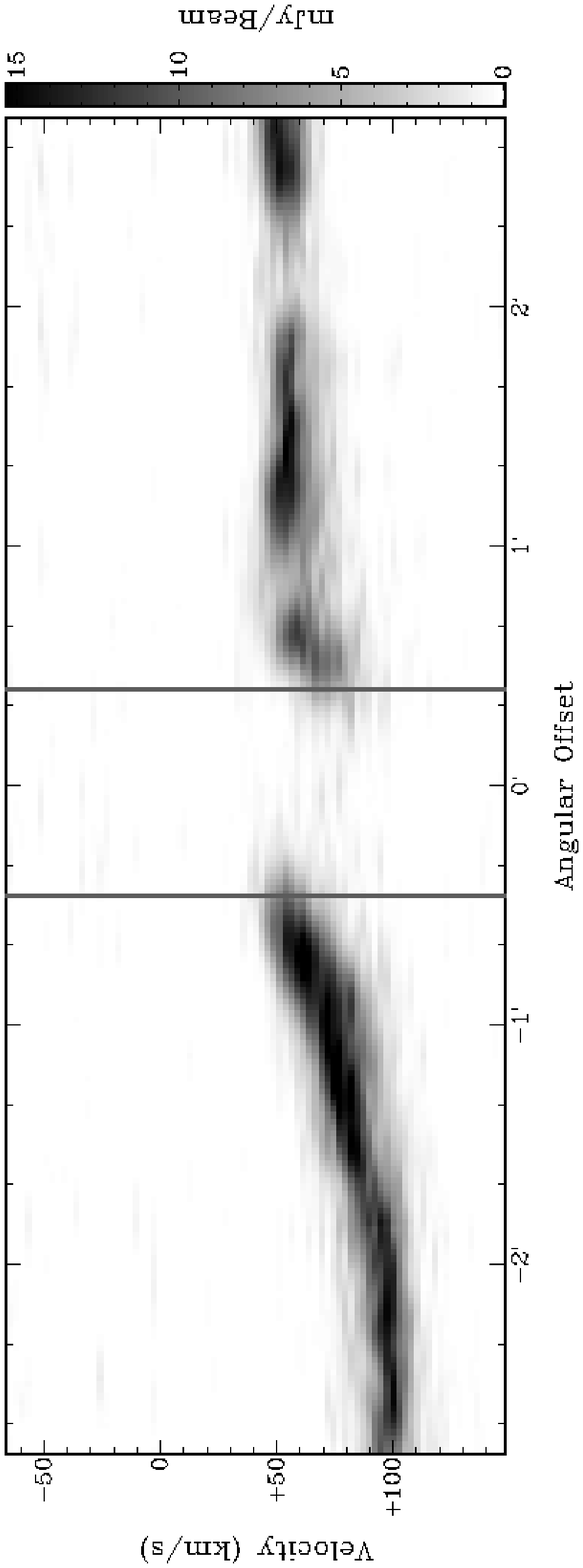}
\includegraphics[angle=-90,scale=.60]{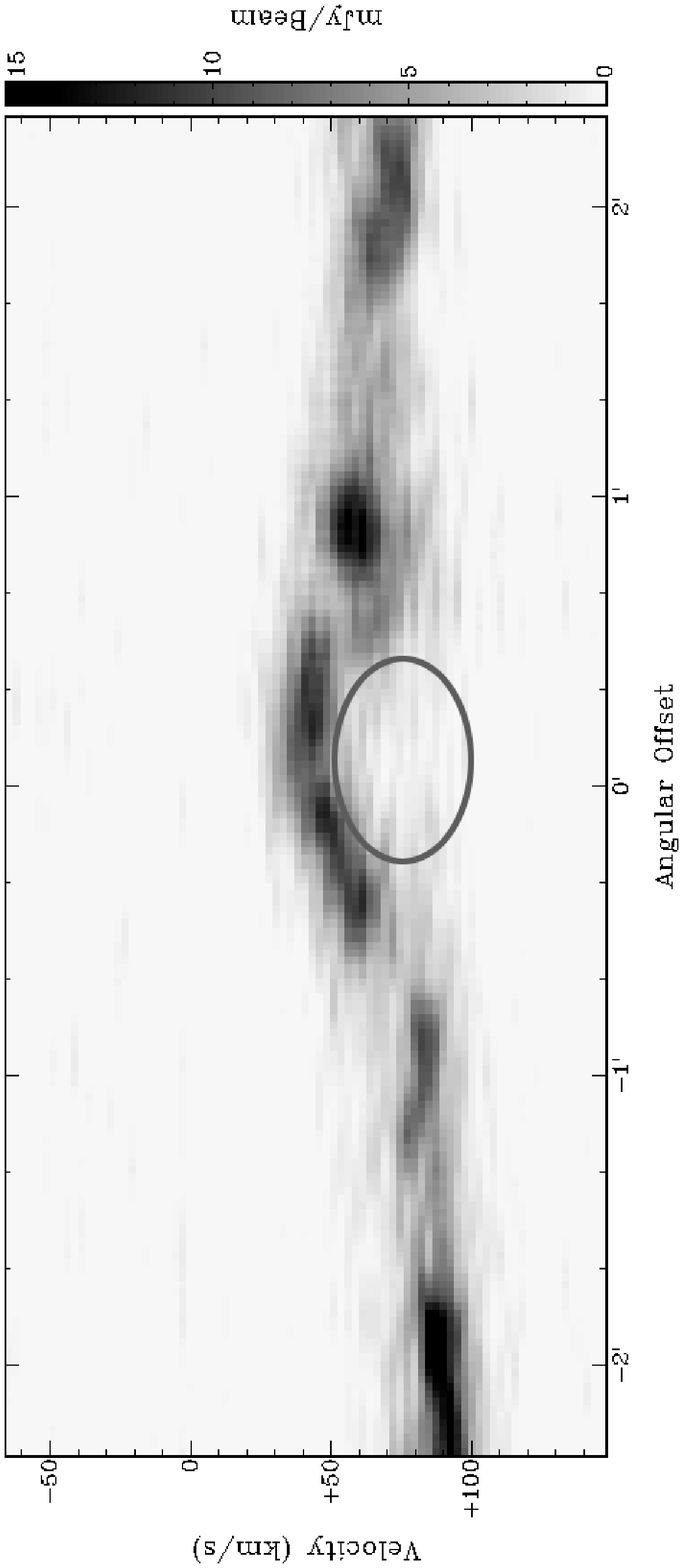}
\includegraphics[angle=-90,scale=.60]{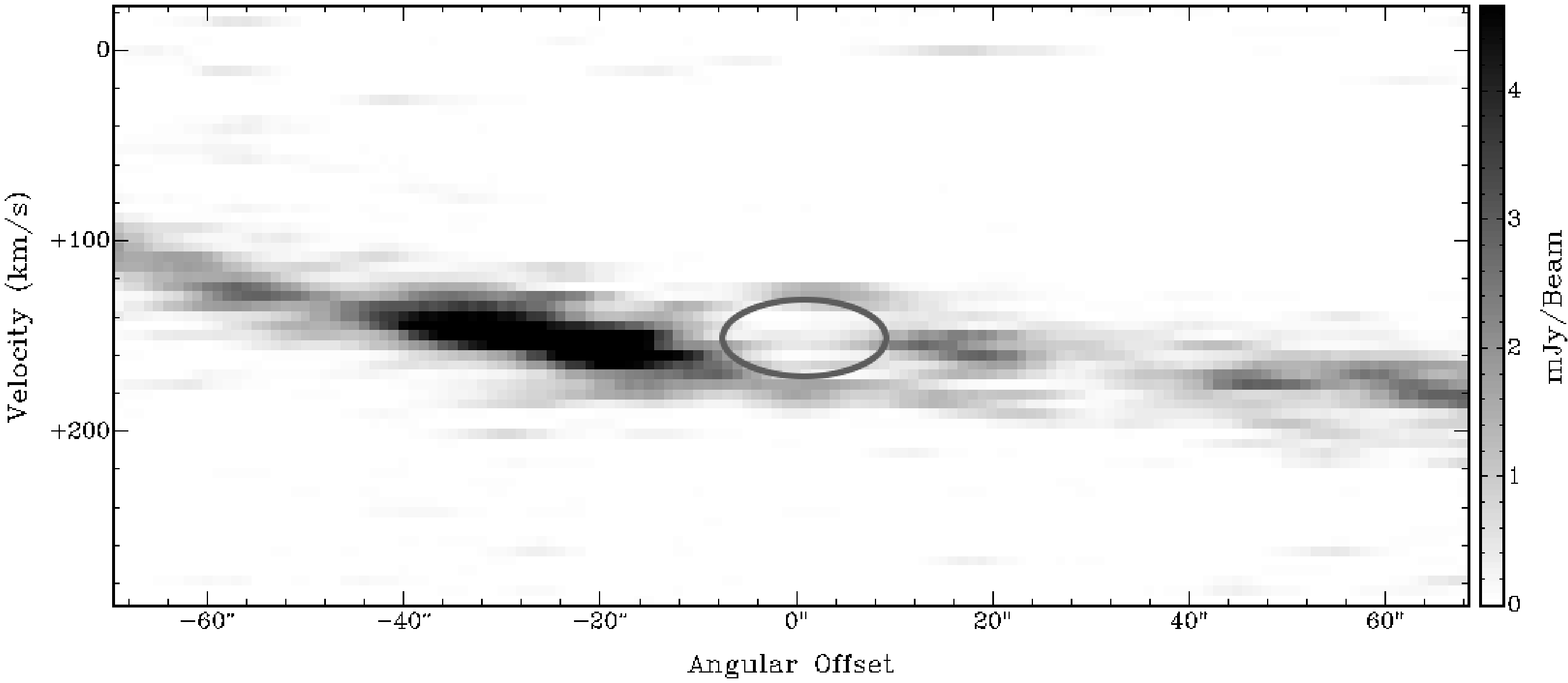}
\end{center}
\caption{\label{fig:types} Examples of the appearance in \textit{pV} diagrams of type 1 (top panel, hole no 21 in IC\,2574), type 2 (middle, no 23 in IC\,2574) and type 3 (bottom, no 90 in NGC\,2403) holes as defined in \citet{bri86}. The solid lines illustrate the position and shape of the holes in \textit{pV} space.}
\end{figure}

\begin{enumerate}

\item The position of the centre of the hole in right ascension and declination (J2000 coordinates) determined by the (average) pixel where the flux density is at a minimum. The uncertainty in the position is taken as half the robust beam size for each galaxy and is of order 3\arcsec.

\item The heliocentric velocity ($V_\mathrm{hel}$) of the hole. This was defined by the velocity channel in which the contrast between the \ion{H}{1} hole and its rim was highest. The uncertainty here is equal to the velocity resolution (2.6 - 5.2 km\,s$^{-1}$). 

\item The semi-major and semi-minor axes ($b_\mathrm{maj}$, $b_\mathrm{min}$) of the hole. To determine the major and minor axes, the approximate orientation of the hole was estimated and the task {\sc kpvslice} was then used to draw cuts along these directions through the centre of the hole, either as seen directly on the integrated \ion{H}{1} map or inferred from the \textit{pV} diagram. These cuts revealed the decrease in \ion{H}{1} at the position of the hole. The axes were taken as the FWHM of this depression. The uncertainty is of order half the robust beam size or 3\arcsec.

\item The axial ratio of the hole was defined by the ratio of minor axis to major axis. The uncertainty of the axial ratio is of order 25\%. 

\item The position angle (\textit{PA}) of the hole on the sky measured from north through east. We estimate the uncertainty to be of order 20 degrees.

\item The type of the hole ($1-3$). Following \citet{bri86}, the holes were assigned to one of three different types, based on their appearance in the \textit{pV} diagrams (Fig.~\ref{fig:types}). A type 1 hole is a hole that has completely blown out of the disk of the galaxy, and therefore neither the approaching nor the receding sides of the hole can be observed as can be seen in Fig.~\ref{fig:types}, top. In a type 2 hole we only observe one side of the hole, which is usually seen in \textit{pV} diagrams as a deviation toward either higher or lower velocities with respect to the velocity of the bulk of the \ion{H}{1} within that area of the galaxy (Fig.~\ref{fig:types}, middle). Finally, in a type 3 hole we see both sides of the hole. Its signature in \textit{pV} diagrams is that of an elliptical structure (Fig.~\ref{fig:types}, bottom). Typically, type 1 holes are larger than type 2 and 3 as they represent a later stage of bubble evolution i.e. type 3 holes are expected to evolve into type 1 (and possibly type 2) after growing. 

\item The expansion velocity ($V_\mathrm{exp}$) of the hole. This was measured from the signature of the hole in the \textit{pV} diagrams. For type 1 holes it is impossible to measure an expansion velocity. However, it is very likely that these have actually stalled and their final expansion velocity will be equal to the velocity dispersion of the gas. For type 2 holes the expansion velocity was taken as the difference between the velocity of the \ion{H}{1} from that side of the hole which is seen to deviate from its measured velocity and the mean velocity of ambient gas surrounding the hole. In the case of type 3 holes, the expansion velocity was taken as half the difference between the velocity of the approaching and receding sides of the hole. We adopt an uncertainty that is equal to the velocity resolution (2.6 or 5.2 km\,s$^{-1}$). We tested the validity of the derivation of the expansion velocity for type 2 holes by applying the method explained for type 2 to the type 3 holes. The derived velocities were the same to within the uncertainties.

\item The average flux density (\textit{I}) around the hole. This was determined by averaging the pixel values on the integrated \ion{H}{1} map over an area around the hole, twice the area of the hole itself. This value is needed in order to derive the approximate density before the creation of the hole. Errors here are estimated to be of order 10\%.
\end{enumerate}

\subsection{Derived properties}
The observed characteristics have been used to determine the derived properties as described in the following:
\begin{enumerate}
\item The diameter (\textit{d}) of the hole. Since most of the holes are not circular but elliptical, the diameter was defined using the geometric mean:
\begin{equation}
d=2\sqrt{b_\mathrm{maj}\,  b_\mathrm{min}}
\end{equation}

where $b_\mathrm{maj}$ and $b_\mathrm{min}$ are the semi-major and semi-minor axis of the hole respectively. The uncertainty in $d$ is of order 10\%.

\item The kinetic age $t$ of the hole. Assuming that over its lifetime the hole was expanding with $V_\mathrm{exp}$, this is:
\begin{equation}
t_\mathrm{kin}\,\mathrm{[Myr]} =0.978\frac{d/2}{V_\mathrm{exp}}
\end{equation}
where $d$ is the diameter of the hole in pc and $V_\mathrm{exp}$ is in km\,s$^{-1}$. Since the expansion velocity was higher at earlier stages of the expansion, the kinetic age is an upper limit. Its statistical uncertainty is around 20\%.

\item The \ion{H}{1} column density ($N_\mathrm{HI}$) around the hole. The column density is given by:
\begin{equation}
N_\mathrm{HI}\,\mathrm{[cm^{-2}]}=1.823\times 10^{18}\sum T_\mathrm{B}\, dV
\end{equation}
where $T_\mathrm{B}$ is the brightness temperature in units of Kelvin, $V$ the velocity in km\,s$^{-1}$ and the summation is over the velocity channels which contain signal. The relation between brightness temperature and flux density at 21--cm wavelength is given by:
\begin{equation}
T_\mathrm{B}\,\mathrm{[K]}=\frac{S}{1.66\times 10^{-3}(B_\mathrm{maj}  B_\mathrm{min})}\,
\end{equation}
where $B_\mathrm{maj}$ and $B_\mathrm{min}$ are the FWHM of the beam respectively in arcsec and $S$ is the flux in mJy\,beam$^{-1}$. We calculated $T_\mathrm{B}$ from the average flux density (\textit{I}) around the hole which we converted to Kelvin using the above equation.

\item The midplane \ion{H}{1} volume density ($n_\mathrm{HI}$). This is given by:
\begin{equation}
n_\mathrm{HI}\,\mathrm{[cm^{-3}]}=\frac{N_\mathrm{HI}}{3.08\times 10^{18}\,l(r)}
\end{equation}
where $N_\mathrm{HI}$ is the column density in cm$^{-2}$ and $l$ is the effective thickness of the \ion{H}{1} layer in pc which is assumed to have a Gaussian profile with a $1 \sigma$ scale height $h$ (see Sec.~\ref{sec:Char}). Thus:
\begin{equation}
l(r)\,\mathrm{[pc]}=\frac{h\,\sqrt{8\ln 2}}{\cos i}
\end{equation}

\item The volume ($V$) of the hole. For holes that are contained within the disk (type 3) this is simply:
\begin{equation}
V\,\mathrm{[pc^3]} = (4/3) \pi (d/2)^{3}
\end{equation}
where $d$ is the diameter of the hole in pc as given in Eq. (3). Type 3 holes are intrinsically smaller (and younger) and for those it is justified to assume that their extent perpendicular to the disk is the geometric mean of the other two axes. As is shown {\em a posteriori} (see Section \ref{RelProp}) this is a very reasonable approximation given that their shape can not be greatly distorted during their short lifetime. For holes that are larger than the thickness of the \ion{H}{1} disk (almost all type 1 and type 2 holes) we assumed that these have taken a cylindrical shape and therefore the volume is:
\begin{equation}
V\,\mathrm{[pc^3]} = 2 \pi (d/2)^{2} \sqrt{8 \ln{2}}\,h
\end{equation}
where $h$ (in pc) is the $1 \sigma$ scale height and the term $ \sqrt{8 \ln{2}}\,h$ expresses the thickness of the disk. 

\item Indicative \ion{H}{1} mass ($M_\mathrm{HI}$). This is an estimate of the mass that filled the hole, assuming the hole is spherical and completely empty and that the gas had a uniform distribution with volume density $n_\mathrm{HI}$ before the creation of the hole. In addition, assuming that only a small fraction of the surrounding \ion{H}{1} is ionized, this value indicates the amount of \ion{H}{1} mass which is expected to have been swept up to form a rim for those holes not blown out of the disk. $M_\mathrm{HI}$ was calculated using: 
\begin{equation}
M_\mathrm{HI}\,\mathrm{[M_{\sun}]} = 2.9 \times 10^{5}\, n_\mathrm{HI}\, V
\end{equation}
where $n_\mathrm{HI}$ is the volume density of the gas around the hole in cm$^{-3}$ and $V$ its volume in pc$^{3}$. The uncertainty of $M_\mathrm{HI}$ is around 30\%.

\item The galactocentric distance (R) of the hole. This was calculated using:
\begin{equation}
R\,\mathrm{[pc]}=D\times10^{6}\sqrt{(x'')^{2}+(y'')^{2}} \,\mathrm{pc}
\end{equation}
where \textit{D} is in Mpc and
\begin{eqnarray}
x''=x\sin \theta +y \cos \theta  \\
y''=\frac{ y\sin \theta - x\cos \theta}{\cos i} 
\end{eqnarray}
and 
\begin{eqnarray}
x = (\alpha-\alpha_\mathrm{0}) \cos \delta_\mathrm{0} \\
y = \delta - \delta_\mathrm{0} 
\end{eqnarray}
where $\theta$ is the position angle of the galaxy defined by the direction of the receding major axis, $i$ its inclination, ($\alpha , \delta $) are the celestial coordinates of the centre of the hole and ($\alpha_\mathrm{0} , \delta_\mathrm{0} $) are the celestial coordinates of the centre of the galaxy (see Table~\ref{tbl-glob}).

\item The estimated energy requirements ($E_\mathrm{E}$) of the hole. \citet{che74} derived an equation for the energy requirements of an expanding shell, assuming this energy is provided by a single explosion: 
\begin{equation}
E_\mathrm{E}\,\mathrm{[erg]} = 5.3 \times 10^{43}\, n_\mathrm{o}^{1.12}\, (d/2)^{3.12}\, V_\mathrm{exp}^{1.4}
\end{equation}
where $V_\mathrm{exp}$ is the expansion velocity in km\,s$^{-1}$, $n_\mathrm{o}$ is the density of the ambient medium in particles\,cm$^{-3}$, and $d$ is the diameter of the hole in pc. As was done in the case of M\,31 \citep{bri81}, Ho II \citep{puc92}, IC\,2574 \citep{wal99} and DDO\,47 \citep{wal01}, $n_\mathrm{o}$ was replaced by $n_\mathrm{HI}$, which means we ignored contributions by He and H$_2$. In order to correct $E_\mathrm{E}$ for a He contribution, a factor of 1.5 would need to be applied. Due to the large number of assumptions, in Chevaliers equation, the uncertainty in $E_\mathrm{E}$ is of the order of 50\%. Despite the large uncertainty, for reasons of consistency with previous papers on this subject, and for want of a better alternative, we decided to derive energies using this equation.
\end{enumerate}

\section{Comparative analysis of the Holes}
\label{sec:Stats}
The results for each individual galaxy are presented in the Appendix, Tables~\ref{tbl-NGC628}--\ref{tbl-NGC7793} and Figures~\ref{NGC628-map}--\ref{NGC7793-map}. Table~\ref{prop-sum} provides a basic summary of the properties of the holes on a per galaxy basis and lists, among other things, the range of \ion{H}{1} hole diameters, expansion velocities, and kinetic ages. The number of holes detected in each galaxy varies substantially. This variation is partly due to the difference in spatial resolution at which each galaxy was observed. Specifically, the number of type 3 holes detected is greatly affected by resolution. As these holes are fully contained within the \ion{H}{1} disk their diameter is typically less than 300 pc, the resolution limit in some of the more distant galaxies. 
In the case of spiral galaxies the lack of \ion{H}{1} in between the spiral arms sometimes appears as an \ion{H}{1} hole. These inter-arm regions in general do not constitute genuine holes and therefore have not been included in any comparative analysis. This is also the case for features that resemble holes in \textit{pV} diagrams but are the result of a warped disk or for holes which are superpositions of 2 or more holes.  

\begin{deluxetable}{lcrrrrrr}
\tabletypesize{\scriptsize}
\tablecaption{Summary of basic properties of \ion{H}{1} holes across the sample of galaxies \label{prop-sum}}
\tablewidth{0pt}
\tablehead{
\colhead{Name} & \colhead{No} & \colhead{d} & \colhead{$V_\mathrm{exp}$} & \colhead{$t_\mathrm{kin}$} & \colhead{type 1} & \colhead{type 2} & \colhead{type 3}\\
\colhead{} & \colhead{} & \colhead{(kpc)} & \colhead{(km\,s$^{-1}$)} & \colhead{(Myr)} & \colhead{(\%)} & \colhead{(\%)} & \colhead{(\%)}}
\startdata
NGC\,628&102&0.32-1.94&5-21&14-135&70&25&5\\
NGC\,2366&26&0.21-0.69&9-24&6-38&54&23&23\\
NGC\,2403&173&0.10-1.78&6-32&2-109&62&24&14\\
Holmberg\,II&39&0.26-2.11&7-20&10-147&64&13&23\\
DDO\,53&3&0.18-0.36&8-10&9-22&33&33&33 \\
NGC\,2841&15&0.89-1.67&9-28&20-91&93&0&7\\
Holmberg\,I&6&0.19-0.74&6-16&6-61&50&17&33\\
NGC\,2976&13&0.13-0.24&9-23&3-13&62&23&15\\
NGC\,3031&306&0.09-0.57&8-20&3-35&43&34&23\\
NGC\,3184&39&0.48-1.64&7-20&21-114&87&13&0\\
IC\,2574&27&0.32-2.02&5-36&9-141&55&15&30\\
NGC\,3521&13&0.60-1.20&10-16&28-59&92&8&0\\
NGC\,3627&17&0.36-0.74&12-30&6-30&59&29&12\\
NGC\,4214&56&0.11-1.23&4-25&4-100&82&11&7\\
NGC\,4449&20&0.23-0.90&10-30&4-44&70&15&15\\
NGC\,4736&23&0.21-0.45&9-30&6-22&48&35&17\\
DDO\,154&9&0.27-0.68&6-13&10-42&67&22&11\\
NGC\,5194&47&0.29-1.34&8-32&8-82&60&38&2\\
NGC\,6946&56&0.20-2.14&8-21&8-131&84&9&7\\
NGC\,7793&27&0.10-1.06&7-16&10-65&59&30&11\\
\enddata
\tablecomments{Col 1: Galaxy name; Col 2: number of HI holes; Cols 3--5: the minimum and maximum, indicating the range of hole diameters, hole expansion velocities, and kinetic age, respectively; Cols 6--8: the percentage of type 1, 2, and 3 holes detected.}
\end{deluxetable}

For the first time we are now able to undertake a comparative study of the small scale structure of the ISM, in a systematic way and across Hubble type. The homogeneity of our sample allows us to perform a number of analyses which we present here in three sections. The first section includes a comparison of the properties of the holes in each galaxy; the second section compares these properties across the galaxies in our sample; the third section searches for global trends and correlations with, e.g., Hubble type, luminosity, etc. We then proceed to test the hypothesis that \ion{H}{1} holes are compatible with being a direct result of massive star formation. In what follows we decided to exclude the dwarf galaxy DDO\,53 as this has only 3 detected \ion{H}{1} holes. We also exclude those few \ion{H}{1} holes which in the discussion of the individual galaxies were flagged as being superpositions or potentially not genuine holes. These holes are marked with an asterisk in Tables~\ref{tbl-NGC628}--\ref{tbl-NGC7793}.

\subsection{Overall Properties of the Holes}
\label{OvProp}

\begin{figure}
\begin{center}
\includegraphics[angle=0,scale=.69]{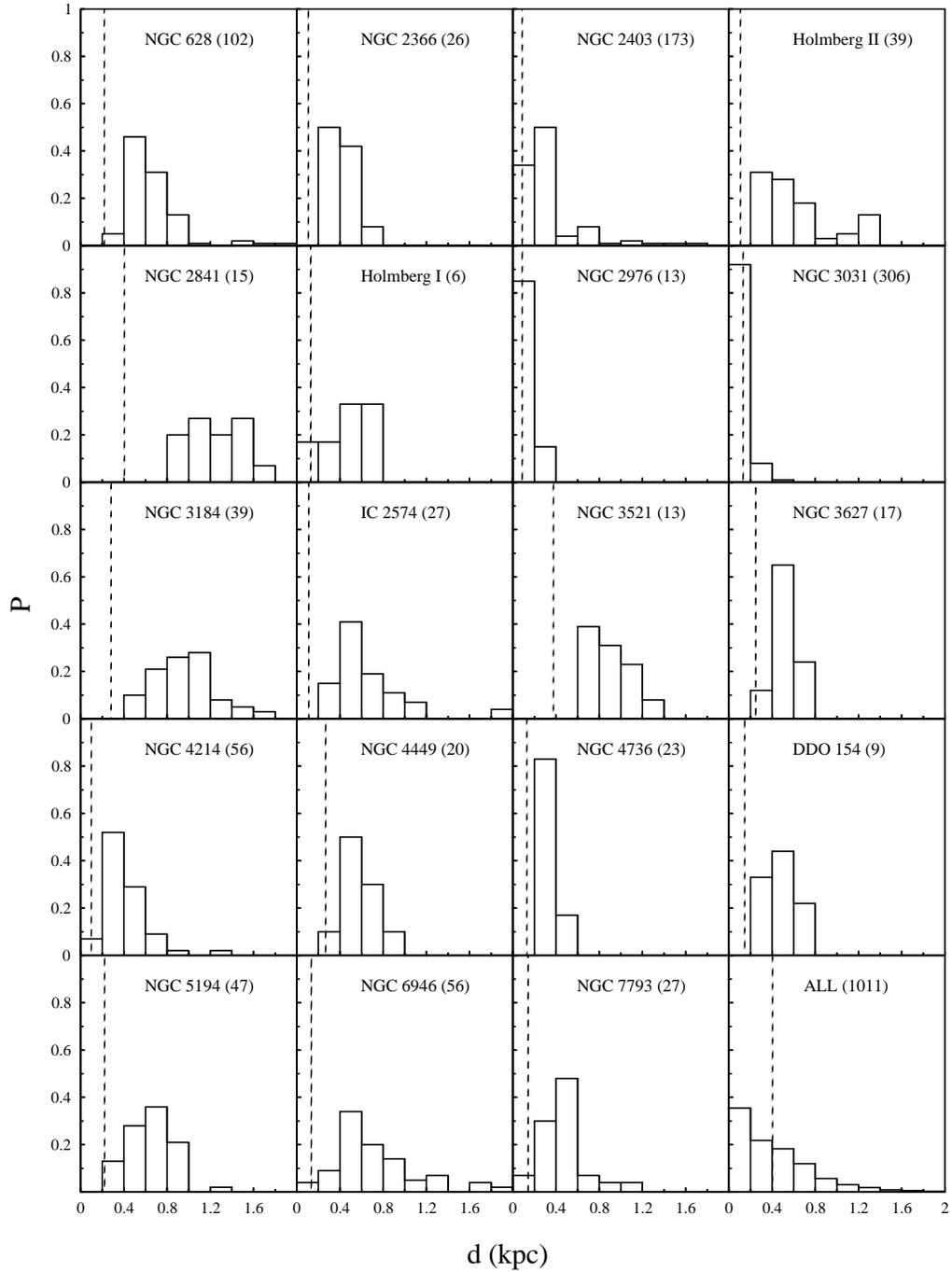}
\end{center}
\caption{Relative number distribution of the diameters of the holes. The vertical dashed line indicates the lower limit set by the linear resolution. Each panel lists the galaxy it refers to and the last panel shows all holes as well as the worst linear resolution across the entire sample (dashed line). The number in parentheses indicates the total number of holes represented by the histogram. \label{dd}}
\end{figure}

The set of figures we will describe next shows the number distributions of the observed and derived properties of the holes. We start with showing the diameter distribution of the holes in order to investigate if there are any obvious trends regarding their size distribution, both within a galaxy as well as from target to target. The result is shown in Fig.~\ref{dd} which shows a mosaic of 20 panels, each panel showing a histogram representing the relative (normalized) frequency distribution as a function of diameter of each of the 19 targets, plus a final panel showing the distribution summing over all targets. The bins are 200\,pc in size. There are obvious differences among the galaxies. However, in this and subsequent figures one has to guard against selection effects. The absence of small diameter holes in, e.g. NGC\,2841, NGC\,3521 and NGC\,3184 is due to their larger relative distance which sets a lower limit to the linear resolution. We therefore add a vertical dotted line to indicate the resolution limit for each galaxy. Also there are other factors influencing the detection of holes such as the signal to noise ratio, and the inclination of the object under study (e.g. NGC\,2841 and NGC\,3521). The large diameter end of the distributions varies greatly among the galaxies. In three cases, NGC\,2976, NGC\,3031, and NGC\,4736, we see a sharp cutoff at around 400 pc which, at first glance, does not seem to correlate with galaxy type or any other global property. On the other hand it appears that Scd type galaxies tend to have small numbers of large ($> 1$ kpc) holes in their \ion{H}{1} distribution. We return to this below.

\begin{figure}
\begin{center}
\includegraphics[angle=0,scale=.69]{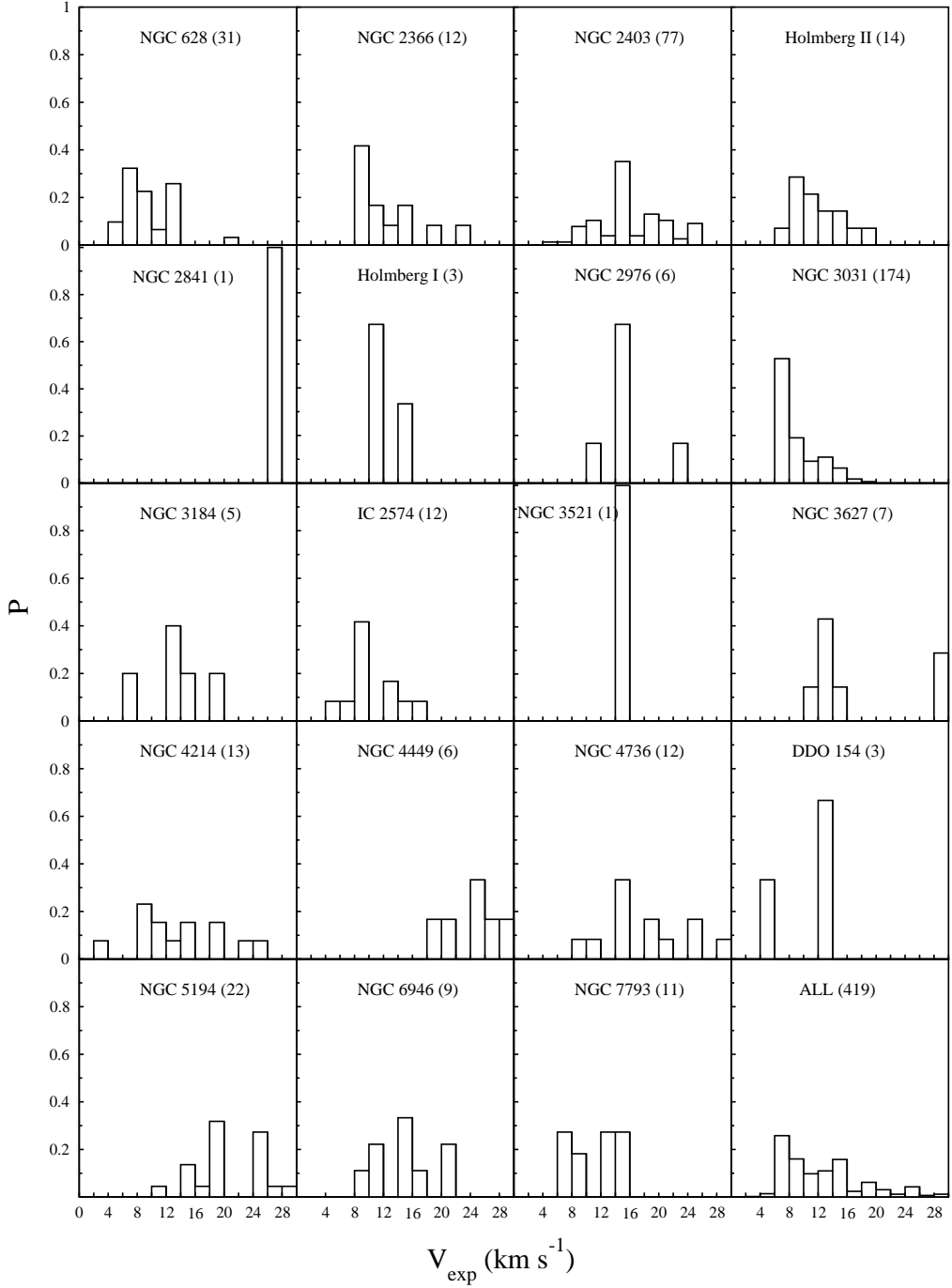}
\end{center}
\caption{Relative number distribution of the expansion velocities of the expanding holes. Each panel lists the galaxy it refers to; the number in parentheses indicates the total number of holes represented by the histogram. The sharp cut--off point at $\sim 5 \,\mathrm{km\,s^{-1}}$ is due to our velocity resolution.\label{dv}}
\end{figure}

The relative number distribution of the expansion velocities is plotted in Figure~\ref{dv}. The bins correspond to $2\,\mathrm{km\,s^{-1}}$ and only holes for which an expansion velocity could be measured are plotted. The distributions appear to vary significantly among the galaxies mainly because of small number statistics. 

\begin{figure}
\begin{center}
\includegraphics[angle=0,scale=.69]{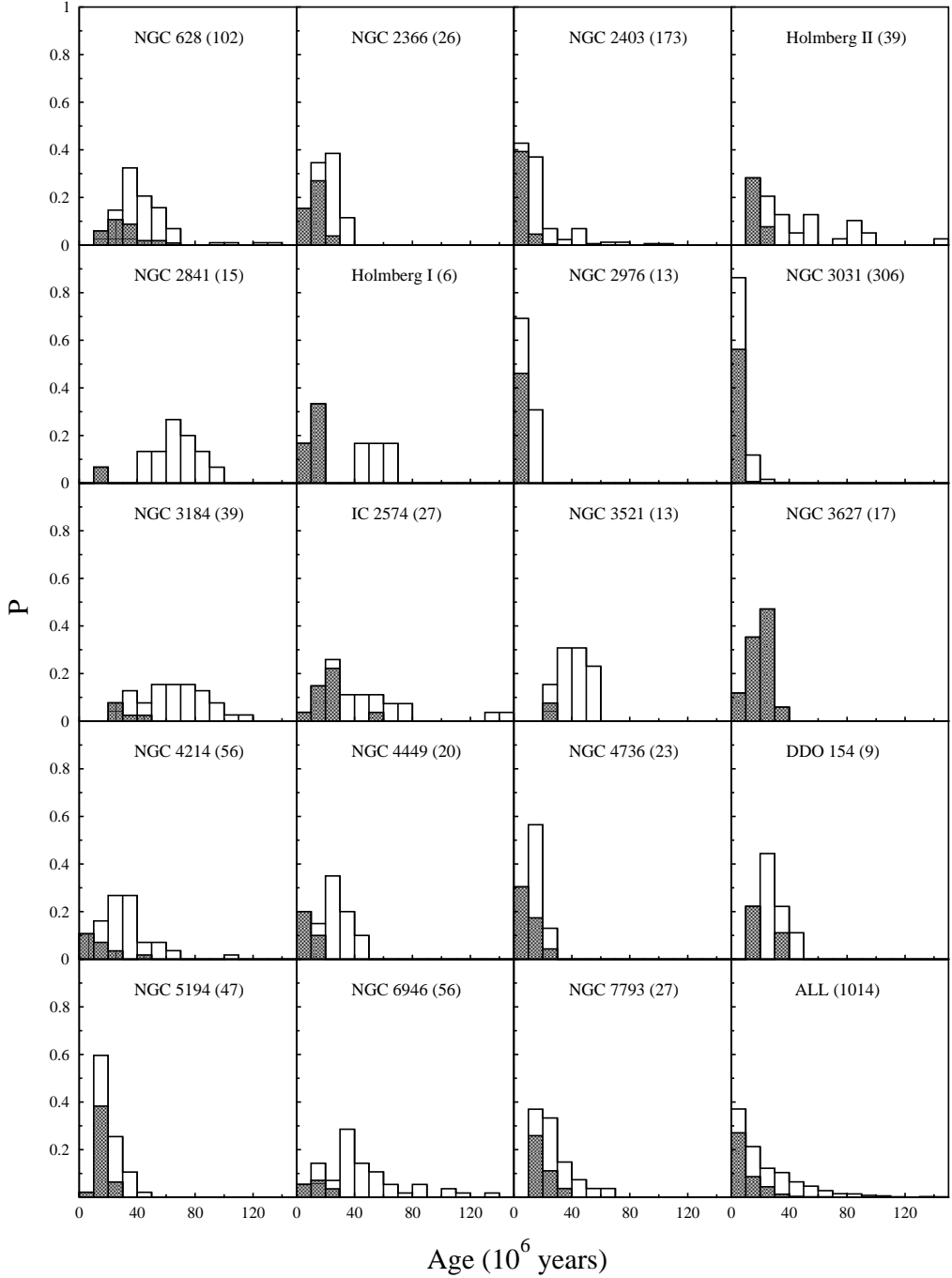}
\end{center}
\caption{Relative number distribution of the kinetic ages of the holes. The shaded areas represent the fraction of holes for which an expansion velocity was measured (types 2 and 3). In the other cases we assumed an expansion velocity equal to the velocity dispersion. Each panel lists the galaxy it refers to; the number in parentheses indicates the total number of holes represented by the histogram.\label{dt}}
\end{figure}

Figure~\ref{dt} shows the relative number distribution of the kinetic ages of the holes in the 19 galaxies in a format similar to Fig.~\ref{dd}. The bins measure $10^7$ yr. The shaded histograms correspond to those holes for which an expansion velocity could be measured, i.e. types 2 and 3. For type 1 holes, which are likely to have stalled, we adopted a lower limit to the expansion velocity which was set to the one--dimensional velocity dispersion of the gas. The kinetic age distributions are similar to those of the diameters of the holes which is to be expected since the kinetic age is derived from the diameter and the expansion velocity which, across the targets, has a small range. M\,81 (NGC\,3031) shows a quite striking distribution. As mentioned in the previous paragraph, this galaxy predominantly has small holes. Their ages are all below 40 Myr. One possible explanation is that M\,81 is actively interacting with M\,82 and NGC\,3077: numerical models by \cite{yun99} provide an estimate of $2.2 \times 10^8$\,yr for the most recent perigalacticon and it is conceivable that this interaction eventually set off a sequence of events which led to a phase of enhanced star formation $0.5 \times 10^8$\,yr ago, leading to the current abundance of small, young \ion{H}{1} holes. Similarly, NGC\,2976 appears to have had a star formation episode $1-2\times10^7$\,yr ago triggered by gas infall from an intergalactic cloud generated during encounters between members of the M\,81 group \citep{bro92}.

\begin{figure}
\begin{center}
\includegraphics[angle=0,scale=.69]{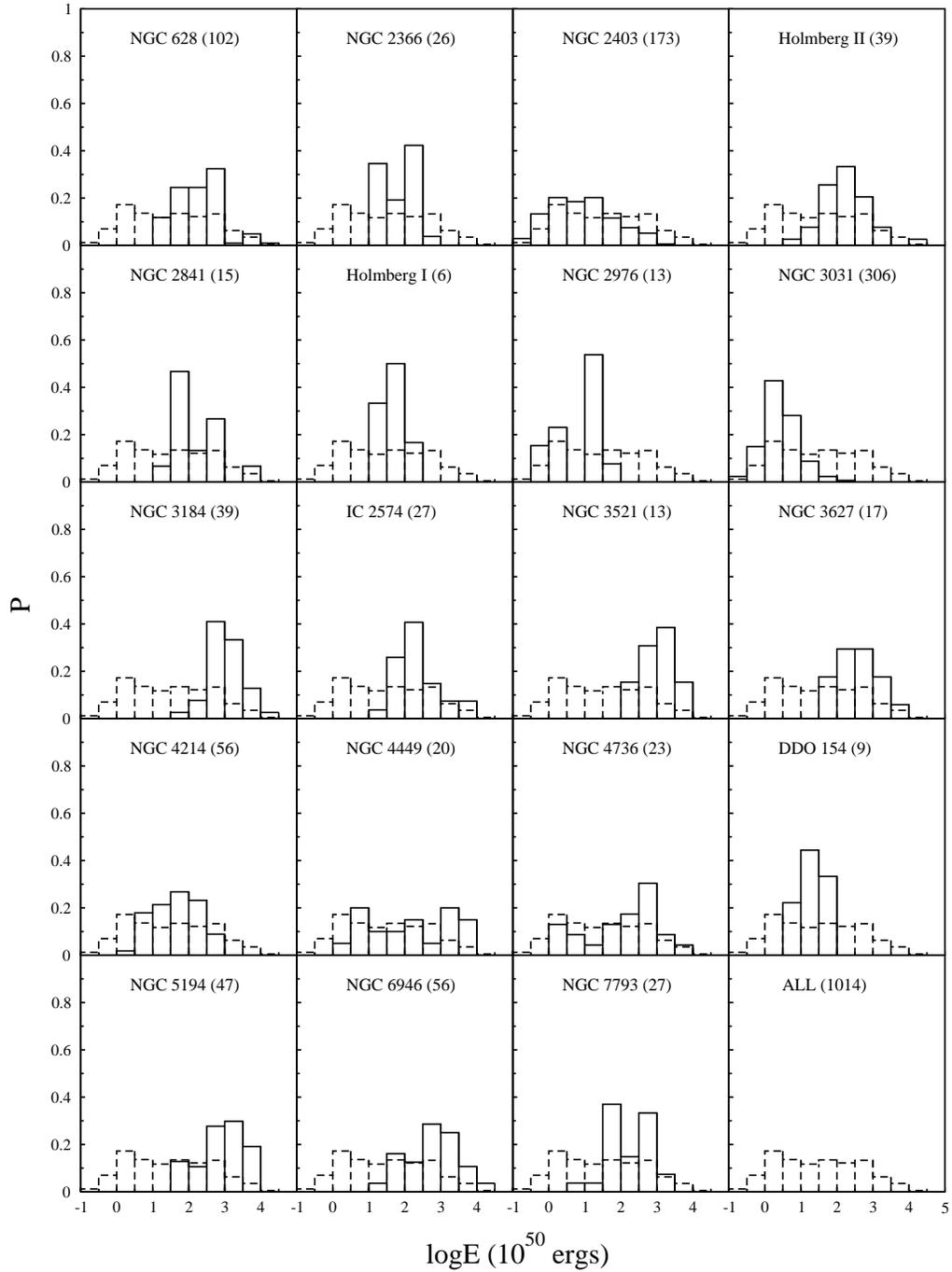}
\end{center}
\caption{Relative number distribution of the energy requirements to create the holes. Note that the horizontal axis is binned logarithmically. The bottom right panel shows the distribution of all holes combined. This dashed histogram is repeated for reference in each of the other panels. Each panel lists the galaxy it refers to; the number in parentheses indicates the total number of holes represented by the histogram.
\label{de}}
\end{figure}

The relative number distribution of the characteristic energies required to create the holes is shown in Fig.~\ref{de}. The energies are binned logarithmically. It is evident that nearly all distributions fall in a range that spans 3 orders of magnitude. The shift towards higher energies for the galaxies NGC\,2841, NGC\,3184 and NGC\,3521 is a selection effect due to their relative large distance which reduces the linear resolution and sensitivity to pick up small holes. In the case of NGC\,3627 the shift towards higher energies is due to the fact that this galaxy has the highest velocity dispersion, which is likely driven by the central bar, and hence the energies of all type 1 holes have probably been overestimated.

\begin{figure}
\begin{center}
\includegraphics[angle=0,scale=.69]{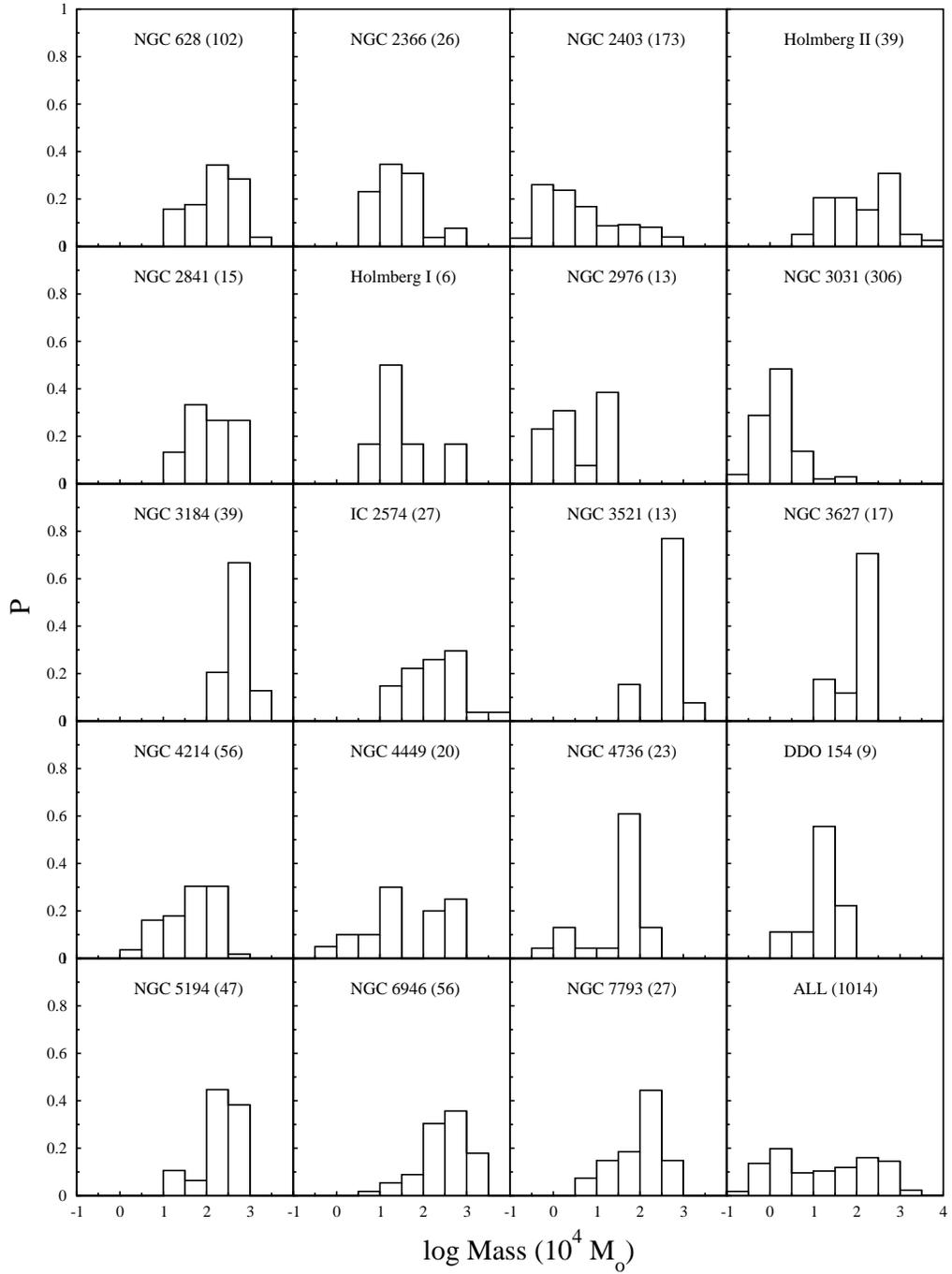}
\end{center}
\caption{Relative number distribution of the indicative \ion{H}{1} masses of the holes in logarithmic bins. Each panel lists the galaxy it refers to; the number in parentheses indicates the total number of holes represented by the histogram. \label{dm}}
\end{figure}

Figure~\ref{dm} shows the relative number distribution of the indicative \ion{H}{1} masses, binned logarithmically. The shape of these distributions is dominated by the diameter distribution of the holes, as the volume densities are showing a much smaller range of values ($2\sigma$ from $\sim 0.1$ to $\sim 0.7\,\mathrm{cm^{-3}}$). The indicative masses span a range of ($2\times 10^{3}\, M_{\odot} -3\times 10^{7}\, M_{\odot}$). Apart from NGC\,2403, NGC\,2976 and NGC\,3031 which have their distributions skewed towards smaller mass values, all other galaxies cover the same range.

The radial location of the holes in the disk of the host galaxy is illustrated in Figure~\ref{fig:rgc-r25}. The horizontal axis represents the galactocentric distance normalized to the maximum extent of the \ion{H}{1} disk, $R_\mathrm{max}$ (given in Tab. \ref{tbl-char}), in bins of size 0.1$R_\mathrm{max}$. From this figure it is evident that wherever there is \ion{H}{1}, holes are found, even in the outskirts of galaxies. The vertical line in the figure indicates the location of $R_{25}$, confirming the finding that holes are found well beyond the main optical body of a galaxy. The number of holes \emph{per area} has its peak near the center, though, as can be seen in Fig.~\ref{fig:Nperkpc2} which shows the log of the number of holes per kpc$^2$ as a function of $R/R_\mathrm{max}$. The surface number density of \ion{H}{1} holes is fairly constant or slightly drops out to $R_{25}$, beyond which radius it drops significantly in most of the galaxies.

Finally, Figure~\ref{fig:axial} illustrates the relative number distribution of the axial ratio of the holes in each galaxy. It appears that dwarf galaxies have rounder holes, on average, than spirals (the mean axial ratio for dwarfs is 0.80 and 0.71 for spirals). This was first observed in Holmberg\,II by \citet{puc92} and explained by \citet{wal99} and \citet{bri02} as follows. Within $R_{25}$ the \ion{H}{1} disk in a dwarf is intrinsically thicker than in spirals (see Table~\ref{tbl-char}) which prevents holes from blowing out at an early stage. Holes will therefore remain intrinsically spherical and the apparent axial ratio of an intrinsically spherical hole is not affected by the inclination of the host galaxy. Holes in spiral galaxies will tend to break out of the disk, leaving an approximately circular feature in the plane of the galaxy which, when seen under an angle, will appear elliptical. As will be discussed below another possible explanation for this trend is the effect that shear plays on the appearance of holes.

\subsection{Relations between Hole Properties}
\label{RelProp}

\begin{figure}
\begin{center}
\includegraphics[angle=0,scale=.69]{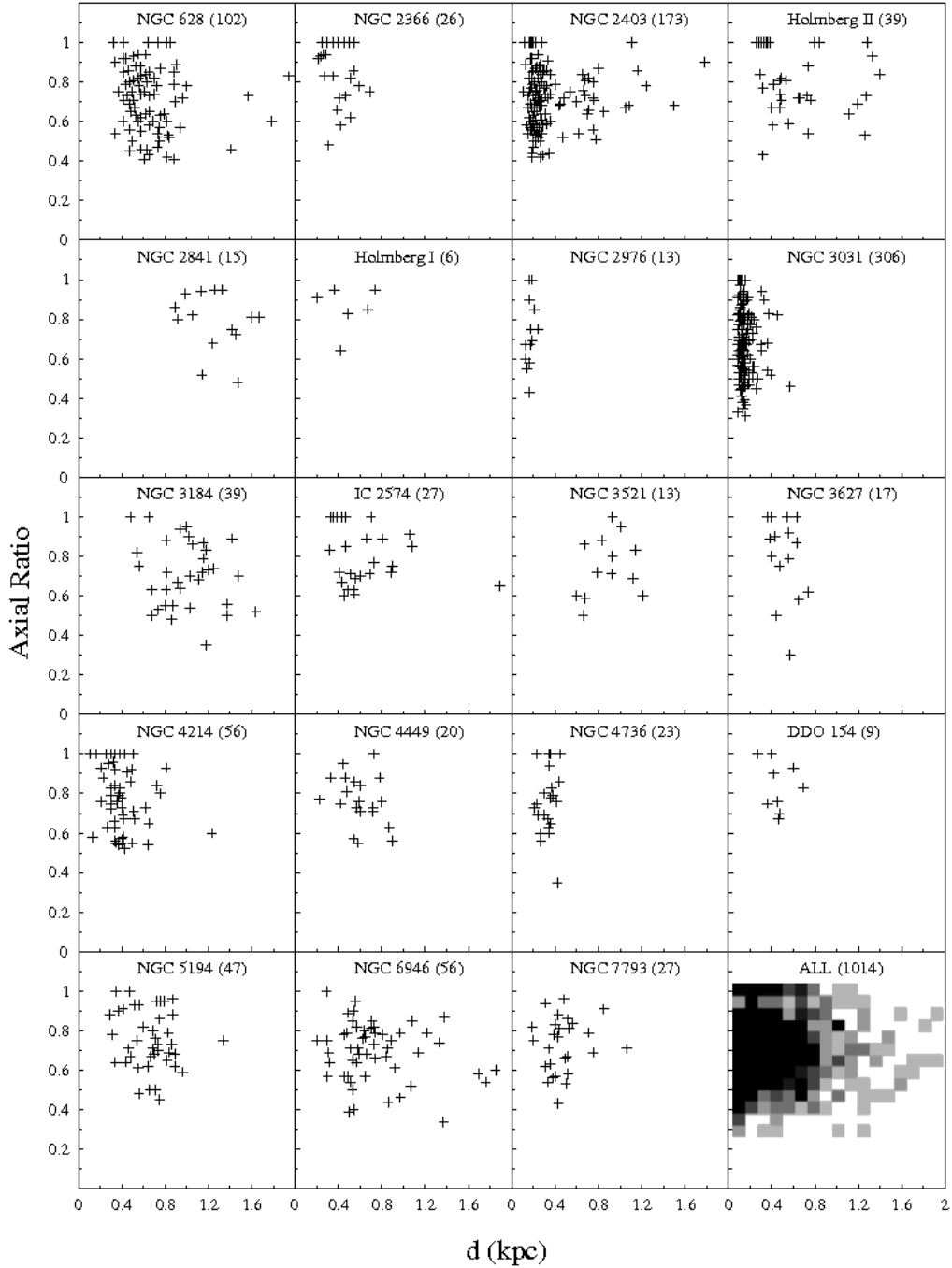}
\end{center}
\caption{Axial ratio vs.\ diameter of the \ion{H}{1} holes. The bottom right panel shows the accumulated result for all galaxies where, rather than plotting individual values we present the data binned and in the form of a density plot. Each panel lists the galaxy it refers to; the number in parentheses indicates the total number of holes represented by the histogram.\label{fig:axial-d}}
\end{figure}

The trend that dwarf galaxies tend to have rounder holes on average is also evident in Figure~\ref{fig:axial-d} which presents the axial ratio vs.\ the diameter of the holes. The last panel in this figure shows the entire sample of holes from which can be seen that there are few, if any, round holes larger than 1 kpc. For holes smaller than 1 kpc we see that the axial ratio takes the whole range of values indicative of the different factors that determine the axial ratio. This is the first of the next series of figures where we search for correlations between two observed and/or derived properties among the holes detected in each of the 19 targets. 

\begin{figure}
\begin{center}
\includegraphics[angle=0,scale=.69]{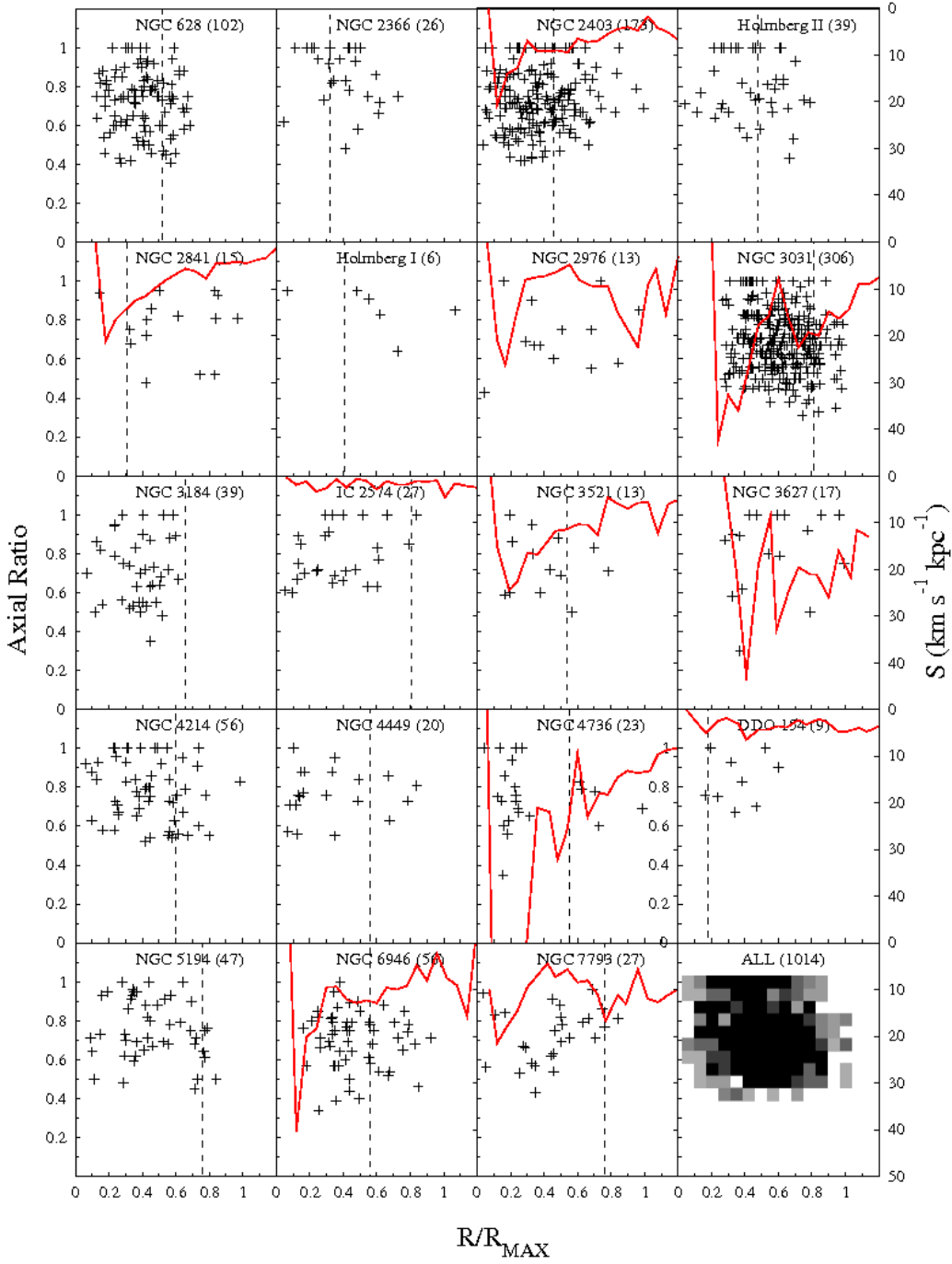}
\end{center}
\caption{Axial ratio vs.\ radial distribution normalized to $R_\mathrm{max}$ with shear overplotted (red line, axis on the right hand side of the plots and maximum value on the bottom). The dashed vertical line indicates for each galaxy the corresponding location of $R_{25}$. The bottom right panel shows the accumulated result for all galaxies where, rather than plotting individual values we present the data binned and in the form of a density plot. Each panel lists the galaxy it refers to; the number in parentheses indicates the total number of holes represented by the histogram. \label{fig:axial-rgc}}
\end{figure}

\begin{figure}
\begin{center}
\includegraphics[angle=0,scale=.69]{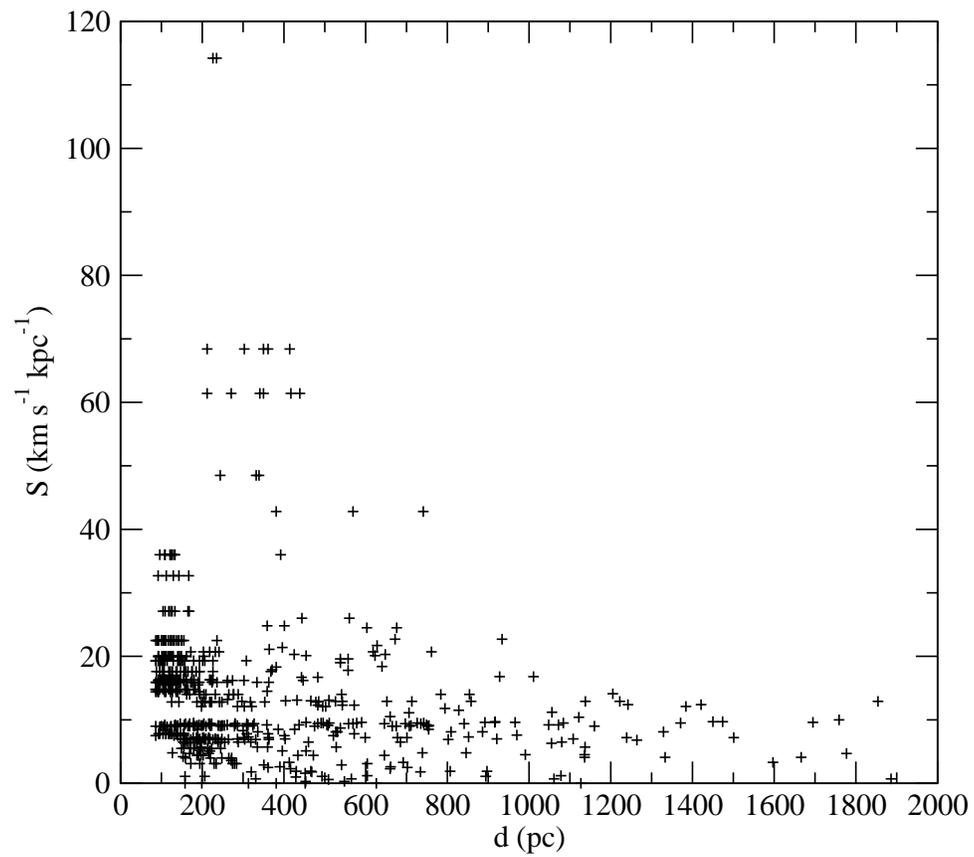}
\end{center}
\caption{Shear vs.\ diameter of the holes for galaxies for which shear could be measured. \label{fig:Shear-d}}
\end{figure}

Figure~\ref{fig:axial-rgc} is a scatter diagram of axial ratio vs.\ galactocentric radius. Shear at a point $R_o$ with respect to a neighbouring point at distance $\Delta R$ in a differentially rotating disk is given by:

\begin{equation}
S[\mathrm{km\,s^{-1}\,kpc^{-1}}]=\frac{1}{2}(\frac{V_o}{R_o}-\frac{\Delta V}{\Delta R})
\end{equation}

where $V_o$ is the velocity at $R_o$, $\Delta R$ is the distance between the points and $\Delta V$ the difference in velocity between these two points. The last term in equation 18 is zero for a flat rotation curve ($\Delta V=0$). Using the rotation curves of galaxies (where applicable) we calculated values for shear between points separated by $\sim  0.05 \% \,R_\mathrm{max} $ as illustrated in Fig.~\ref{fig:axial-rgc} (red line). The shear gradient is typically 10--50 km\,s$^{-1}$\,kpc$^{-1}$, the upper boundary being typical for Sa galaxies, later types reaching half this value. This is enough to turn an originally circular structure of 500 pc diameter, into an elliptical feature with axial ratio of $\sim 0.5$ within 50 Myr. The range in shear gradient listed is what one expects to find between the outer and inner regions of a spiral galaxy, the inner regions having the highest values (not considering the very inner solid body part). Most dwarf galaxies are dominated by solid body rotation, so shear is absent. One would therefore na\"ively expect holes at larger radii to be rounder than holes at smaller radii. However no such trend is apparent in the galaxies shown in Fig.~\ref{fig:axial-rgc}. This null result does not imply, though, that shear is not important in spirals. Holes are supposedly continuously formed, and more so within $R_{25}$, which means that there will be many young, and therefore round \ion{H}{1} holes populating the inner regions of galaxies. In the outer regions, some holes might be sufficiently long--lived for the more modest shear there to have an observable effect. One would therefore expect a relation of axial ratio with kinetic age in the sense that on average older holes in shear dominated, i.e., spiral galaxies, are less round than younger ones \citep{sil96,elm02}. This is investigated in the next figure, Fig.~\ref{fig:t-axial}. Except for some of the dwarf galaxies, where shear is not that important, only 5\% of holes older than 50 Myr are round (average axial ratio is 0.70). In the case of younger holes, the entire range of axial ratio is found, as expected, as one gets a mix of young, round holes in low shear environments with elliptical ones that find themselves in regions of high shear. The destructive role of shear is also illustrated in Fig.~\ref{fig:Shear-d} where the diameter of holes is plotted against shear for galaxies for which it could be measured. There is a clear negative correlation between shear and the diameter of holes i.e. in environments where shear is higher than $\sim$ 15 km\,s$^{-1}$\,kpc$^{-1}$ no holes larger than 1 kpc were found. 

It is not just shear which defines the observed axial ratio of an \ion{H}{1} hole. The axial ratio of those holes which have blown out and those that have lost their initial spherical shape due to shear depends crucially on viewing angle. The inclination should cause the major axis of holes to be aligned with the major axis of the disk. We also know that holes do not live in isolation and that two neighbouring spherical holes can merge to form a larger elliptical one. All these effects will tend to wash out whatever trend of axial ratio there might be with radius. However, there are some conclusions which can be drawn from Figs.~\ref{fig:axial}--\ref{fig:t-axial}: i) \ion{H}{1} holes in dwarfs are rounder than in spirals (see Sec.~\ref{GlobProp}); ii) blow--out holes in spirals are not circular as we do not observe any alignment between the major axis of the holes and the major axis of the disk (if anything, cursory inspection of the figures in the Appendix suggests the holes align with the spiral arm structure); iii) there is a sharp cut--off to the axial ratio at a level of 0.4. Regarding iii), in the identification process there was no requirement for holes to be less flattened than a certain value. The sharp lower envelope at 0.4 suggests then, that structures more elongated than that become difficult to identify as holes.

\begin{figure}
\begin{center}
\includegraphics[angle=0,scale=.69]{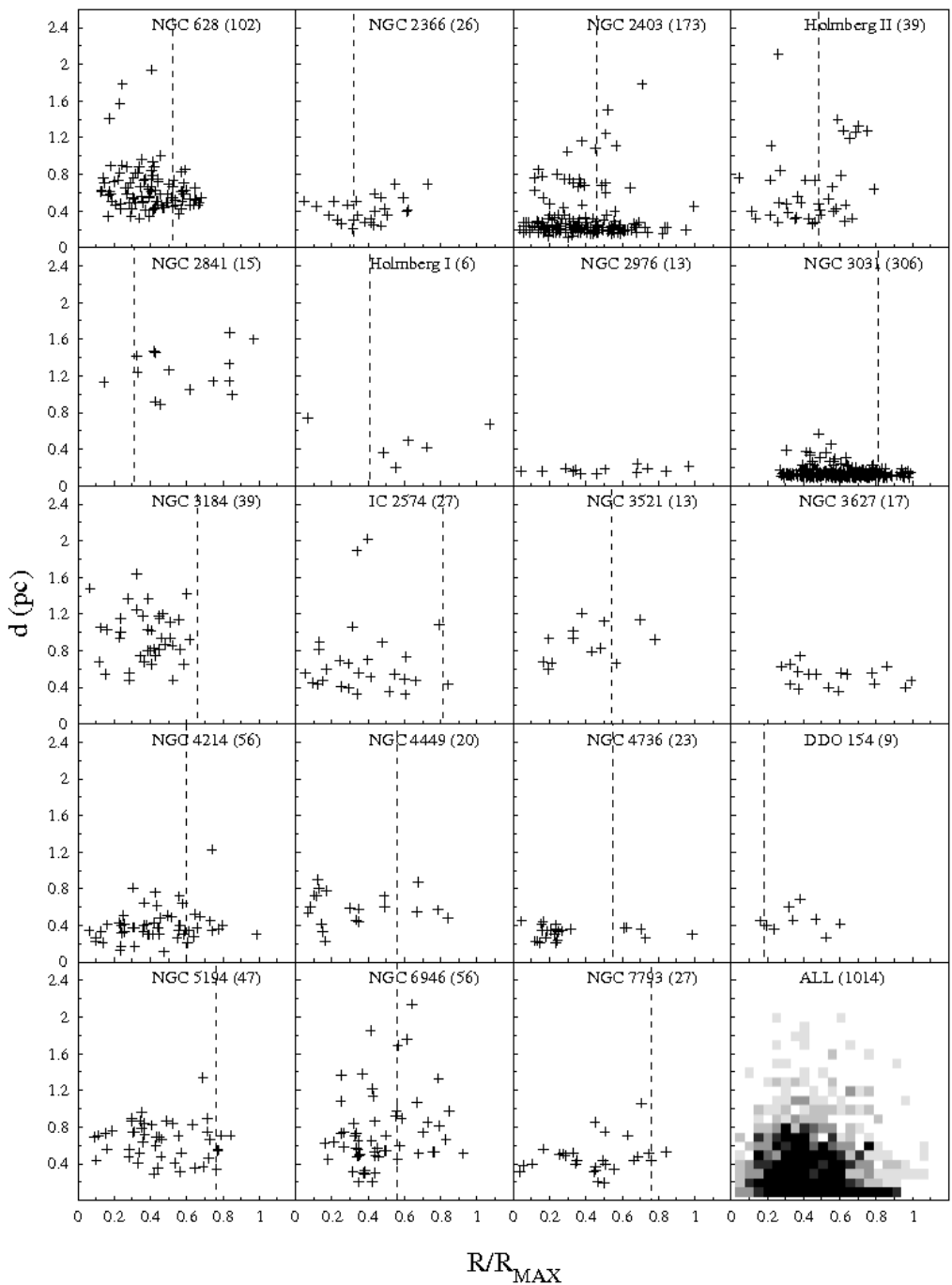}
\end{center}
\caption{As Fig.~\ref{fig:axial-rgc} but showing diameter vs.\ normalized radial distribution. The dashed vertical line indicates for each galaxy the corresponding location of $R_{25}$. The bottom right panel shows the accumulated result for all galaxies where, rather than plotting individual values we present the data binned and in the form of a density plot. \label{fig:d-rgc}}
\end{figure}

The diameter of a hole versus its position in the host galaxy is shown in Figure~\ref{fig:d-rgc} where the horizontal axis represents the galactocentric distance as a fraction of the \ion{H}{1} disk radius. In the case of dwarf galaxies there is a tendency for large holes to be located at larger galactocentric radii. In spiral galaxies, no such trend is seen. This is somewhat contrary to expectations. As we have explained in Sec.~\ref{sec:Char}, galaxy disks flare at large radius. One would therefore expect holes to be able to grow to larger diameters before breaking out and subsequently stalling, considering that the ambient density also drops with radius. The lack of finding a corresponding trend in the diameters of holes could be due to a variety of factors, one being that there is a growing incompleteness of finding holes in the very low \ion{H}{1} surface brightness outer regions of spirals. Alternatively, we know that star formation is inefficient at large radii \citep[e.g.,][]{ken89,mar01,sch04,tho06,big08,ler08} and consequently the number of stars more massive than 8\, M$_{\odot}$ which are formed cospatially decreases, reducing the total energy deposited after their demise as supernovae, resulting in holes of more modest diameter. Another possible explanation is that the ambient density drops significantly perpendicular to the disk of the galaxy and therefore holes will tend to be elongated in this direction. Consequently, even though these chimney--like structures will be intrinsically larger in the direction perpendicular to the disk, when viewed inclined, they might appear to have a lower ellipticity and mimic spherical features. 

\begin{figure}
\begin{center}
\includegraphics[angle=0,scale=.69]{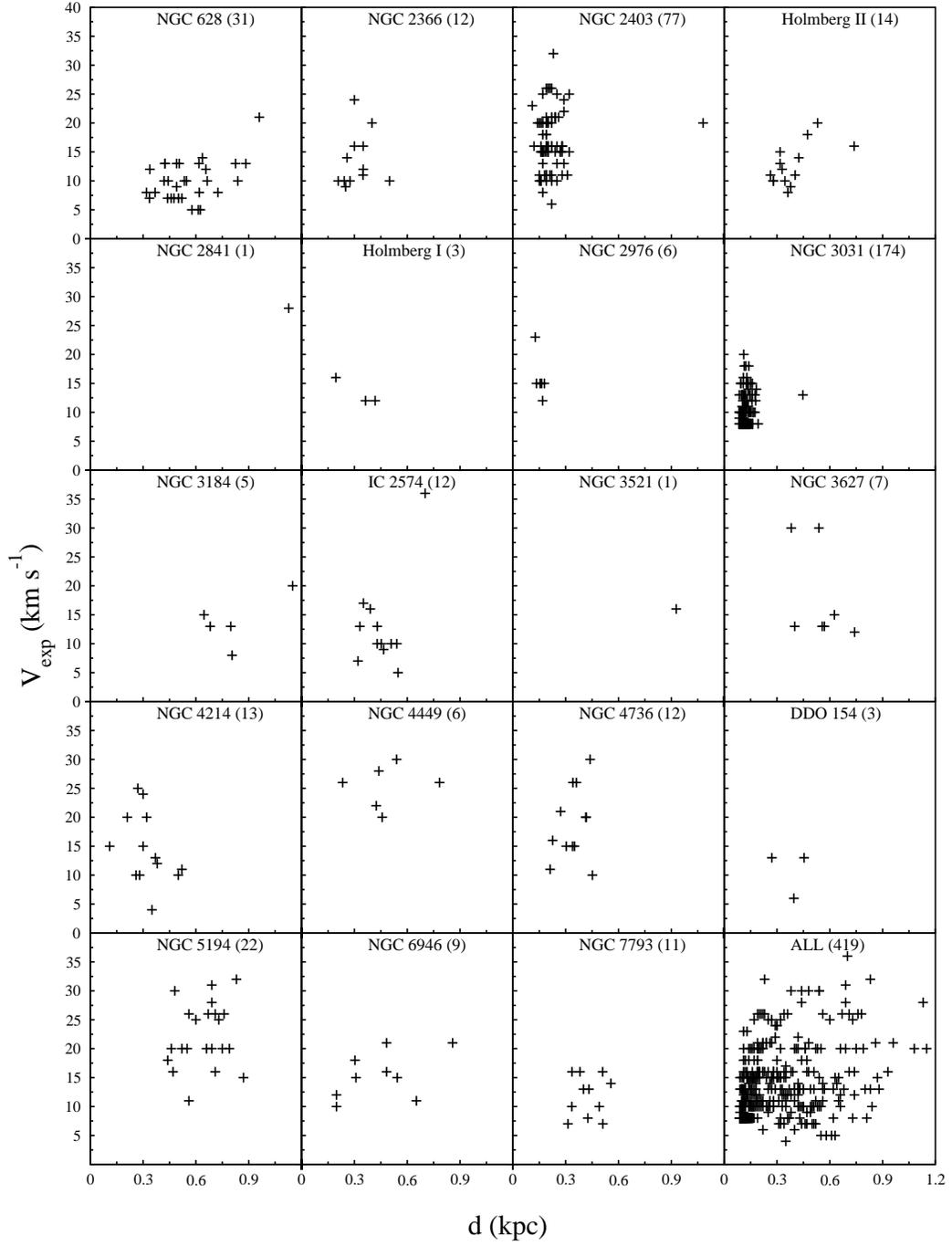}
\end{center}
\caption{Diameter vs.\ expansion velocity of the expanding \ion{H}{1} holes.\label{fig:d-v}}
\end{figure}

\begin{figure}
\begin{center}
\includegraphics[angle=0,scale=.50]{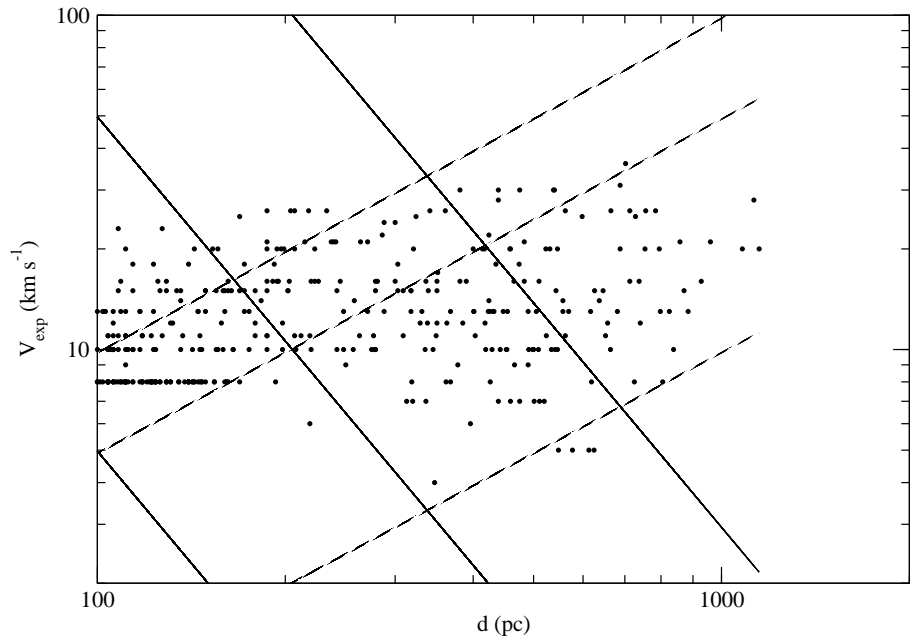}
\end{center}
\caption{Diameter vs.\ expansion velocity of the expanding \ion{H}{1} holes across the entire sample of galaxies. The dotted lines represent kinetic ages of 5 (top line), 10 and 50 (bottom line) Myr. The solid lines represent energies of $10^{49}$ (leftmost line), $10^{50}$ and $10^{51}$ (outermost line) erg.\label{fig:A-d-v}}
\end{figure}

Figure~\ref{fig:d-v} shows the diameter of a hole versus its expansion velocity for those holes for which we could measure this, i.e., types 2 and 3. Contrary to the findings by \citet{sta07} we do not find a trend of increasing expansion velocity with increasing size. However, \citet{sta07} were able to detect holes down to 15 pc where this trend is strongest. We should also like to note that their result is based on just one interacting system, the LMC/SMC. There are examples in Figure~\ref{fig:d-v} of individual galaxies which show a similar trend, such as Ho~II and NGC\,4736. However, this trend does not dominate across galaxies. This is also illustrated in Fig.~\ref{fig:A-d-v} where we plot on a log--log scale expansion velocity versus hole diameter for all expanding holes across the entire sample of galaxies. In this figure the straight lines correspond to lines of constant age, drawn at (0.5, 1, and $5) \times 10^7$\,yr. Assuming Chevalier's single blast approximation and, for simplicity a constant ambient density of $n_\circ = 1.0$\,cm$^{-3}$, we can draw curves of constant required energy. There are two competing mechanisms which both contribute to the way the points are distributed in Figs.~\ref{fig:d-v} and~\ref{fig:A-d-v}. Once a supernova has gone off, the ejecta which initially move at high speeds will slow down as they interact with the surrounding circumstellar medium and ISM, eventually entering the snowplough phase which is when we tend to identify them as \ion{H}{1} holes. One would therefore expect larger holes to have lower expansion velocities. On the other hand, as long as stars from the parent OB association go off as SNe, energy is continuously deposited and, if the expanding shell has not yet suffered blow--out, will maintain an over--pressure within the shell which in turn will keep up an accelerated flow. The latter mechanism is more likely to dominate in the smaller, younger holes. It is then interesting to note that the distribution of points in our Fig.~\ref{fig:A-d-v} agrees with the corresponding graph in \citet{sta07} if one restricts the comparison to holes larger than 100\,pc. In other words we hypothesize that the relation found by these authors is possibly a consequence of the smaller holes still harboring supernova activity.

\begin{figure}
\begin{center}
\includegraphics[angle=0,scale=.69]{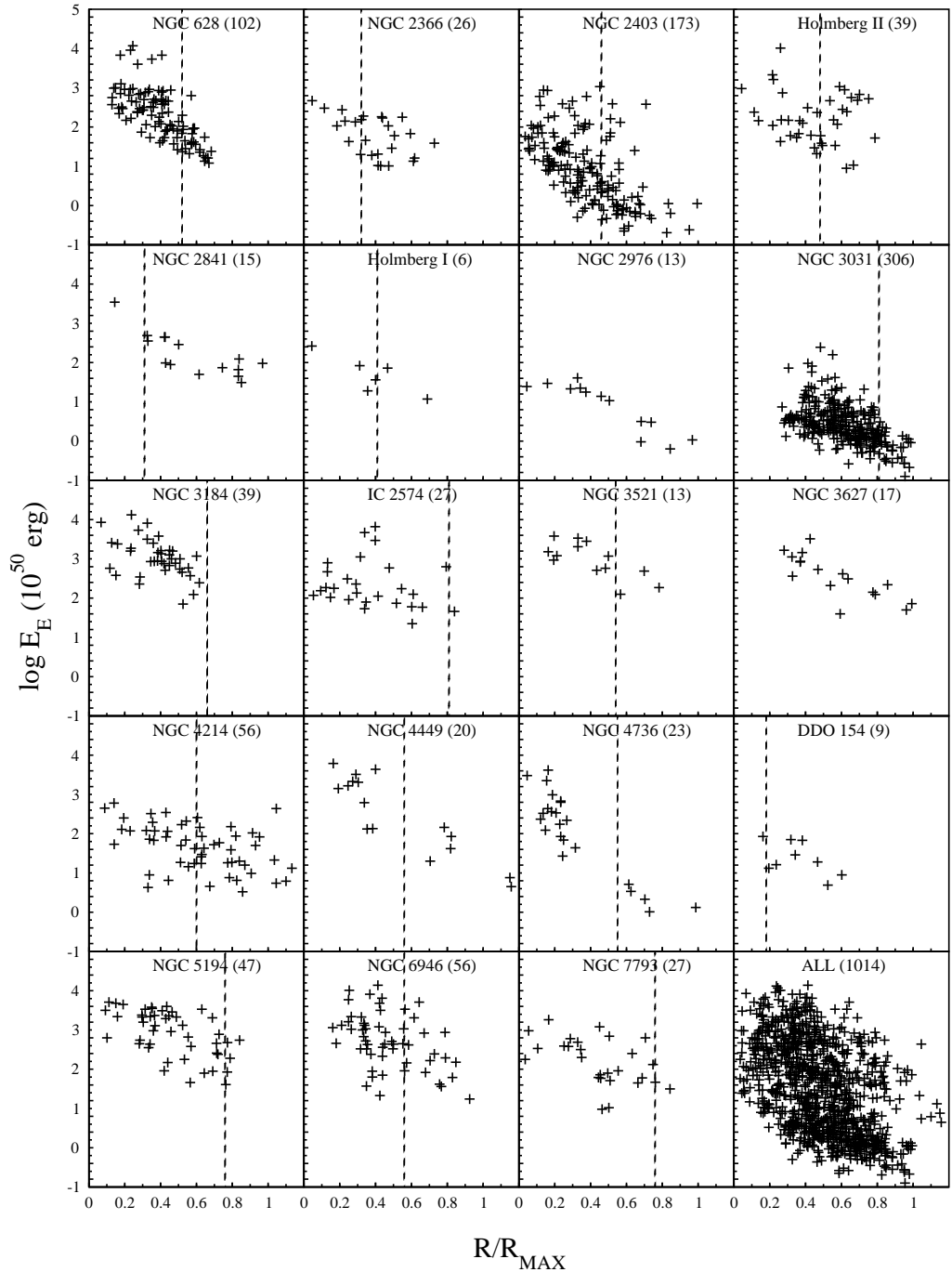}
\end{center}
\caption{Energy requirement, plotted logarithmically, vs.\ normalized radial distribution of the \ion{H}{1} holes. The dashed vertical line indicates for each galaxy the corresponding location of $R_{25}$. The bottom right panel shows the distribution of all holes combined. \label{fig:e-rgc}}
\end{figure}

In Figure~\ref{fig:e-rgc} the energy requirement of each hole versus its radial distance is presented. The radial distribution is normalized by the radius of the \ion{H}{1} disk, $R_\mathrm{max}$. In ten, mostly spiral galaxies there appears to be a trend of higher energy holes to be located at smaller galactocentric radius. Further analysis showed that in all galaxies, apart from NGC\,4214, the average energy requirement is higher for holes lying within $R_{25}$ ($E_\mathrm{E}=2.1\times10^{52}\,\mathrm{erg}$) than for those outside this radius ($E_\mathrm{E}=1.0\times10^{52}\,\mathrm{erg}$). This might be due to the fact that the external pressure is higher within $R_{25}$. First of all the gas density is higher near the center, as can be verified in the paper by \citet{ler08}. Secondly, the SFR is higher in the central regions than in the outskirts which translates into more holes being created (cf., Fig.~\ref{fig:Nperkpc2}), increasing the velocity dispersion and hence pressure of the gas.

We also investigated the following relations, but there do not seem to be any trends of expansion velocity with either of axial ratio, displaced mass, energy required to produce the hole, or with galactocentric radius. Neither is there any relation linking kinetic age with position from the center in a galaxy.

\subsection{Comparison of Properties across the THINGS sample}
\label{GlobProp}

\begin{figure}
\begin{center}
\includegraphics[angle=0,scale=.69]{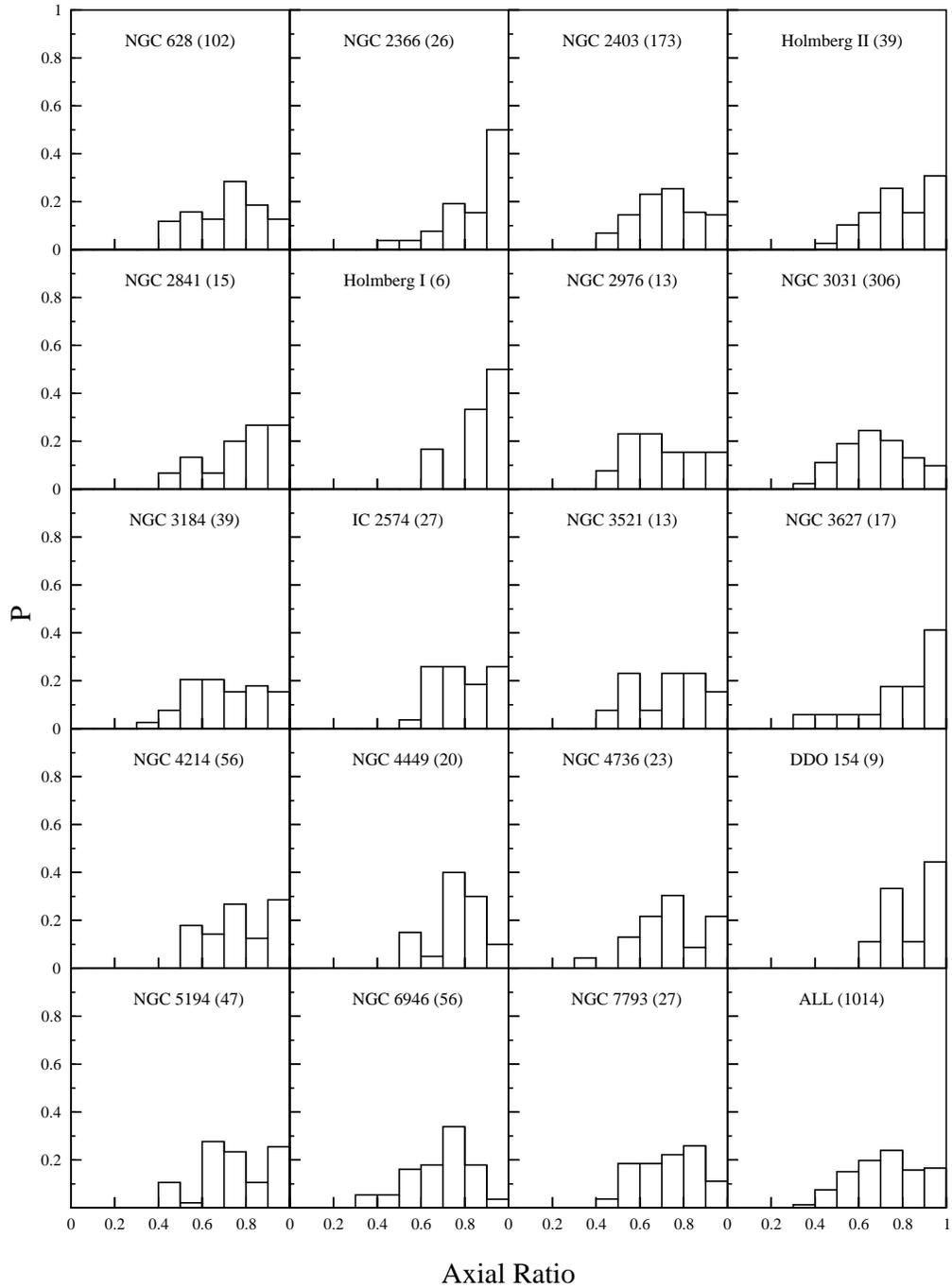}
\end{center}
\caption{Relative number distribution of the axial ratios of the holes. The bottom right panel shows the distribution of all holes combined. Each panel lists the galaxy it refers to; the number in parentheses indicates the total number of holes represented by the histogram. \label{fig:axial}}
\end{figure}

\begin{figure}
\begin{center}
\includegraphics[angle=0,scale=.30]{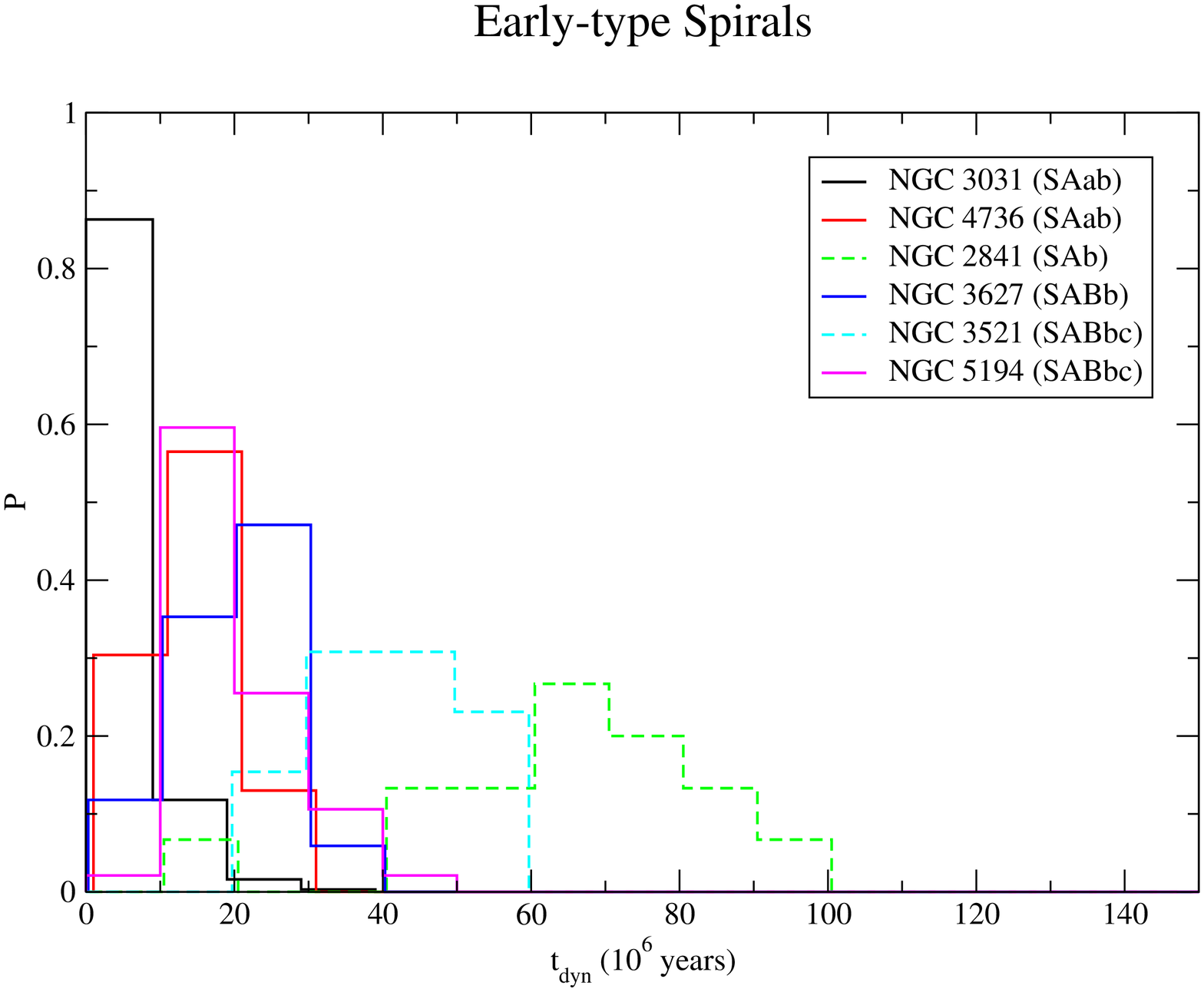}
\includegraphics[angle=0,scale=.30]{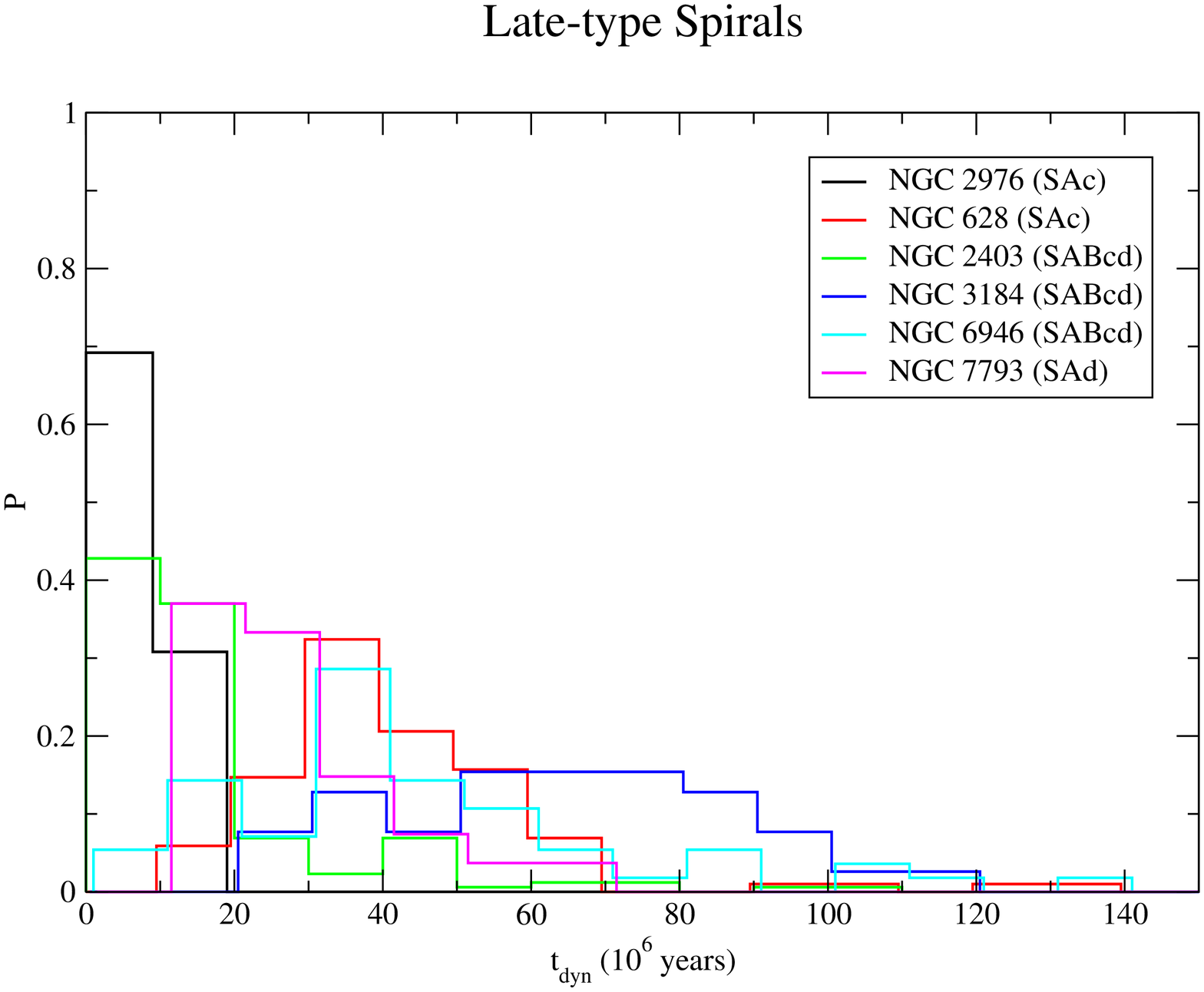}
\includegraphics[angle=0,scale=.30]{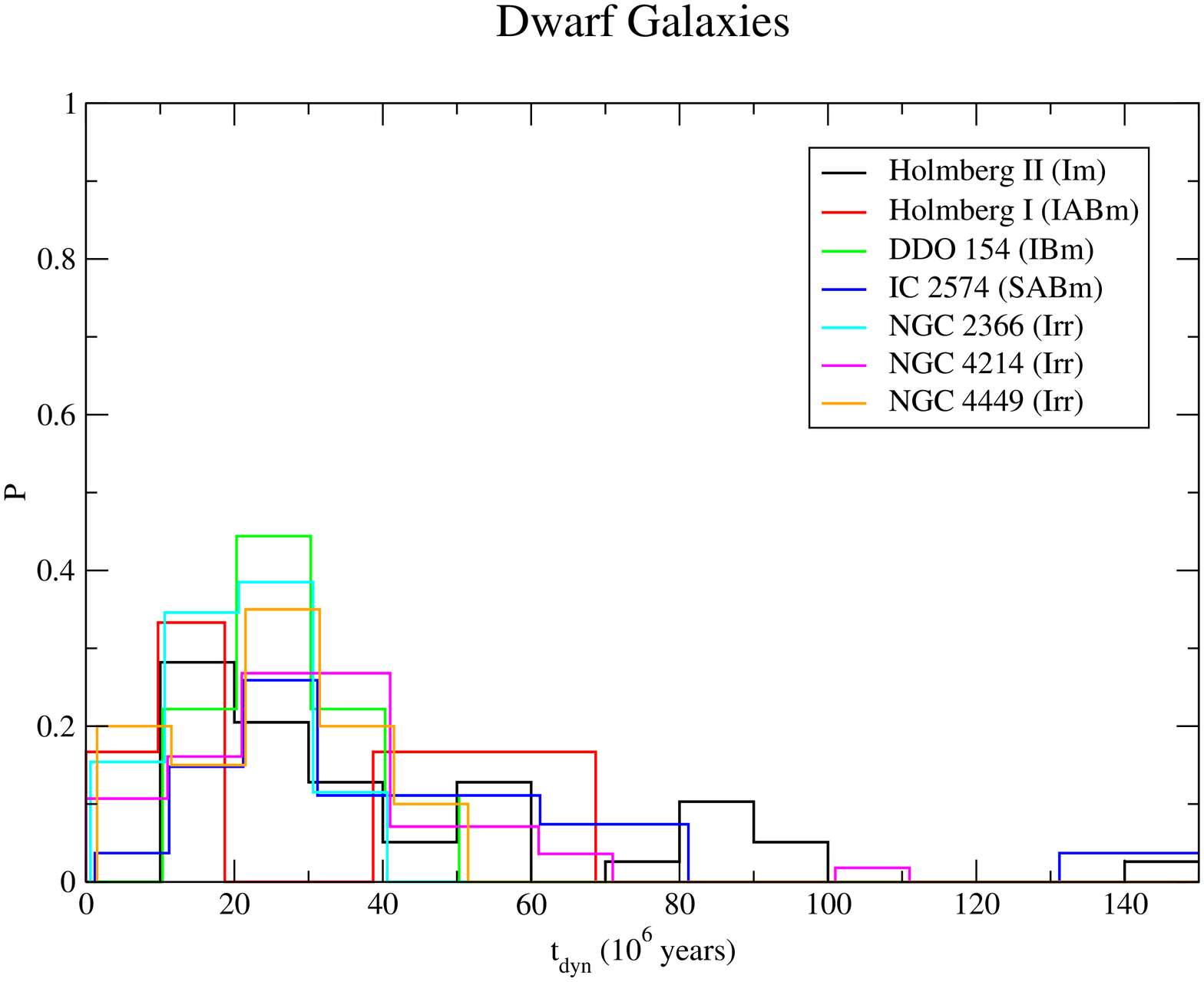}
\end{center}
\caption{\label{fig:ELD} The distributions of the kinetic ages of the \ion{H}{1} holes divided into 3 different groups of galaxies (from early--type spirals to dwarf galaxies). The dashed lines represent galaxies with the worst resolution hence their distributions are skewed to the right. The $y-$axis shows the relative number distribution, \textit{P}.}
\end{figure}

As was mentioned earlier, we selected 20 galaxies out of the 34 available. We ensured that we had 2 galaxies per Hubble type. Within each Hubble type, we tried to select galaxies which differed in some property, such as SFR, to explore as wide a parameter space as possible. If we look at the distribution of the kinetic age of the holes across the sample, we find a trend which becomes more obvious if we group the galaxies in terms of early-type, late-type and dwarf galaxies as illustrated in Fig.~\ref{fig:ELD}. The mean kinetic age shifts towards higher values when we move from early--type spirals ($\overline{t_\mathrm{kin}}=13.9\,\mathrm{Myr}$) to late--type ($\overline{t_\mathrm{kin}}=31.6\,\mathrm{Myr}$) to dwarf galaxies ($\overline{t_\mathrm{kin}}=32.5\,\mathrm{Myr}$). This can be understood as follows. The kinetic age is driven by the diameter distribution, as the expansion velocities cover a limited range, as mentioned earlier. Disks of earlier type spirals present shear and spiral density waves, both mechanisms actively rearranging the gas within the disks and limiting the time an \ion{H}{1} hole may survive. Dwarf irregulars lack both shear and spiral arms. In spiral galaxies where shear is strong (NGC\,2976, NGC\,3031, NGC\,3627, and NGC\,4736) the maximum kinetic age is observed to be 40\,Myr in contrast to the dwarf galaxies where we find holes of up to 140\,Myr. Overall the kinetic age ranges from 10 to 140\,Myr, the upper limit reflecting the fact that holes would blend and possibly be destroyed after one full rotation of the disk. A second explanation for the observed trend is the same argument that applies as used when discussing Fig.~\ref{fig:axial}, namely that the scaleheights of the disks in early--type spirals are small in comparison with those in dIrrs which means that shells will break out of the disk sooner in the former and the remnant shell will stall, leading to \ion{H}{1} holes in spirals having on average smaller diameters, hence leading to systematically lower age estimates.

In order to make this analysis more quantitative we decided to split the sample up in spiral and dwarf galaxies. The subsamples consist of 12 and 7 members, respectively, with a total of 831 versus 183 \ion{H}{1} holes. The comparison between the subsamples was carried out via Mann-Whitney (M--W) and Kolmogorov-Smirnov (K--S) tests. Both are non--parametric tests for assessing whether two samples can be drawn from the same parent distribution based on differences in the location of the medians (M--W) and differences in the shape of the two distributions (K--S). In order to accept or reject a hypothesis we took a 3--sigma level as the minimum significance level on both tests. In the following we compare the two subsamples of spirals and dwarfs in graphical form. The results of the M--W and K--S tests are summarised in Table~\ref{stats}.

\begin{figure}
\begin{center}
\includegraphics[angle=0,scale=.69]{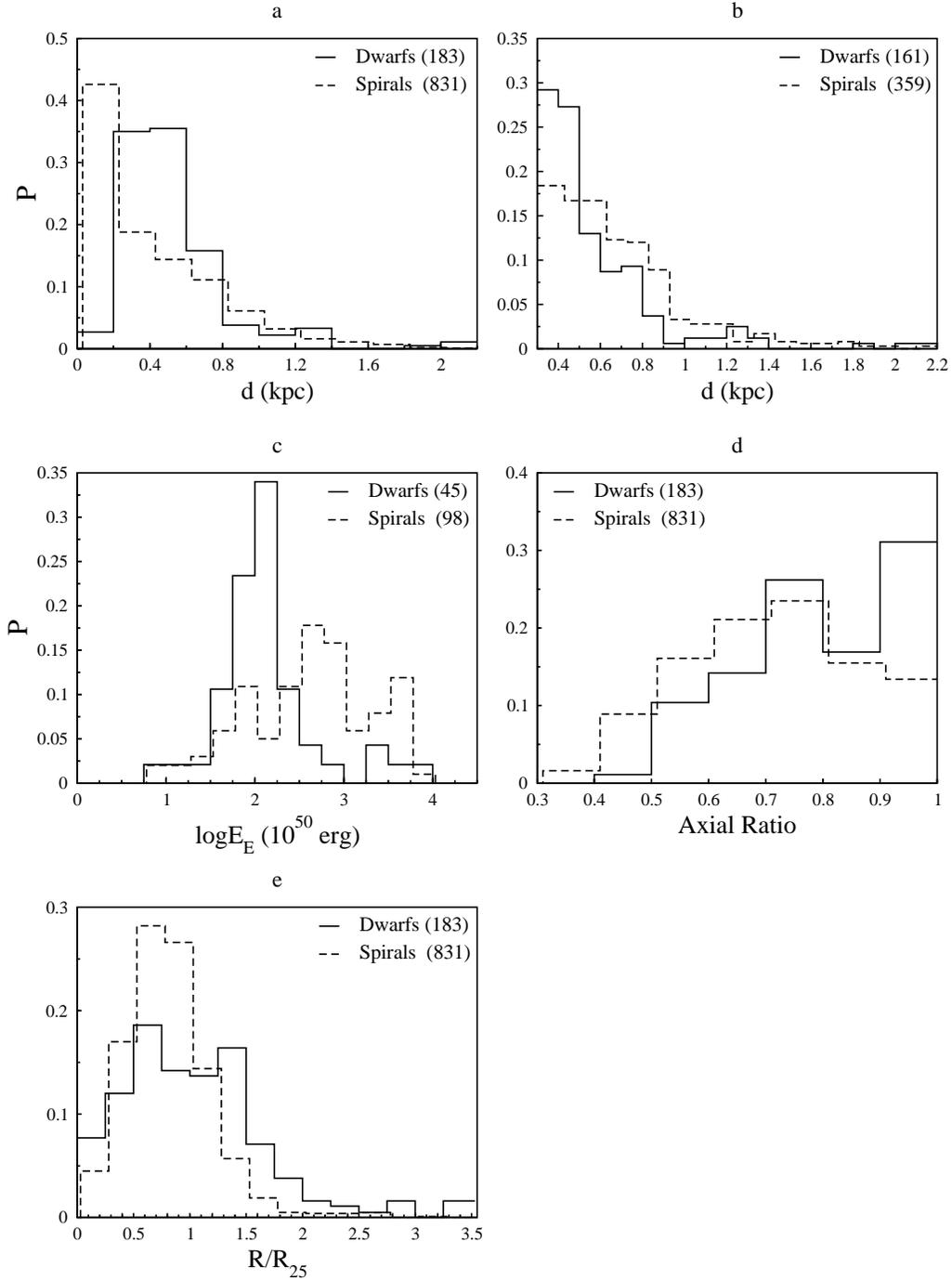}
\end{center}
\caption{Histogrammes of various observed and derived properties of the \ion{H}{1} holes, split over two subsamples: spiral and dwarf galaxies. The numbers in parentheses indicate the number of holes in each group. Panels a,d and e include all holes, panel b includes holes larger than 280 pc and panel c includes all expanding holes larger than 280 pc. See Sec.~\ref{GlobProp}, 3rd par. for details. \label{fig:Dwarf-spirals}}
\end{figure}

Fig.~\ref{fig:Dwarf-spirals}a shows the difference in the distribution of diameters of the holes between dwarf and spiral galaxies. It appears that dwarf galaxies have larger holes on average than spiral galaxies. A Mann-Whitney test confirms this as well. This is of course not surprising in the light of the current discussion. What is surprising is that dwarf galaxies appear to have fewer small holes ($d < 200$ pc) even though the spatial resolution is better on average. This can be understood by realising that whereas star formation is a continuous process in spiral galaxies, it tends to occur in bursts in dwarfs. The time between bursts is longer than the duration of the star formation episodes (e.g., \citealt{ger80}) so it is likely that the dwarf galaxies whose ISM is currently riddled with \ion{H}{1} holes are in a post--starburst phase. There will therefore be a relative lack of small holes, most holes having had the time to grow to typically $\sim 100$ pc and beyond.

In this comparative analysis we included all holes. However, the galaxies are spread over distances which cover a factor of $\sim 3$ between the closest and most distant one (in addition, the angular resolution varies slightly across the sample). Therefore, small holes in those galaxies that are further away will escape detection. If we exclude NGC\,2841 and NGC\,3521 from our sample then for the rest of the galaxies there is a minimum common spatial resolution of 280 pc, i.e. we should have a complete sample of holes larger than 280 pc in 17 galaxies.

Fig.~\ref{fig:Dwarf-spirals}b illustrates the distribution of diameters of holes larger than 280 pc between dwarf and spiral galaxies. We now see that spiral galaxies have slightly larger holes than dwarf galaxies. We can understand this as follows: the energy requirement of expanding holes larger than 280 pc, Fig.~\ref{fig:Dwarf-spirals}c, shows a clear differentiation in the distribution of the two types of galaxies (K--S test). Holes in spiral galaxies appear to be significantly more energetic. Based on the way the energy is derived this difference could be due to the fact that holes in spiral galaxies are intrinsically more energetic, for example because OB associations in spirals are more massive. The effect of spirals having a larger energy input, on average, translates into spiral galaxies having larger \ion{H}{1} holes.

As was mentioned earlier, it appears that dwarf galaxies have rounder holes than spirals. This trend is apparent in Fig.~\ref{fig:Dwarf-spirals}d, where we plot the distributions of the axial ratio for the two types of galaxies. It is again clear that the two distributions differ (Mann-Whitney test).

The spatial distribution of the holes appears to be dependent on the type of the host galaxy as illustrated in Fig.~\ref{fig:Dwarf-spirals}e. In spiral galaxies holes appear to be concentrated within a smaller area of the disk, whereas in dwarf galaxies holes are distributed over a larger area. This is  confirmed by a Mann-Whitney test. Even though the difference between the two distributions is subtle the result can be understood as being due to the fact that star formation in spiral galaxies mostly happens in the spiral arms which are located within the $R_{25}$ radius. In dwarf galaxies star formation is  spread across the entire disk. 

\begin{figure}
\begin{center}
\includegraphics[angle=0,scale=.69]{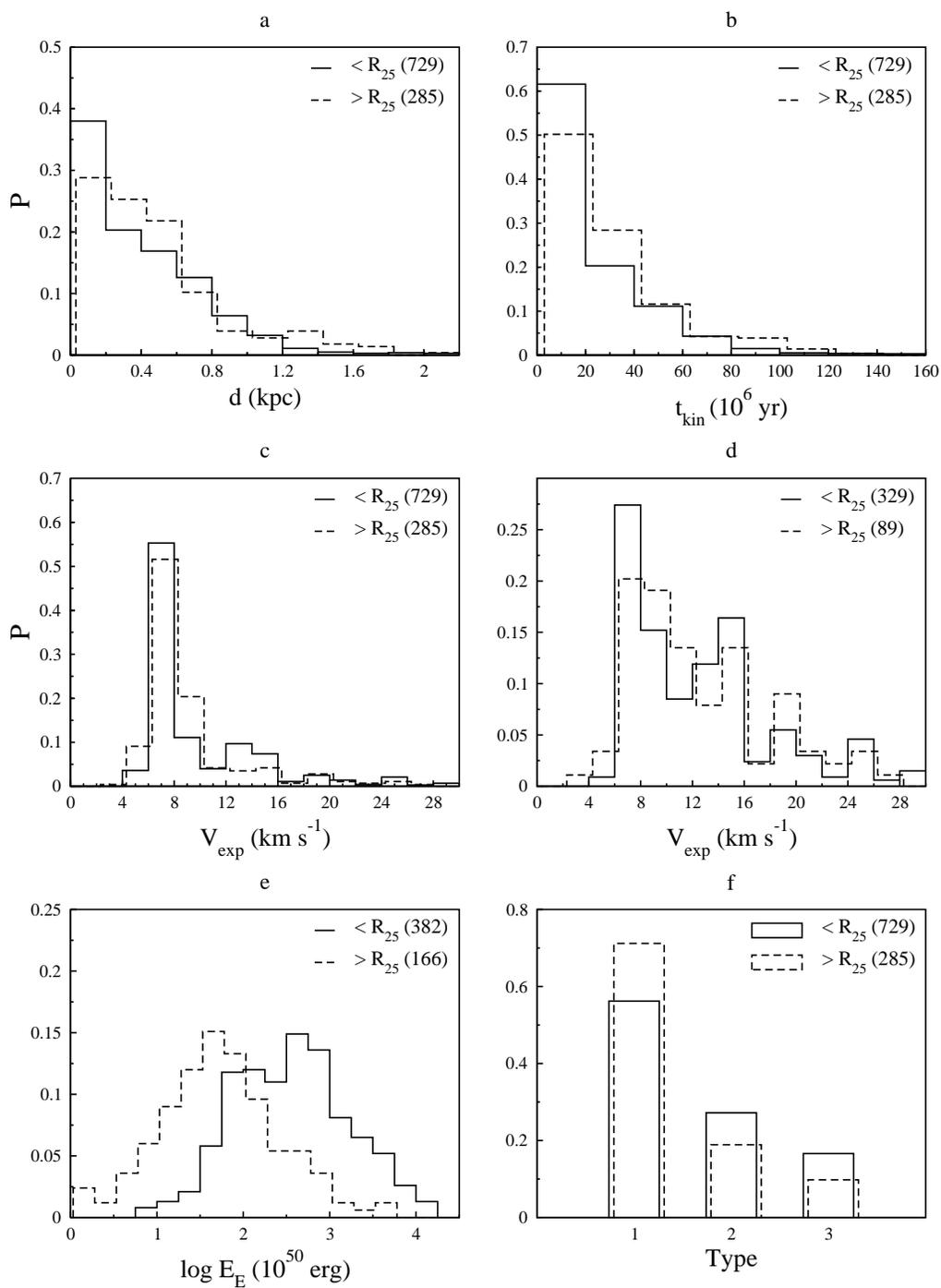}
\end{center}
\caption{ Histogrammes of various observed and derived properties of the \ion{H}{1} holes, split over two subsamples: holes located within or outside $R_{25}$. See Sec.~\ref{GlobProp}, 8th par. for details.\label{fig:InOutR25}}
\end{figure}

To investigate this further we divided our sample of holes based on $R_{25}$, i.e., holes located inside and outside of this radius. We find that regardless of the type of the galaxy, holes outside $R_{25}$ are larger and older, as illustrated in Fig.~\ref{fig:InOutR25}a and b. This is confirmed by the relevant Mann-Whitney test. This is in agreement with holes outside $R_{25}$ being less affected by processes such as shear. When we look at the sample of holes larger than 280 pc no significant differences are found in the diameter and kinetic age distributions within and outside $R_{25}$.

There is no discernible difference in expansion velocity (Fig.~\ref{fig:InOutR25}c) even after we remove the stalled holes from our sample (Fig.~\ref{fig:InOutR25}d). Looking at the distribution of the energy requirement of holes larger than 280 pc with respect to $R_{25}$ we find that on average holes have higher energies when located inside this radius (Fig.~\ref{fig:InOutR25}e). This holds true even when we break down our sample according to galaxy type (dwarf or spiral). Given that the diameter and expansion velocity distributions are similar in these two regions we conclude that this is driven by the difference in volume density.

Regarding the types of the holes we found that in spiral galaxies there are more partially and fully contained holes (type 2 and 3) within $R_{25}$ (Fig.~\ref{fig:InOutR25}f). This is in agreement with the fact that holes inside $R_{25}$ are smaller and younger, on average.

\begin{deluxetable}{lccccc}
\tabletypesize{\scriptsize}
\tablecaption{Significance values of Mann--Whitney and Kolmogorov--Smirnov tests across a number of hole properties}
\tablewidth{0pt}
\tablehead{
 \colhead{} & \colhead{} & \multicolumn{2}{c}
 {Dwarf--Spiral} & \multicolumn{2}{c}
 {$R_{25}$}\\
\colhead{Property} & \colhead{Sample} & \colhead{M--W} & \colhead{K--S} & \colhead{M--W} & \colhead{K--S} }
\startdata
d&All&$<0.0005$&$<0.0005$&0.015&0.002\\
d&$d>280$ pc&$<0.0005$&$<0.0005$&0.078&0.025\\
$V_\mathrm{exp}$&All&$<0.0005$&$<0.0005$&0.036&0.003\\
$V_\mathrm{exp}$&Expanding&0.049&0.006&0.541&0.543\\
$t_\mathrm{kin}$&All&$<0.0005$&$<0.0005$&$<0.0005$&0.001\\
$t_\mathrm{kin}$&$d>280$ pc&0.354&0.227&0.408&0.127\\
$E_\mathrm{E}$&$d>280$ pc&$<0.0005$&$<0.0005$&$<0.0005$&$<0.0005$\\
$E_\mathrm{E}$&$d>280$ pc, expanding&$<0.0005$&$<0.0005$&$<0.0005$&$<0.0005$\\
Axial Ratio&All&$<0.0005$&$<0.0005$&0.573&0.578\\
Axial Ratio&$d>280$ pc&0.001&0.017&0.851&0.397\\
R&All&$<0.0005$&$<0.0005$&\nodata &\nodata\\
R&$d>280$ pc&$<0.0005$&$<0.0005$&\nodata &\nodata\\
\enddata
\label{stats}
\end{deluxetable}

\begin{figure}
\begin{center}
\includegraphics[angle=0,scale=.69]{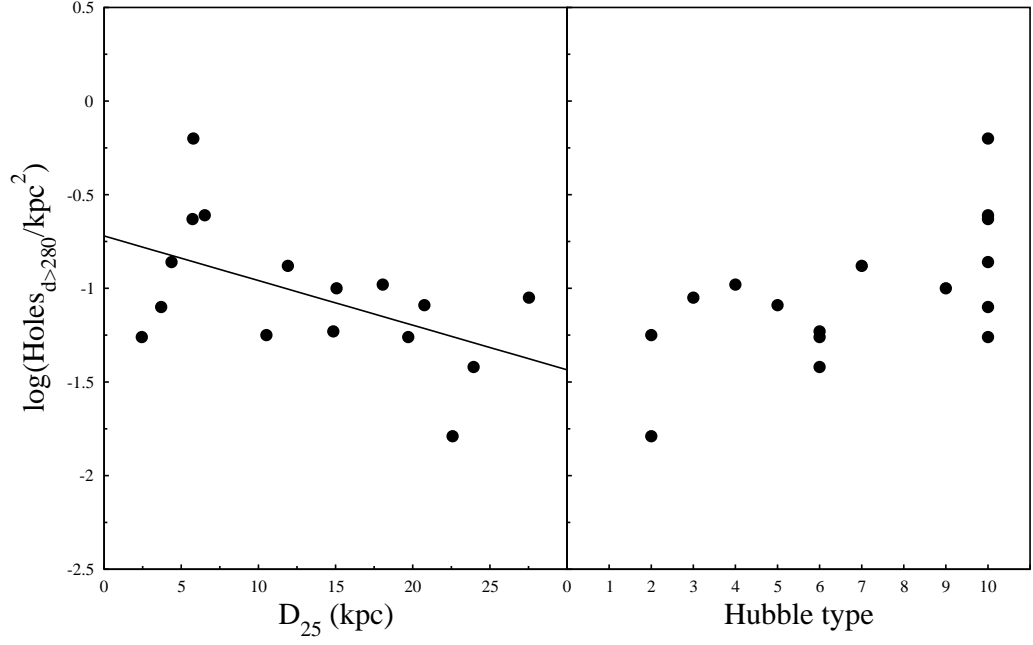}
\end{center}
\caption{Log of the number of holes ($d > 280\, \mathrm{pc}$) per square kpc versus a) $ \mathrm{D}_{25}$, and b) galaxy Type. Each point represents one galaxy and the solid lines represent a linear fit to the data. \label{fig:ss}}
\end{figure}

Because we are essentially complete (to our detection limit) for holes larger than 280 pc in diameter, we looked for relations of this subset of holes across our galaxy sample (excluding NGC\,2841 and NGC\,3521). The number per kpc$^2$ of these holes is presented in Figure~\ref{fig:ss} where this number density is plotted on a log scale against a) optical diameter of the host galaxy and b) against galaxy (Hubble) type. It is clear that despite the scatter there is a clear trend in the sense that the number density of holes decreases with increasing galaxy size. 

\begin{figure}
\begin{center}
\includegraphics[angle=0,scale=.50]{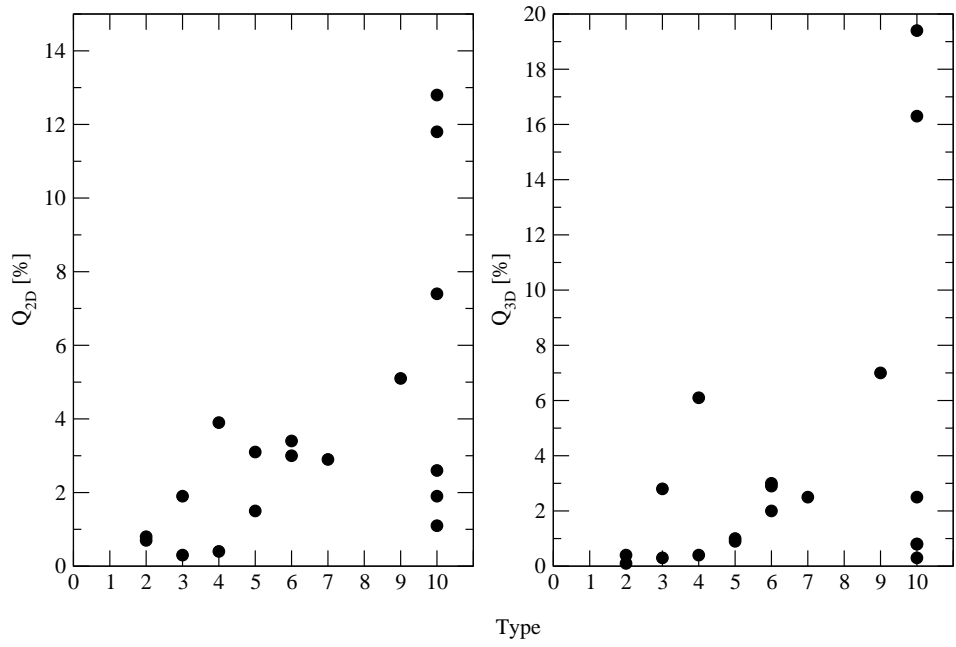}
\end{center}
\caption{Surface (Q$_\mathrm{2D}$) and volume (Q$_\mathrm{3D}$) porosities of the \ion{H}{1} holes versus Hubble Type. \label{fig:poros}}
\end{figure}

\begin{figure}
\begin{center}
\includegraphics[angle=0,scale=.69]{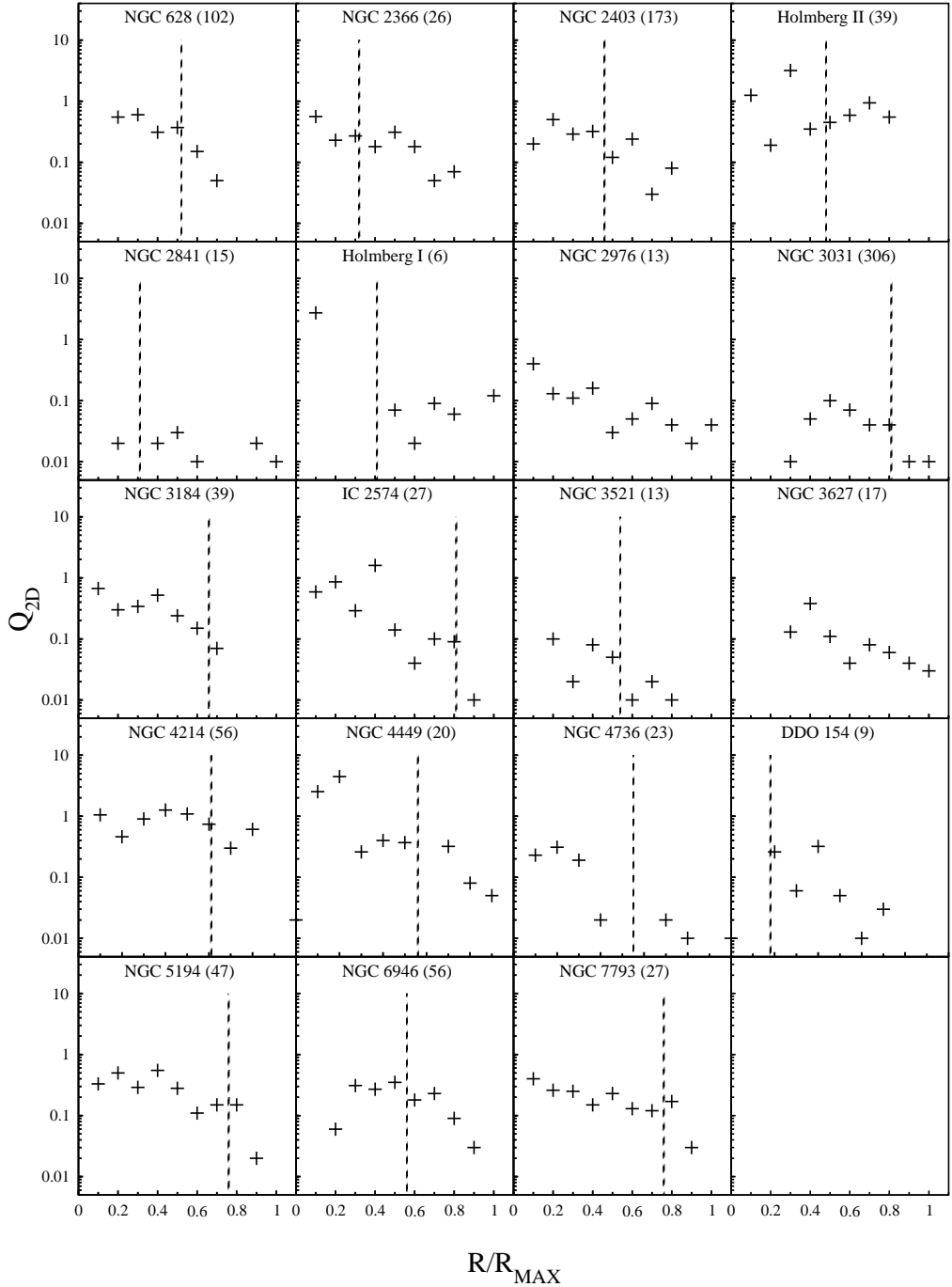}
\end{center}
\caption{The 2--dimensional surface porosity distribution (per annulus) as a function of normalized radius. The vertical, dashed line is drawn at $R_{25}$.\label{fig:poros2D}}
\end{figure}

In the plot of number density against Hubble Type there is a weak trend, the number density remaining more or less constant except for the class of dwarf galaxies (which are dominated by large holes). This trend is also apparent in Figure~\ref{fig:poros} which shows the average porosity of all the holes in a galaxy, versus the type of the host galaxy. The porosity parameter, \textit{Q} \citep{cox74}, is the ratio of total volume (or area) occupied by holes to the total volume (or area) of the host galaxy. For the total volume of the galaxy we used $R_\mathrm{max}$ and the radial scale height as the thickness of the galaxy. The dwarf galaxies stand out as being on average more porous than spirals. \citet{sil97} predicted that the porosity should anti--correlate with star formation surface density. As illustrated in Figure~\ref{fig:poros}, later Hubble types have higher porosity, and these same late--type galaxies form stars at a lower rate per unit mass than earlier types \citep{ler08}. Fig.~\ref{fig:poros2D} shows the 2--D porosity as a function of radius. Although similar to the radial \ion{H}{1} hole surface density distribution (Fig.~\ref{fig:Nperkpc2}), it is not identical as the 2--D porosity distribution takes into account not just the number of holes, but also their size. What we find is that the porosity generally decreases as a function of radius with, as an exception perhaps, several of the dwarf galaxies such as Ho\,I and Ho\,II, where the histograms stay flat. The low porosity in the outskirts is likely a reflection of the generally lower star formation rate (SFR) density in these regions, i.e., well beyond the optical radius of a galaxy.

\begin{deluxetable}{lrrrrrrrrr}
\tabletypesize{\scriptsize}
\tablecaption{Supernova and star formation rates \label{tbl:SFR}}
\tablewidth{0pt}
\tablehead{
\colhead{Galaxy} & \colhead{$log(SFR_\mathit{H\alpha})$} & \colhead{$log(SFR_\mathit{Holes})$} & \colhead{$log(SFR_\mathit{corr})$} & \colhead{$log(E_\mathrm{E})$} & \colhead{$log(\nu_\mathrm{SN})$} &
 \colhead{$log(\nu_\mathrm{SN}^\mathit{corr})$} & \colhead{$a_{\mathrm{\nu}}$} & \colhead{$Q_{2D}$} & \colhead{$Q_{3D}$} \\
\colhead{} & \colhead{$(\mathrm{M_{\sun}\,yr^{-1}})$} & \colhead{($\mathrm{M_{\sun}\,yr^{-1}})$} & \colhead{$(\mathrm{M_{\sun}\,yr^{-1}})$} & \colhead{$(\mathrm{10^{50}\,erg})$} & \colhead{$(\mathrm{SN\,yr^{-1}})$} &
 \colhead{$(\mathrm{SN\,yr^{-1}})$}& \colhead{} & \colhead{} & \colhead{}
}
\startdata
NGC\,628&0.1&$-2.0$&1.3&4.9&$-4.2$&$-0.8$&$-3.6$&3.1\%&4.8\%\\
NGC\,2366&$-1.0$&$-3.2$&$-1.2$&3.4&$-5.3$&$-3.3$&$-2.6$&2.6\%&0.8\%\\
NGC\,2403&$-0.1$&$-2.8$&$-0.6$&4.0&$-4.9$&$-2.7$&$-2.7$&2.7\%&2.9\%\\
Holmberg\,II&$-1.1$&$-2.7$&$-2.2$&4.4&$-4.8$&$-4.3$&$-0.8$&12.8\%&16.3\%\\
DDO\,53&$-2.1$&$-4.2$& \nodata &2.0&$-6.1$& \nodata & \nodata & \nodata &\nodata\\
NGC\,2841&$-0.7$&$-3.0$& \nodata &3.8&$-5.1$& \nodata & \nodata &0.3\%&0.3\%\\
Holmberg\,I&$-2.2$&$-4.3$& \nodata &2.7&$-6.4$& \nodata & \nodata &1.9\%&0.8\%\\
NGC\,2976&$-1.0$&$-5.1$& \nodata &2.3&$-7.2$& \nodata & \nodata &1.5\%&0.9\%\\
NGC\,3031&0.0&$-3.5$&0.3&3.3&$-5.6$&$-1.8$&$-3.9$&0.8\%&0.4\%\\
NGC\,3184&0.2&$-2.2$&$-0.4$&4.9&$-4.3$&$-2.5$&$-2.7$&3.0\%&2.0\%\\
IC\,2574&$-0.9$&$-3.1$&$-1.8$&4.3&$-5.3$&$-4.0$&$-1.8$&5.1\%&7.0\%\\
NGC\,3521&0.5&$-2.3$& \nodata &4.3&$-4.4$& \nodata & \nodata &0.4\%&0.4\%\\
NGC\,3627&0.4&$-2.7$&0.6&4.1&$-4.8$&$-1.6$&$-3.5$&1.9\%&2.8\%\\
NGC\,4214&$-1.3$&$-3.1$&$-0.2$&3.7&$-5.2$&$-2.3$&$-3.1$&11.8\%&19.4\%\\
NGC\,4449&$-0.3$&$-2.4$&$-0.1$&4.3&$-4.5$&$-2.2$&$-3.1$&7.4\%&2.5\%\\
NGC\,4736&$-0.4$&$-2.7$&0.7&4.2&$-4.8$&$-1.4$&$-3.8$&0.7\%&0.1\%\\
DDO\,154&$-2.4$&$-4.1$& \nodata &2.5&$-6.2$& \nodata & \nodata &1.1\%&0.3\%\\
NGC\,5194&0.8&$-1.7$&0.4&4.9&$-3.8$&$-1.7$&$-3.1$&3.9\%&6.1\%\\
NGC\,6946&0.7&$-2.1$&0.3&5.0&$-4.2$&$-1.8$&$-2.8$&3.4\%&3.0\%\\
NGC\,7793&$-0.3$&$-2.6$&$-1.2$&4.0&$-4.8$&$-3.3$&$-2.4$&2.9\%&2.5\%\\
\enddata
\tablecomments{Col 1: Galaxy name; Col 2: SFR based on H$\alpha$; Col 3: SFR based on observed holes; Col 4: Corrected SFR derived from hole distribution function; Col 5: Total energy of holes; Col 6: SN rate based on observed holes; Col 7: Corrected SN rate derived from hole distribution function; Col 8: the slope of the hole distribution function; Col 9: surface porosity; Col 10: volume porosity}
\end{deluxetable}

\subsection{\ion{H}{1} holes and the link with Star Formation}
\label{sec:SFR}

The ISM of galaxies is dominated by structures which are thought to originate from local energetic events. Even though OB associations, through the effects of stellar winds and supernovae, seem to be the obvious mechanism behind the creation of these structures \citep{oey97}, a direct correlation between OB associations and \ion{H}{1} holes is often lacking \citep[e.g.,][]{kam93,vdh96,rho99,pas08,wei09b}. Furthermore, there have been cases of holes with energy requirements that surpass the energy input from OB associations, or of holes which are found in regions, such as the Magellanic Bridge \citep{sta07} where star formation is essentially absent. This has often been used as an argument to question the supernova origin leading to the many alternative explanations listed in the introduction. There are cases, notably the LMC/SMC where a clear link is found between massive stars and the location of \ion{H}{1} holes \citep{sta07}, although even there in some 10\% of the cases no such link is evident \citep{hat05}.

\begin{figure}
\begin{center}
\includegraphics[angle=0,scale=.69]{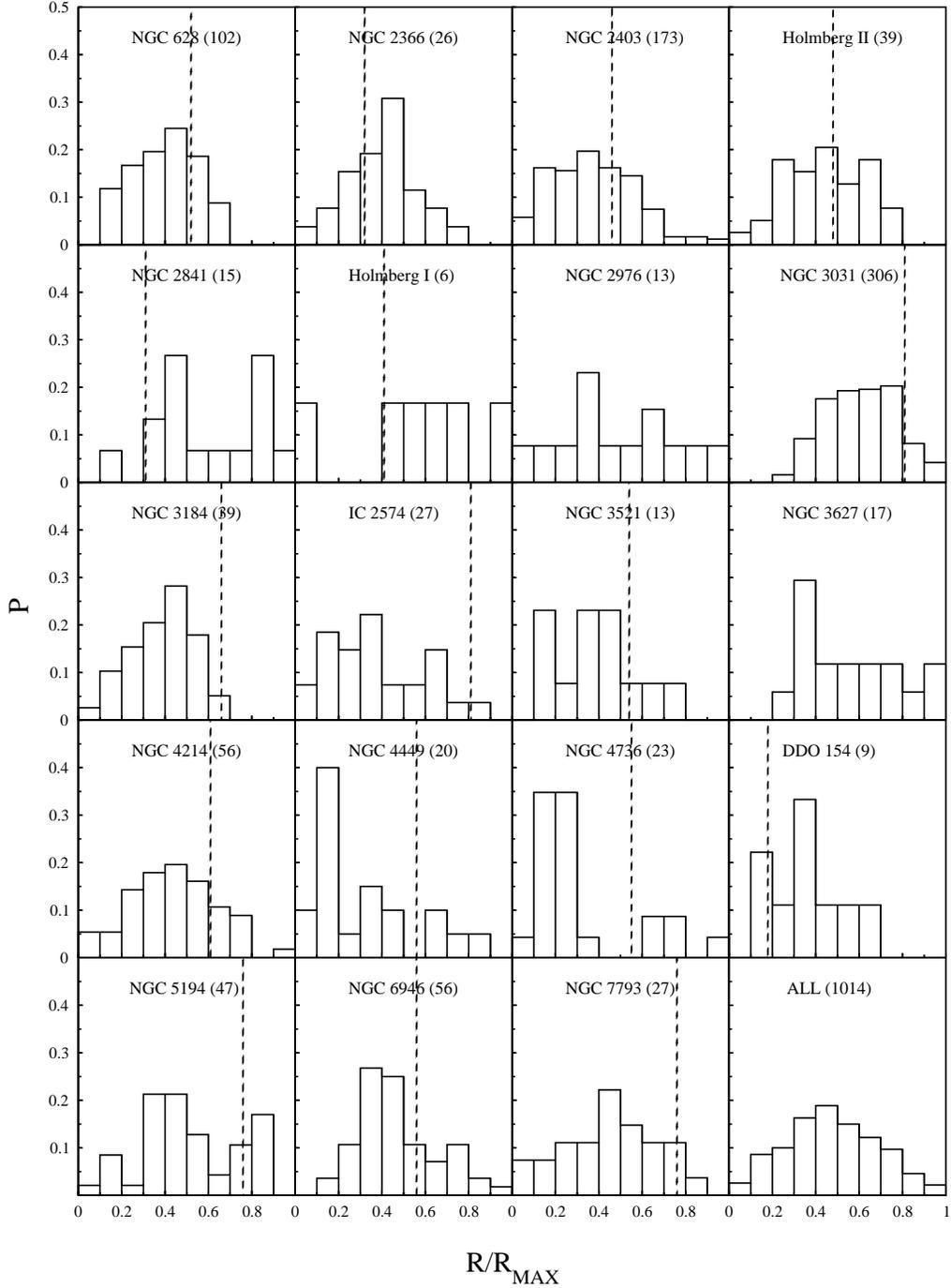}
\end{center}
\caption{Relative radial number distribution of the \ion{H}{1} holes. The x--axis represents the relative galactocentric radius of an \ion{H}{1} hole with respect to the largest extent of the \ion{H}{1} disk of the host galaxy, $R_\mathrm{max}$. The dashed vertical line indicates for each galaxy the corresponding location of $R_{25}$. The bottom right panel shows the distribution of all holes combined. Each panel lists the galaxy it refers to; the number in parentheses indicates the total number of holes represented by the histogram. Overall 77\% of the holes are located within $R_{25}$ and 23\% outside.\label{fig:rgc-r25}}
\end{figure}

\begin{figure}
\begin{center}
\includegraphics[angle=0,scale=.69]{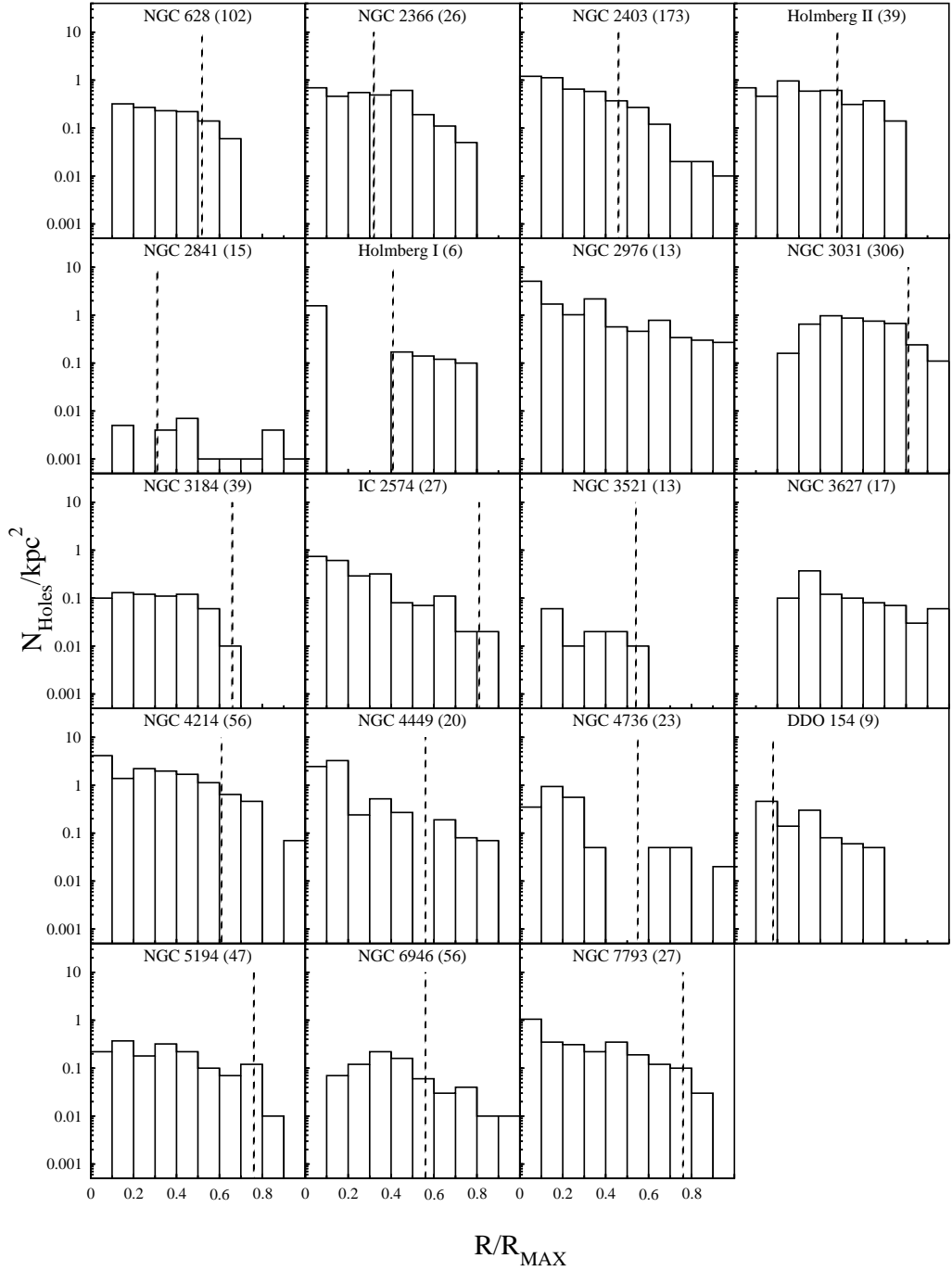}
\end{center}
\caption{As Fig.~\ref {fig:rgc-r25} but showing the number of \ion{H}{1} holes per kpc$^2$ as a function of galactocentric radius normalised by $R_\mathrm{max}$. The dashed vertical line indicates for each galaxy the corresponding location of $R_{25}$. The y--axis is plotted on a logarithmic scale.\label{fig:Nperkpc2}}
\end{figure}

As Fig.~\ref{fig:rgc-r25} illustrates, in all but three galaxies many holes are found beyond the optical disk ($R_{25}$) of the parent galaxy. If the holes are indeed due to OB associations this picture may be problematic as the SFR density drops steeply beyond $R_{25}$. It has been suggested that a ``galactic fountain" \citep{bre80}, i.e., material expelled into the halo of the galaxy by supernova explosions, could be invoked. Hot fountain gas would eventually cool and fall back on to the disk, creating holes away from the original location of the OB association. If this is indeed the case then one would expect to find extraplanar gas in abundance surrounding the \ion{H}{1} disk. Evidence of extraplanar gas expelled from galaxy disks, as expected in the ``galactic fountain" model has recently been found by \citet{oos07} and \citet{boo08}. In addition, models of extraplanar gas by \citet{fra06} and deep \ion{H}{1} studies by \citet{san08} do show a substantial supply of gas within a thick \ion{H}{1} disk and halo, but their analysis seems to favour an external origin for this gas.

Recent results from the {\em GALEX} Nearby Galaxies Survey \citep{ngs, thi07} have obviated the need for these alternative explanations. Many galaxies show extended UV (XUV) disks. {\em HST} ACS observations resolve these XUV-disk sources into loosely clustered complexes, likely evolved OB associations \citep{thi08} and it thus seems safe to assume for the moment that the majority of \ion{H}{1} holes, even those well beyond $R_{25}$ are the result of massive star formation. If this is the case we can derive an independent estimate of the SFR of the host galaxy based on the energy requirements of the holes in a manner similar to how HII regions are related to the SFR based on measurements of their H$\alpha$ flux; for an instantaneous burst and as long as the SF region is younger than 20 Myr, the Lyman--continuum output is constant and scales linearly with the SFR. An instantaneous burst does not have a SF rate  but taking an ensemble average over an entire galaxy, this becomes equivalent to seeing a continuous SFR. The conversion between Ly-cont flux and SFR thus becomes independent of burst age. In the case of \ion{H}{1} holes we are looking at individual and instantaneous events which, if averaged over an entire galaxy can similarly be linked to a SFR, under the assumption that every SF event will result in a hole. This assumption is likely to lead to an underestimate of the true SFR, of course.

If we assume that the supernovae that create \ion{H}{1} holes are all of Type II and that the kinetic energy released by a Type II supernova is $10^{51}$ erg \citep{mcc87} we can estimate for each hole the number of stars in the parent stellar association with masses above 8 M$_{\odot}$ \citep[the minimum mass a star must have to go off as a Type II supernova,][]{bie74}.

\begin{figure}
\begin{center}
\includegraphics[angle=0,scale=.69]{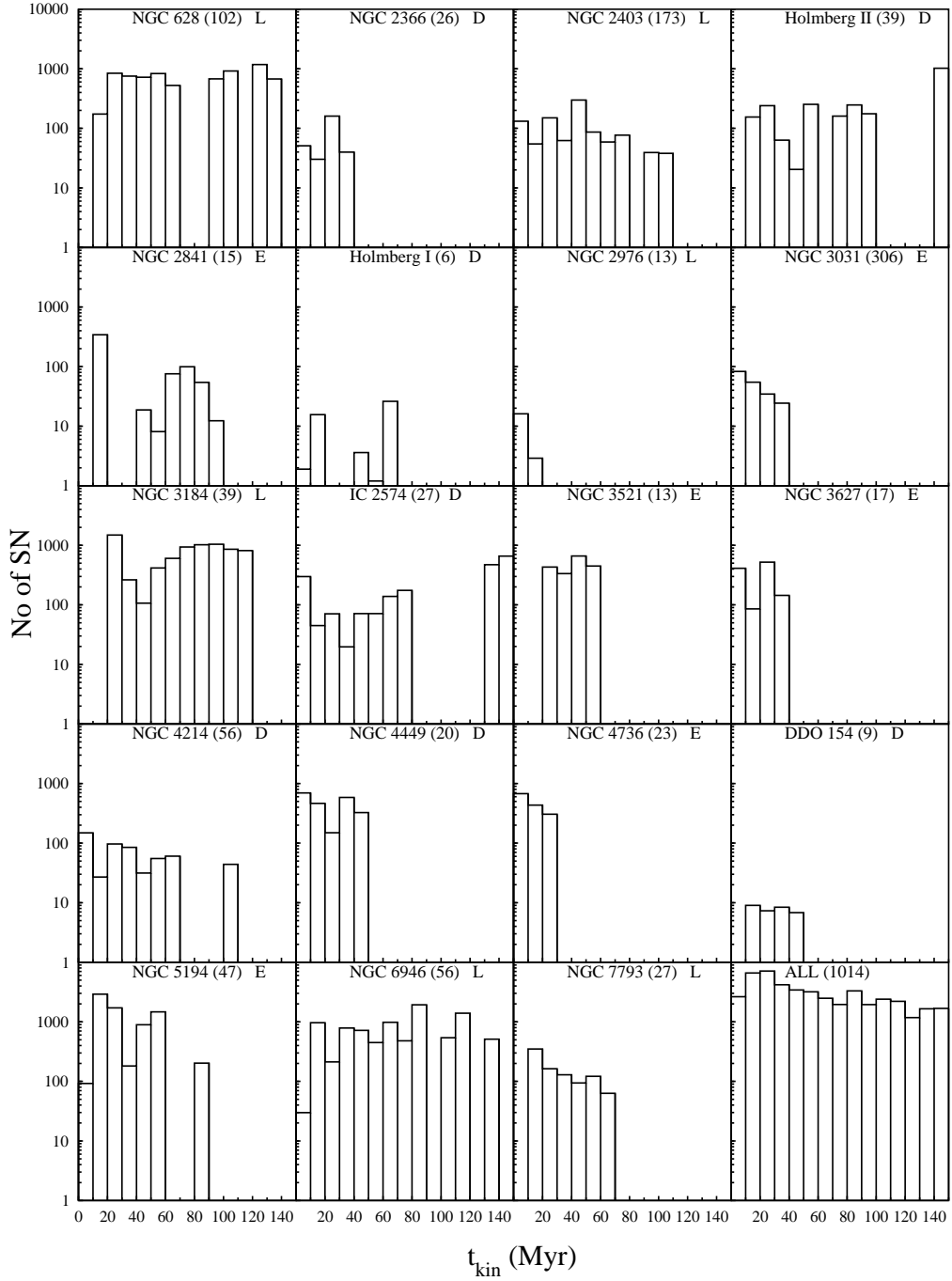}
\end{center}
\caption{The number of SNe needed to create the \ion{H}{1} holes as a function of the kinetic age in bins of 10 Myr. Note that the y--axis is on a logarithmic scale.\label{fig:SN-t}}
\end{figure}

Fig.~\ref{fig:SN-t} shows the number of SN per 10 Myr bin i.e the star formation history of each galaxy by plotting a histogram of the number of SN needed per hole against age of the hole for all holes in a galaxy. The letters E, L, D in the upper right corner stand for early--type, late--type and dwarf galaxy. For most of the galaxies (and specially all early-types) there is a cut--off at $\sim 60$ Myr indicating there is large incompleteness due to shear and diffusion. The lack of SNe younger than $\sim 10$ Myr in some of the galaxies is a reflection of the lower resolution obtained for these galaxies. So, overall, holes trace the supernova history from 10 to 60 Myr, a time frame of 50 Myr. 
From Fig.~\ref{fig:SN-t} we learn that for most galaxies the SN history as derived from the \ion{H}{1} holes is more or less constant. What this means is that by taking an ensemble average over a galaxy, and assuming a typical lifetime for an \ion{H}{1} hole of 60 Myr, we can convert the supernova history into a supernova rate. The values for the calculated SN rates are listed in Table~\ref{tbl:SFR}.

\begin{figure}
\begin{center}
\includegraphics[angle=0,scale=.50]{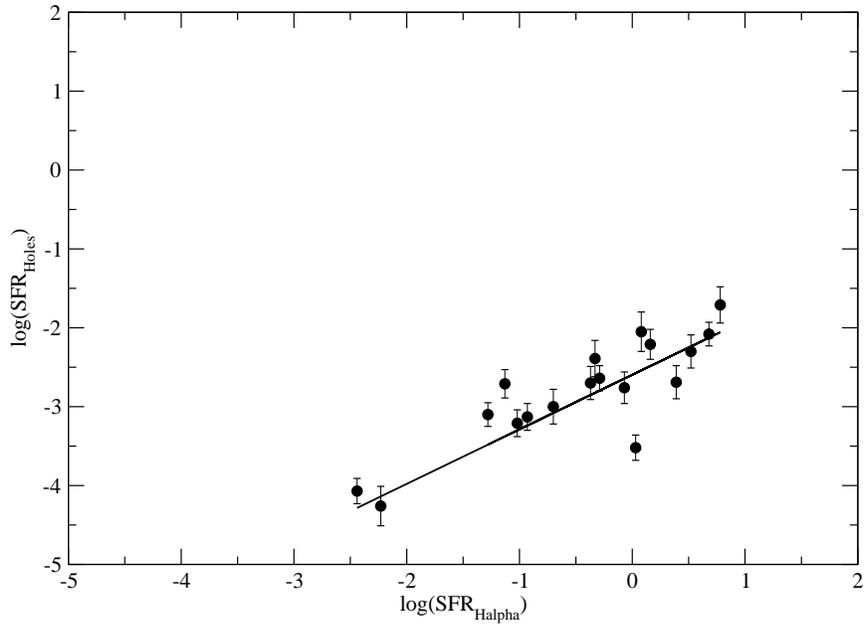}
\end{center}
\caption{SFR derived from the \ion{H}{1} holes plotted against the literature SFR based on observations of the H$\alpha$ flux. The fitted solid line has a correlation coefficient of 0.78 and corresponds to $\mathrm{log(SFR_{Holes})}=-2.60(\pm0.15) +0.69(\pm0.15)\,\mathrm{log(SFR_{H\alpha})}$ \label{fig:SFR-SFRHist10-60}}
\end{figure}

The supernova rate is proportional to SFR(M$>8$M\,$_\odot$), the star formation rate for stars more massive than 8 M$_{\odot}$. We can derive the total SFR by assuming a Salpeter IMF and integrating this between upper and lower stellar mass limits of 120 and 0.1\,M$_\odot$. Given the fairly large uncertainty in the energy requirements of the holes the choice of IMF is not particularly critical. Using a Salpeter IMF we find that SFR $\sim 134 \times $SFR(M$>8$M\,$_\odot$). The calculated SFR over 50 Myr, hereafter referred to as SFR$_\mathit{Holes}$, correlates very well (Spearman's rho=0.78) with the current (H$\alpha$) SFR as illustrated in Fig.~\ref{fig:SFR-SFRHist10-60} and Table~\ref{tbl-glob} but our values fall short by 2--3 orders of magnitude. The error bars are formal errors using the uncertainties quoted in the observed and derived quantities. 

\begin{figure}
\begin{center}
\includegraphics[angle=0,scale=.50]{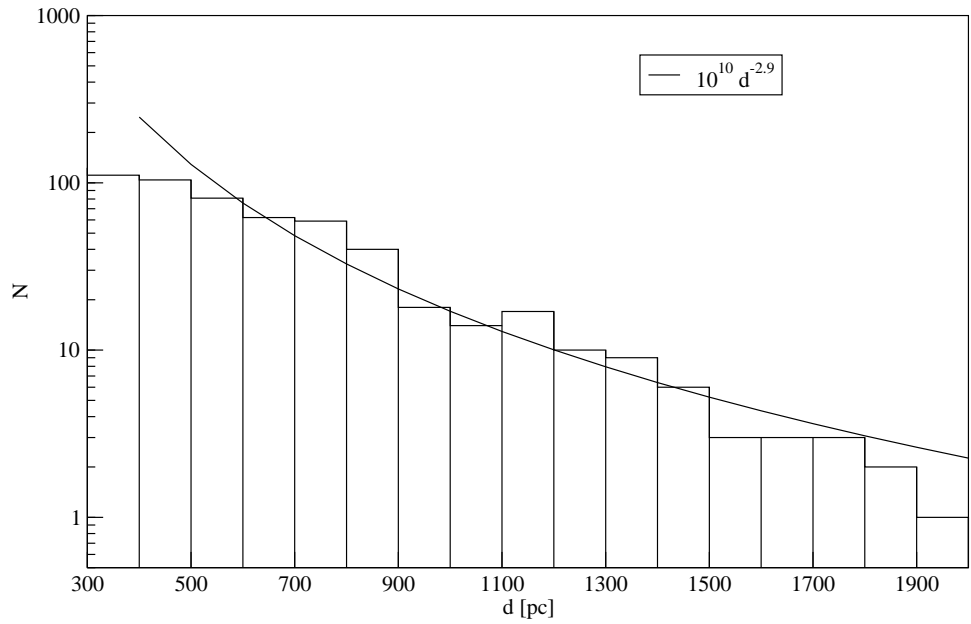}
\end{center}
\caption{The distribution function of the \ion{H}{1} holes across the entire sample of galaxies. Only holes larger than 280 pc were used. The solid line represents the best power law fit to the data. \label{fig:slope}}
\end{figure}

\begin{figure}
\begin{center}
\includegraphics[angle=0,scale=.50]{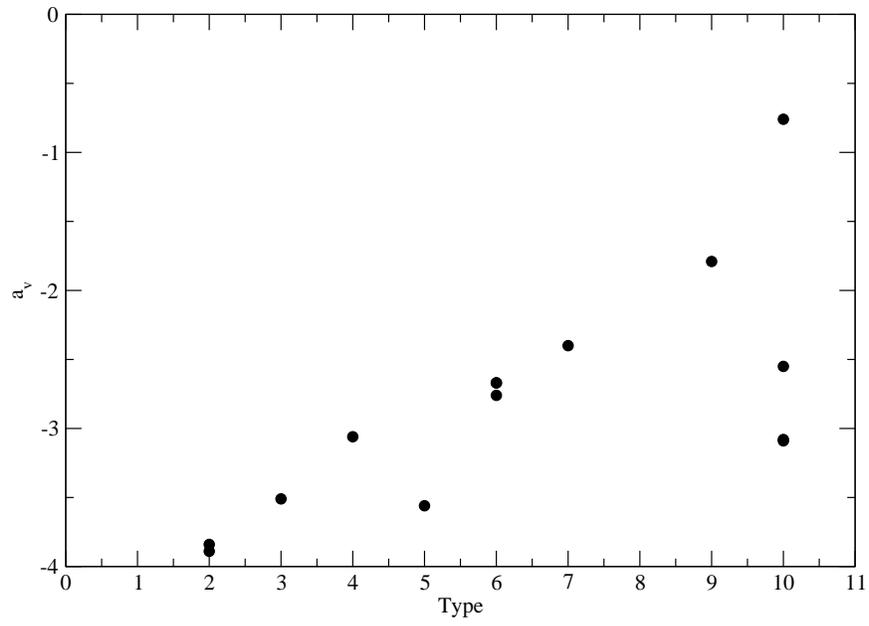}
\end{center}
\caption{The slope of the distribution function of the \ion{H}{1} holes vs.\ the type of the galaxy. Early--type spirals have a steeper slope compared to late--type and dwarf galaxies. \label{fig:slope-type}}
\end{figure}

\begin{figure}
\begin{center}
\includegraphics[angle=0,scale=.50]{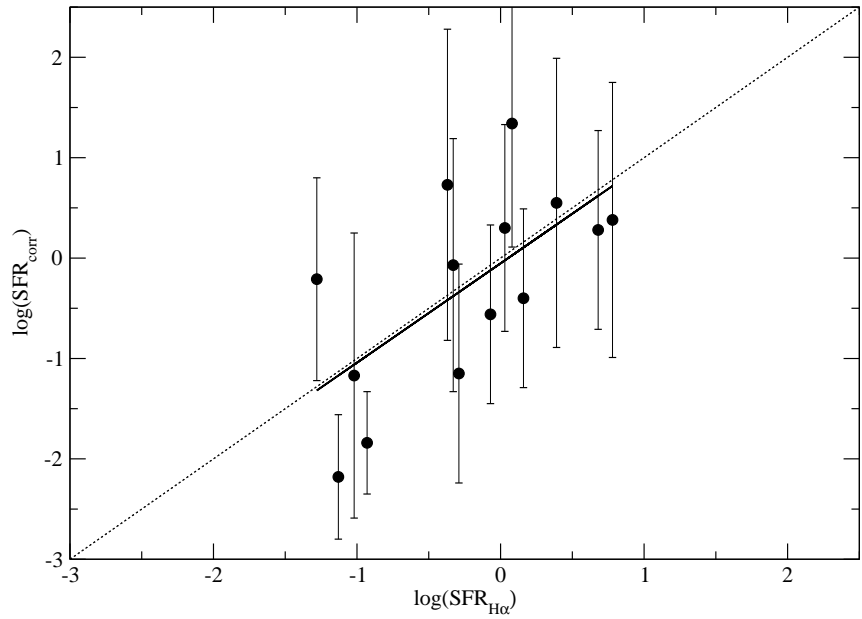}
\end{center}
\caption{Corrected SFR derived from the \ion{H}{1} holes plotted against the literature SFR based on observations of the H$\alpha$ flux. The fitted solid line has a correlation coefficient of 0.78 and corresponds to $\mathrm{log(SFR_{corr})}=-0.05(\pm0.23) +1(\pm0.34)\, \mathrm{log(SFR_{H\alpha})}$. The dotted line corresponds to a one to one correlation. \label{fig:SFR-SFRcorr}}
\end{figure}

The SFR based on the \ion{H}{1} holes is a lower limit and depends critically on the energies we estimated. Various factors conspire to make this a lower limit. Our estimate for the kinetic age is an upper limit therefore the SFR derived here is a lower limit by default. In addition, our limited spatial resolution means that we are missing holes smaller than 100--200 pc in size. Given that these dominate the \ion{H}{1} distribution (compare, e.g. the higher linear resolution data on the LMC/SMC) they account for a large fraction of the star formation. From theoretical models (\citealt{oey97}) and from studies of the Magellanic Clouds we know that the size distribution of holes follows a power--law. This is also evident from our data for the galaxies with the highest resolution (e.g., NGC\,3031, NGC\,2403 and Holmberg\,II). We derived an \ion{H}{1}--hole distribution function for each of the galaxies and fitted a power--law to the high end part of each galaxy. On average we find an exponent of $a_{\mathrm{\nu}} =-2.9$, in good agreement with \citet{oey97} who predict $-4 < a_{\mathrm{\nu}} < -2$. This is illustrated in Fig.~\ref{fig:slope} for the integrated distribution function over all galaxies, for holes with sizes larger than 280 pc. The average error of $a_{\mathrm{\nu}}$ is of order $25\%$ although in cases where the sample of holes is small the error is of order $50\%$. As Fig.~\ref{fig:slope-type} illustrates there is a correlation between $a_{\mathrm{\nu}}$ and the type of the galaxy. Early--type spirals exhibit a steeper slope than late--type spirals and dwarf galaxies. This is a consequence of what we discussed above, namely that early--type spirals are dominated by a large number of small holes whereas dwarf galaxies have proportionally larger holes. We can extrapolate this power--law to the smallest size hole that has been discovered (Milky Way and SMC). Integrating from 10 pc to the spatial resolution in a particular galaxy we can derive corrected supernova and SF rates, $\nu_\mathrm{SN}^\mathit{corr}$ and SFR$_\mathit{corr}$ for all galaxies, assuming that every star forming region will create a hole. These values (tabulated again in Table~\ref{tbl:SFR}) are likely to be representing the current SF since the vast majority of holes will now be small and young. We plot SFR$_\mathit{corr}$ in Fig.~\ref{fig:SFR-SFRcorr} and find that our rather bold extrapolation indeed brings the corrected SFR to within the range found based on other SFR indicators, although the scatter has increased considerably as well. We do not wish to push this exercise too far but if anything, the hypothesis that holes are a consequence of massive star formation is compatible with our observations. Moreover, future studies at higher linear resolution and increased sensitivity should allow us to probe the size distribution down to \ion{H}{1} holes created by a single SN event, leading to a near complete census of such features and a thorough testing of this assertion. Deep optical and UV observations that are able to resolve individual stars will be instrumental in characterising the expected remnant stellar population inside the holes.

\section{Summary}\label{sec:Summary}
The present study is the first attempt to identify and compare in a systematic and consistent way \ion{H}{1} holes across a range of nearby galaxies. Our sample consists of 20 spiral and dwarf galaxies which have been observed as part of the THINGS project. A detailed examination of the neutral ISM reveals a wealth of structures, the dominant one being formed by \ion{H}{1} holes; out of more than 4000 candidate holes we catalogued the 1000 highest quality ones for which we determined a set of observed and derived properties. The number of these structures varies from galaxy to galaxy, partly because of the different linear resolution of our observations at the distance of the targets, and partly due to intrinsic differences between the galaxies, such as viewing angle. The target selection, and the wide range of observed characteristics, preclude us from performing any type of statistical analysis on a ``complete" sample. Instead, we base our study on a comparative analysis, identifying broad trends.

Diameters of \ion{H}{1} holes range from 100 pc (the resolution limit for the closest targets) to $\sim 2$\,kpc with the majority being smaller than 200\,pc. Holes for which we could measure an expansion velocity (types 2 \& 3, typically making up 10\% to 50\% of the holes) list values in the range of $4-36$\,km\,s$^{-1}$. Their kinetic age varies from 3 to 150\,Myr. \ion{H}{1} holes are found throughout the \ion{H}{1} disk of the parent galaxy, all the way to the edge of the observed atomic gas distribution. Across the sample, 77\% of the holes fall within $R_{25}$, the main optical body of a galaxy, the remainder populate the outskirts. Their number surface density is fairly constant within $R_{25}$ but drops at larger radii. A comparative analysis shows that:

\begin{enumerate}
\item \ion{H}{1} holes in dwarfs are rounder, on average than in spirals;
\item blow--out holes in spirals are not circular as, otherwise, the inclination under which the disks are viewed would cause the major axis of holes to be aligned with the major axis of the disk; 
\item there is a lower limit to the axial ratio at 0.4. 
\end{enumerate}

The above results lead us to propose that in spiral galaxies, shear is an important mechanism that not only modifies the shape of the holes but also limits the time they are observable, i.e., their kinetic age. This age limit is of order 50 Myr. Dwarf galaxies predominantly rotate more like solid bodies, hence shear does not play a major role there and consequently holes in dwarf galaxies are rounder, larger and older than in spirals. Here we note that diameter and kinetic age are closely linked as the expansion velocity covers a modest range (typically a factor of a few) whereas the diameter of the holes covers an order of magnitude in range.

Dividing the sample into dwarf and spiral galaxies, we were able to draw the following broad conclusions.

\begin{enumerate}
\item we find no difference in expansion velocity between dwarfs and spirals;
\item the radial distribution of \ion{H}{1} holes is more extended in dwarf galaxies;
\item holes outside $R_{25}$ are larger and older;
\end{enumerate}

The fact that holes outside $R_{25}$ are larger and older is likely again a consequence of shear which is most effective within $R_{25}$. We did not find a correlation between the diameter and expansion velocity as found in the LMC/SMC \citep{sta07}. We note that their correlation is strongest for small holes and speculate that this relation is due to small holes not having reached their final size yet, but that they are still being powered by ongoing SN explosions.

The porosity of the ISM was found to be higher in later Hubble types. We speculate that the lack of destruction mechanisms in dwarf galaxies contributes to this trend. In addition, higher porosity values translate to a more dynamic ISM where star formation becomes inefficient. The surface and volume porosity ($Q_\mathrm{2D}$ and $Q_\mathrm{3D}$) correlate with the type of the host galaxy as predicted by \citet{sil97}: later Hubble types tend to be more porous which reflects their current lower star formation rate per unit mass. 

The size distribution of the holes in our sample follows a power law with a slope of $a_{\mathrm{\nu}}\sim -2.9$ in broad agreement with \citet{oey97}. Individual galaxies have slopes ranging from $-2$ to $-4$. Since our spatial resolution is at best 100 pc, we are missing a substantial number of small holes. Assuming a power law shape for the size distribution, and a minimum hole size of 10 pc, expected to be the size of a shell created by a single O star, we can extrapolate our observations and predict the total size distribution of holes.

Assuming that the holes are the result of massive star formation, we derive values for the SFR over a time--frame of 50 Myr that correlate with the SFR derived based on other tracers but at the same time underestimating the true SFR if we restrict ourselves to the observed holes. When we apply an extrapolation to holes created by a single supernova, we find a surprisingly close agreement. This agreement does not mean that one would able to estimate the SFR of a galaxy based on the holes, but it does show that \ion{H}{1} holes can be understood as being the result of star formation. 

\acknowledgements
We are grateful to Martin Hardcastle for helpful discussions on the statistical methods used.
AU has been supported through a grant by the UK Science \& Technology Facilities Council. EB gratefully acknowledges financial support through an EU Marie Curie International Reintegration Grant (Contract No. MIRG-CT-6-2005-013556). The work of WJGdB is based upon research supported by the South African Research Chairs Initiative of the Department of Science and Technology and National Research Foundation.

\clearpage

\clearpage

\clearpage

\clearpage

\clearpage

\clearpage

\clearpage

\clearpage

\clearpage

\clearpage

\clearpage

\clearpage

\clearpage
\begin{figure}
\begin{center}
\includegraphics[angle=0,scale=.69]{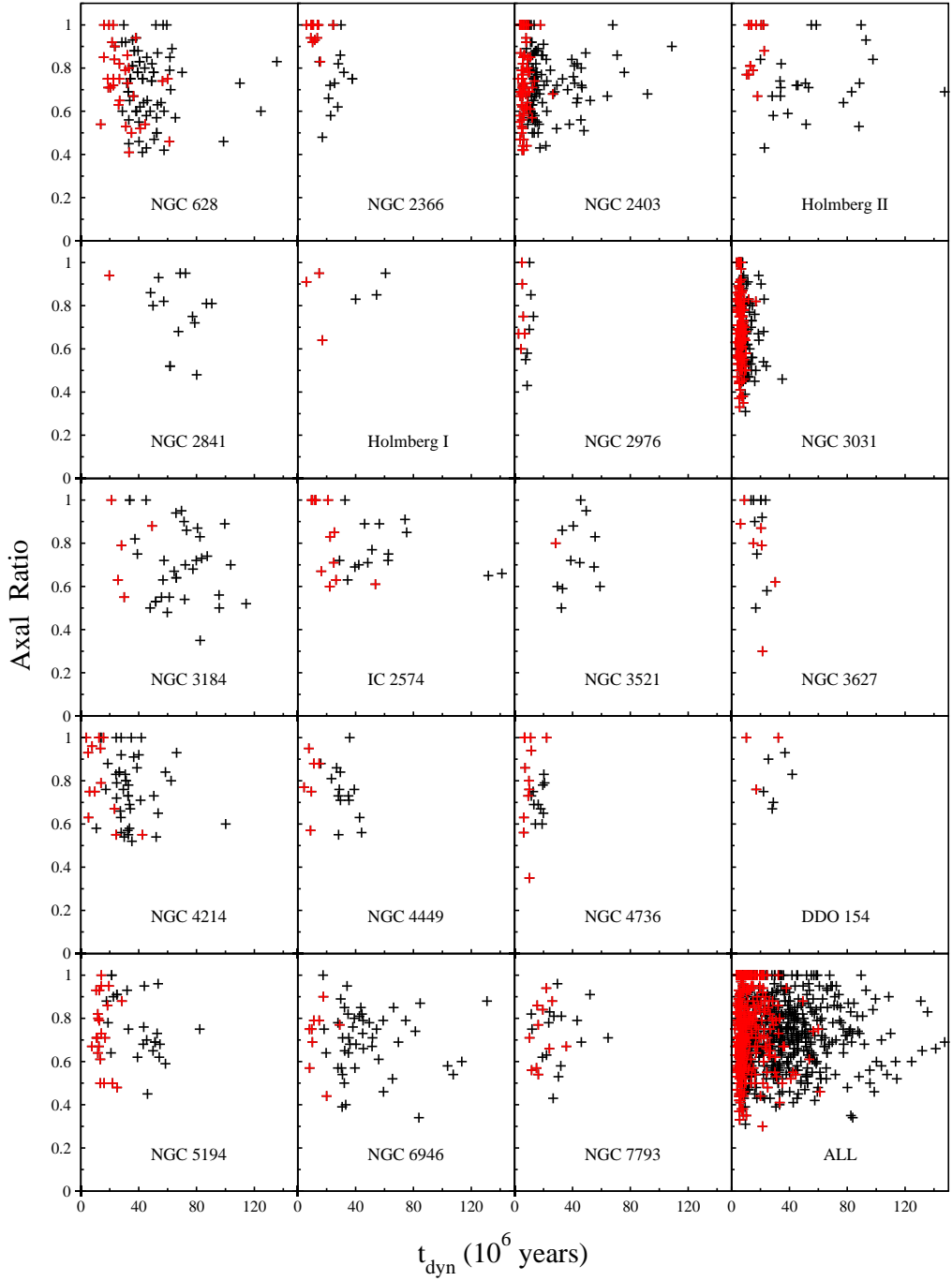}
\end{center}
\caption{Axial ratio vs.\ kinetic age in units of $10^6$\,yr of the stalled (black) and expanding (red) \ion{H}{1} holes.\label{fig:t-axial}}
\end{figure}

\clearpage

\clearpage

\clearpage

\clearpage

\clearpage

\clearpage

\clearpage

\clearpage

\clearpage

\clearpage

\clearpage

\clearpage

\clearpage

\clearpage

\clearpage

\clearpage

\clearpage
\appendix

\section{Description}
\label{descr}

In this appendix, we present each galaxy individually. We give a short description of each galaxy and an outline of our results. The complete hole catalogues are presented in section~\ref{tables}, Tables~\ref{tbl-NGC628}--\ref{tbl-NGC7793}. Each table has 14 columns: Column 1 is the number of each hole; Columns 2 and 3 the position (in $\alpha$ and $\delta$ in J2000.0); Column 4 the heliocentric velocity, $V_\mathrm{hel}$, of the centre; Column 5 the type; Column 6 the diameter, $d$; Column 7 the expansion velocity, $V_\mathrm{exp}$; Column 8 the position angle, $P.A.$; Column 9 the axial ratio of the hole, $b_\mathrm{min}/b_\mathrm{maj}$; Column 10 the galactocentric radius, R; Column 11 the volume density, $n_\mathrm{HI}$, around the hole; Column 12 the kinetic age, $t_\mathrm{kin}$; Column 13 the estimated energy requirement, $log(E_\mathrm{E})$; and Column 14 the estimated \ion{H}{1} mass, $log(M_\mathrm{HI})$, missing from the hole.
Figures~\ref{NGC628-map}--\ref{NGC7793-map} in section~\ref{figures} show the detected \ion{H}{1} holes overlaid as ellipses on the \ion{H}{1} surface brightness maps. 

\subsection{NGC\,628}
NGC\,628 is a grand-design Sc spiral galaxy located at a distance of 7.3 Mpc and is seen almost completely face--on. We detected 104 holes in its \ion{H}{1} layer, with diameters ranging from 0.24 kpc (the resolution limit) to 2 kpc. Only a small fraction of these holes are type 3 ($\approx 5\%$). Several supershells were detected, however 4 cases (holes number 24, 45, 47 and 99) are probably superpositions of 2 or more holes. 

\subsection{NGC\,2366}
NGC\,2366 is a dwarf galaxy member of the M\,81 group. It is considered to be a typical irregular galaxy and is characterized by a supergiant \ion{H}{2} complex (NGC\,2363) in the southwest \citep{you99}. We detected 26 holes in NGC\,2366 but since this is a highly inclined system ($i=63.8^{\circ}$) we might be missing out on some features. All of the detected holes appear to be relatively young as their kinetic age was found to be between 6--38 Myr. In a previous \ion{H}{1} study \citet{hun01} searched for but did not detect any holes, however their spatial resolution was $\approx 260$ pc, almost 2.5 times worse than our resolution.

\subsection{NGC\,2403}
Another member of the M\,81 group, NGC\,2403, is a late-type spiral galaxy located at a distance of 3.2 Mpc. It has previously been studied by \citet{thi98} and by \citet{mas99} using an automated object recognition package who detected 50 and 601 holes respectively. Even though this galaxy is highly inclined ($i=62.9^{\circ}$), its close proximity and good resolution allowed us to detect 173 holes in its \ion{H}{1} distribution. A few supershells were detected but the majority (81\%) of the holes are smaller than 350 pc in diameter and younger than 20 Myr. There are also several special cases: some of the holes are most likely a superposition of two or more holes (holes number 51, 157 and 173); holes number 10, 43 have a peculiar signature in \textit{pV} space.

\subsection{Holmberg\,II}
Holmberg\,II is a gas-rich dwarf galaxy located at a distance of 3.4 Mpc and is a member of the M\,81 group of galaxies as well. An \ion{H}{1} hole analysis was carried out by \citet{puc92} in which they found 51 holes across the entire face of this galaxy. We detected 39 holes in Holmberg\,II the majority of which were also detected by \citet{puc92}, all with very good agreement with respect to the position and size of the holes. The difference in the number of holes detected is due to our strict criteria of what constitutes a genuine hole as well as to the fact that several holes in our catalog were classified as a superposition of two (or more) holes from the Puche et al. catalog. As Figure~\ref{HoII-map} illustrates, these holes dominate the \ion{H}{1} distribution of Holmberg\,II. In addition several supershells were detected, the largest one (no 17) being over 2 kpc, one of the largest and oldest ($\sim 150$ Myr) across the entire sample of holes. Note that hole no 37 is probably a superposition of two or more holes.

\subsection{DDO\,53}
The dwarf irregular galaxy DDO\,53 lies at a distance of 3.6 Mpc and also belongs to the M\,81 group. It is the smallest galaxy in our sample with an \ion{H}{1} radius of only 2 kpc. It is therefore not surprising that we detected only 3 holes in its \ion{H}{1} distribution, one of each type, with diameters in the range of 180-360 pc.

\subsection{NGC\,2841}
NGC\,2841 is an early-type Sc spiral and is the most distant galaxy in our sample located at 14.1 Mpc. The limited spatial resolution and the high inclination of this galaxy meant that we were only able to detect supershells of type 1. We found 15 \ion{H}{1} holes and only one of these shows signs of expansion (number 10). In addition, holes number 4, 12 and 13 are possibly superpositions of holes. Hole no 5 is probably not a genuine hole but the result of the warp in the \ion{H}{1} disk.

\subsection{Holmberg\,I}
Holmberg\,I is a dwarf irregular galaxy which forms part of the M\,81 group of galaxies. Its \ion{H}{1} distribution is dominated by a supergiant shell as was noted by \citet{ott01} which does not coincide with the kinetic centre of the galaxy. We detected 6 holes in this galaxy including the one detected by \citet{ott01}. There is good agreement with respect to the position of this hole between both studies. 

\subsection{NGC\,2976}
The dwarf galaxy NGC\,2976 is located at 3.6 Mpc and belongs to the M\,81 group of galaxies. In NGC\,2976 we found only 10 holes, which is not surprising given that this is one of the smallest galaxies in our sample with an \ion{H}{1} disk radius of $R_\mathrm{max}=2.5 \, \mathrm{kpc}$. No supershells were detected. 

\subsection{NGC\,3031}
NGC\,3031, also known as M\,81, is a grand design spiral galaxy located at a distance of 3.6 Mpc. The analysis of NGC\,3031 revealed 321 \ion{H}{1} holes, the largest number of holes detected in any of the galaxies in this sample. This is due to various factors such as its relatively large size, close proximity and high spatial resolution of the observations. We were able to detect holes down to $\sim 80$ pc which led to the detection of a wealth of small holes; almost 90\% of the total holes detected have a size less than 200 pc. Only a handful of supershells were detected. The mean kinetic age is 7.8 Myr with 80\% of the holes being younger than 9 Myr. Several features detected (holes no: 34, 72, 88, 93, 94, 97, 99, 102, 106, 109, 113, 114, 170, 255 and 273) are probably not genuine holes but the result of the warp in the \ion{H}{1} disk. Due to its large angular size, M\,81 was divided into 6 areas (Figures~\ref{M81-areas} --~\ref{M81-SW}) to be able to show all detected holes. 

\subsection{NGC\,3184}
NGC\,3184 is a spiral galaxy located at $D=11.1$ Mpc and is viewed almost face-on. Its analysis revealed 40 \ion{H}{1} holes, the majority of those being supershells (80\%). This could be attributed to the large distance of this galaxy which only allowed us to detect holes with diameters larger than 400 pc. We failed to detect any type 3 holes something which is expected since the scale height of the disk was found to be 250 pc, significantly smaller than our spatial resolution. Nevertheless, this is the largest number of supershells detected in any one galaxy. The mean kinetic age of the holes is also high ($\approx 64$ Myr). Note that hole number 13 is a superposition of 2 holes.

\subsection{IC\,2574}
Another nearby ($D=4.0$ Mpc) dwarf galaxy member of the M\,81 group, IC\,2574 hosts a relatively large number of \ion{H}{1} holes \citep{wal99}. One particular hole (our hole no 21) has been extensively studied at different wavelengths from radio to X--ray (\citealt{ste00}, \citealt{can05}, \citealt{wei09a}). These studies showed that there is a remnant star cluster in the center of this hole and revealed evidence of propagating star formation along its rim. Here we report on 29 \ion{H}{1} holes compared to 48 holes detected by \citet{wal99} and 22 detected by \citet{ric08} the largest one being larger than 2 kpc in diameter. Almost all of the 29 holes we present here have been detected by \citet{wal99}. The difference in the total number of holes detected is due to the strict criteria we used to classify a structure as a genuine hole. Almost half of the holes (45\%) show signs of expansion.

\subsection{NGC\,3521}
Another distant galaxy in our sample, NGC\,3521 is a spiral galaxy with a prominent bar located at a distance of 10.7 Mpc and viewed almost edge-on ($i=72.7^{\circ}$). The analysis of NGC\,3521 revealed 13 holes in the \ion{H}{1} distribution. The small number of holes detected and the lack of type 3 holes can be attributed to the large distance of this galaxy as well as its high inclination. All of the holes detected were consequently classified as supershells.

\subsection{NGC\,3627}
NGC\,3627 is a spiral galaxy at an adopted distance of $D=9.3$ Mpc and is a member of the Leo Triplet. Its high inclination ($i=61.8^{\circ}$) meant we were only able to detect 18 holes in NGC\,3627, two of them being type 3. One holes worth noting is 7 which has unusually high expansion velocities of 65 km\,s$^{-1}$ respectively, the highest observed in the entire sample. 

\subsection{NGC\,4214}
NGC\,4214 is a nearby ($D=2.9$ Mpc) dwarf irregular galaxy classified as a starburst galaxy. We found 56 \ion{H}{1} holes in NGC\,4214 with a wide range of sizes, from 0.11 to 1.2 kpc. Given the close proximity of this galaxy and consequently the high spatial resolution it is surprising that only about 7\% of the holes detected are type 3. One possible explanation is that NGC\,4214 has a low scale height and therefore holes can break out of the disk faster.

\subsection{NGC\,4449}
Another dwarf irregular galaxy, NGC\,4449, is located at a distance of $D=4.2$ Mpc. It has a very extended \ion{H}{1} structure reaching out to 6 times the Holmberg radius. The $R_\mathrm{max}$ radius for this galaxy was taken at the edge of the \ion{H}{1} disk thus excluding this structure. What is also unusual about NGC\,4449 is that it has two counter--rotating gas systems \citep{hun98}. We detected 20 \ion{H}{1} holes in NGC\,4449, most of them located within $R_{25}$ and 6 located in the extended structure outside the $R_\mathrm{max}$ radius.

\subsection{NGC\,4736}
NGC\,4736 is a ringed spiral galaxy located at $D=4.7$ Mpc in which we encountered 23 holes in its \ion{H}{1} distribution. The range of their diameters is quite small (0.21-0.45 kpc) and they are young holes ($t_\mathrm{kin}<22 $ Myr). As Figure~\ref{NGC4736-map} illustrates, the majority of the holes are located at small galactocentric radii and only type 1 holes lie at larger galactocentric radii. Finally note that hole number 11 is probably a superposition of 2 holes.

\subsection{DDO\,154}
The gas-rich dwarf galaxy DDO\,154 is a nearby galaxy with a very extended and warped \ion{H}{1} disk. \citet{hof01} found 2 holes in the \ion{H}{1} distribution which were also detected in our analysis. We report here on a total of 9 holes. The scale height of DDO\,154 was found to be $\approx$ 500 pc which explains why holes can grow to a relatively large diameter before breaking out of the disk, as is the case with shell number 6 ($\approx 400$ pc). 

\subsection{NGC\,5194}
NGC\,5194 is a grand-design spiral galaxy at a distance of $D=8.0$ Mpc which is undergoing interaction with NGC\,5195. We report 47 holes in NGC\,5194 with only one type 3 hole detected. Most of holes are in fact supershells, with one of those (no 5) being superposition of two or more holes.

\subsection{NGC\,6946}
Located at a distance of 5.9 Mpc, NGC\,6946 is a large spiral galaxy seen almost face-on. Previous \ion{H}{1} studies by \citet{kam93} and \citet{boo07} revealed 19 and 121 \ion{H}{1} holes respectively. We detected 56 holes, a subset of those detected by \citet{boo07}. There is excellent agreement for holes larger than 700 pc between our samples but not for smaller ones. This is due to the strict criteria ($Q>5$) we applied on holes whose size is comparable to the resolution limit. NGC\,6946 hosts the largest hole in our sample (no 58) with a size of 2.1 kpc and energy requirements in excess of $4\times10^{54}\,\mathrm{erg}$, which corresponds to $\sim4000$ SNe. Most of the holes detected (84\%) are type 1, with two holes (no 15 and 16) appearing as a superposition of more than one holes.

\subsection{NGC\,7793}
A member of the Sculptor group, NGC\,7793 is an Sd spiral galaxy at $D=3.9$ Mpc. A detailed analysis revealed 27 \ion{H}{1} holes distributed almost uniformly across the \ion{H}{1} disk with sizes ranging from 0.19 to 1.05 kpc. 

\section{Tables}
\label{tables}

\setcounter{table}{0} 
\renewcommand{\thetable}{B\arabic{table}}

\begin{deluxetable}{cccccccccccccc}
\tabletypesize{\scriptsize}
\rotate
\tablecaption{NGC\,628 hole properties}
\tablewidth{0pt}
\tablehead{
\colhead{\textit{No}} & \colhead{\textit{$\alpha$}} & \colhead{\textit{$\delta$}} & \colhead{$V_\mathrm{hel}$} & \colhead{\textit{Type}} &
\colhead{$d$} & \colhead{$V_\mathrm{exp}$} &\colhead{\textit{P.A.}} & \colhead{\textit{Axial Ratio}} & 
\colhead{R} & \colhead{$n_\mathrm{HI}$} & \colhead{$t_\mathrm{kin}$} & \colhead{$log(E_\mathrm{E})$} & 
 \colhead{$log(M_\mathrm{HI})$} \\
\colhead{} & \colhead{($h\ \ \ m\ \ \ s$)} & \colhead{($\arcdeg \ \ \  \arcmin \ \ \  \arcsec$)} & \colhead{(km\,s$^{-1}$)} & \colhead{($1-3$)} &
\colhead{(pc)} &\colhead{(km\,s$^{-1}$)} & \colhead{($^{\circ}$)} & \colhead{} & 
\colhead{(kpc)} & \colhead{(cm$^{-3}$)} & \colhead{(Myr)} & \colhead{($10^{50}\, \mathrm{erg}$)} & 
\colhead{($10^{4}\, \mathrm{M_{\odot}}$)} }
\startdata
1&1 36 20.1&15 43 45.2&645&1&520&7&109&0.7&13.1&0.1&36&1.2&1.2\\
2&1 36 20.2&15 45 07.7&650&1&849&7&0&1.0&11.8&0.1&59&2.0&1.9\\
3&1 36 23.4&15 43 43.6&639&1&575&7&149&0.5&11.7&0.1&40&1.6&1.6\\
4&1 36 24.0&15 44 30.1&639&2&546&10&126&0.8&10.5&0.2&27&1.9&1.6\\
5&1 36 24.1&15 50 55.6&665&2&459&7&0&0.9&12.4&0.2&32&1.4&1.3\\
6&1 36 24.3&15 44 13.6&639&2&520&7&114&0.7&10.7&0.2&36&1.7&1.6\\
7&1 36 24.6&15 47 15.1&650&1&720&7&13&0.8&8.9&0.4&50&2.4&2.5\\
8&1 36 24.8&15 44 51.1&632&1&582&7&81&0.8&9.9&0.2&41&1.9&2.3\\
\enddata
\label{tbl-NGC628}
\end{deluxetable}
\clearpage

\begin{deluxetable}{cccccccccccccc}
\tabletypesize{\scriptsize}
\rotate
\tablecaption{NGC\,2366 hole properties}
\tablewidth{0pt}
\tablehead{
\colhead{\textit{No}} & \colhead{\textit{$\alpha$}} & \colhead{\textit{$\delta$}} & \colhead{$V_\mathrm{hel}$} & \colhead{\textit{Type}} &
\colhead{$d$} & \colhead{$V_\mathrm{exp}$} &\colhead{\textit{P.A.}} & \colhead{\textit{Axial Ratio}} & 
\colhead{R} & \colhead{$n_\mathrm{HI}$} & \colhead{$t_\mathrm{kin}$} & \colhead{$log(E_\mathrm{E})$} & 
 \colhead{$log(M_\mathrm{HI})$} \\
\colhead{} & \colhead{($h\ \ \ m\ \ \ s$)} & \colhead{($\arcdeg \ \ \  \arcmin \ \ \  \arcsec$)} & \colhead{(km\,s$^{-1}$)} & \colhead{($1-3$)} &
\colhead{(pc)} &\colhead{(km\,s$^{-1}$)} & \colhead{($^{\circ}$)} & \colhead{} & 
\colhead{(kpc)} & \colhead{(cm$^{-3}$)} & \colhead{(Myr)} & \colhead{($10^{50}\, \mathrm{erg}$)} & 
\colhead{($10^{4}\, \mathrm{M_{\odot}}$)} }
\startdata
1&7 28 23.3&69 12 03.9&58&1&412&9&133&0.7&4.2&0.1&22&1.2&1.0\\
2&7 28 30.7&69 10 13.0&61&1&418&9&31&0.6&3.3&0.2&23&1.5&1.2\\
3&7 28 31.0&69 08 50.5&58&1&693&9&31&0.8&5.0&0.1&38&1.6&1.4\\
4&7 28 33.2&69 10 34.0&66&3&250&9&0&1.0&2.9&0.3&14&1.0&0.8\\
5&7 28 35.4&69 11 20.6&63&1&515&9&16&0.8&2.3&0.5&28&2.3&2.0\\
6&7 28 37.2&69 09 32.6&56&1&391&9&48&0.7&4.2&0.1&21&1.1&0.9\\
7&7 28 41.6&69 12 07.1&94&2&500&10&0&1.0&1.4&0.7&24&2.4&2.5\\
8&7 28 42.8&69 11 04.1&94&2&300&24&0&1.0&2.2&0.6&6&2.2&1.3\\
\enddata
\label{tbl-NGC2366}
\end{deluxetable}
\clearpage

\begin{deluxetable}{cccccccccccccc}
\tabletypesize{\scriptsize}
\rotate
\tablecaption{NGC\,2403 hole properties}
\tablewidth{0pt}
\tablehead{
\colhead{\textit{No}} & \colhead{\textit{$\alpha$}} & \colhead{\textit{$\delta$}} & \colhead{$V_\mathrm{hel}$} & \colhead{\textit{Type}} &
\colhead{$d$} & \colhead{$V_\mathrm{exp}$} &\colhead{\textit{P.A.}} & \colhead{\textit{Axial Ratio}} & 
\colhead{R} & \colhead{$n_\mathrm{HI}$} & \colhead{$t_\mathrm{kin}$} & \colhead{$log(E_\mathrm{E})$} & 
 \colhead{$log(M_\mathrm{HI})$} \\
\colhead{} & \colhead{($h\ \ \ m\ \ \ s$)} & \colhead{($\arcdeg \ \ \  \arcmin \ \ \  \arcsec$)} & \colhead{(km\,s$^{-1}$)} & \colhead{($1-3$)} &
\colhead{(pc)} &\colhead{(km\,s$^{-1}$)} & \colhead{($^{\circ}$)} & \colhead{} & 
\colhead{(kpc)} & \colhead{(cm$^{-3}$)} & \colhead{(Myr)} & \colhead{($10^{50}\, \mathrm{erg}$)} & 
\colhead{($10^{4}\, \mathrm{M_{\odot}}$)} }
\startdata
1&7 35 13.9&65 37 04.3&61&1&272&15&26&0.8&12.8&0.0&9&0.5&0.0\\
2&7 35 17.0&65 36 53.2&66&2&159&15&174&0.6&12.5&0.0&5&-0.2&-0.6\\
3&7 35 18.3&65 38 04.2&56&1&179&8&110&0.9&11.3&0.0&11&-0.5&-0.6\\
4&7 35 20.7&65 40 32.1&20&1&232&8&105&0.8&10.1&0.1&14&0.0&0.0\\
5&7 35 20.8&65 38 26.1&46&1&223&8&79&0.7&10.7&0.1&14&-0.1&-0.2\\
6&7 35 20.8&65 41 34.1&25&1&232&8&32&0.8&10.2&0.1&14&0.0&-0.1\\
7&7 35 23.4&65 36 08.0&71&2&220&10&166&0.6&12.4&0.1&11&0.1&-0.2\\
8&7 35 29.1&65 39 25.8&30&1&302&8&85&0.8&9.2&0.1&18&0.4&0.4\\
\enddata
\label{tbl-NGC2403}
\end{deluxetable}
\clearpage

\begin{deluxetable}{cccccccccccccc}
\tabletypesize{\scriptsize}
\rotate
\tablecaption{Holmberg\,II hole properties}
\tablewidth{0pt}
\tablehead{
\colhead{\textit{No}} & \colhead{\textit{$\alpha$}} & \colhead{\textit{$\delta$}} & \colhead{$V_\mathrm{hel}$} & \colhead{\textit{Type}} &
\colhead{$d$} & \colhead{$V_\mathrm{exp}$} &\colhead{\textit{P.A.}} & \colhead{\textit{Axial Ratio}} & 
\colhead{R} & \colhead{$n_\mathrm{HI}$} & \colhead{$t_\mathrm{kin}$} & \colhead{$log(E_\mathrm{E})$} & 
 \colhead{$log(M_\mathrm{HI})$} \\
\colhead{} & \colhead{($h\ \ \ m\ \ \ s$)} & \colhead{($\arcdeg \ \ \  \arcmin \ \ \  \arcsec$)} & \colhead{(km\,s$^{-1}$)} & \colhead{($1-3$)} &
\colhead{(pc)} &\colhead{(km\,s$^{-1}$)} & \colhead{($^{\circ}$)} & \colhead{} & 
\colhead{(kpc)} & \colhead{(cm$^{-3}$)} & \colhead{(Myr)} & \colhead{($10^{50}\, \mathrm{erg}$)} & 
\colhead{($10^{4}\, \mathrm{M_{\odot}}$)} }
\startdata
1&8 18 22.2&70 44 13.1&148&1&1260&7&162&0.5&4.7&0.1&88&2.8&2.7\\
2&8 18 28.9&70 42 20.3&161&1&1400&7&89&0.8&4.0&0.2&98&3.0&3.0\\
3&8 18 34.5&70 40 32.0&166&1&287&7&168&0.8&4.3&0.2&20&0.9&1.0\\
4&8 18 40.1&70 41 36.3&164&1&731&7&112&0.7&3.2&0.3&51&2.4&2.6\\
5&8 18 42.4&70 40 58.6&166&1&412&7&89&0.6&3.3&0.3&29&1.6&1.5\\
6&8 18 48.8&70 43 40.3&156&1&486&7&84&0.7&1.8&0.7&34&2.4&2.2\\
7&8 18 50.3&70 45 59.8&143&2&279&10&0&1.0&3.1&0.4&14&1.0&1.3\\
8&8 18 52.0&70 42 46.2&156&3&345&10&0&1.0&1.5&0.9&17&2.2&2.0\\
\enddata
\label{tbl-HoII}
\end{deluxetable}
\clearpage

\begin{deluxetable}{cccccccccccccc}
\tabletypesize{\scriptsize}
\rotate
\tablecaption{DDO\,53 hole properties}
\tablewidth{0pt}
\tablehead{
\colhead{\textit{No}} & \colhead{\textit{$\alpha$}} & \colhead{\textit{$\delta$}} & \colhead{$V_\mathrm{hel}$} & \colhead{\textit{Type}} &
\colhead{$d$} & \colhead{$V_\mathrm{exp}$} &\colhead{\textit{P.A.}} & \colhead{\textit{Axial Ratio}} & 
\colhead{R} & \colhead{$n_\mathrm{HI}$} & \colhead{$t_\mathrm{kin}$} & \colhead{$log(E_\mathrm{E})$} & 
 \colhead{$log(M_\mathrm{HI})$} \\
\colhead{} & \colhead{($h\ \ \ m\ \ \ s$)} & \colhead{($\arcdeg \ \ \  \arcmin \ \ \  \arcsec$)} & \colhead{(km\,s$^{-1}$)} & \colhead{($1-3$)} &
\colhead{(pc)} &\colhead{(km\,s$^{-1}$)} & \colhead{($^{\circ}$)} & \colhead{} & 
\colhead{(kpc)} & \colhead{(cm$^{-3}$)} & \colhead{(Myr)} & \colhead{($10^{50}\, \mathrm{erg}$)} & 
\colhead{($10^{4}\, \mathrm{M_{\odot}}$)} }
\startdata
1&8 34 10.0&66 10 38.0&17&1&363&8&44&0.7&0.4&0.6&22&1.8&1.6\\
2&8 34 11.7&66 10 42.5&22&2&309&8&148&0.8&0.6&0.4&19&1.4&1.2\\
3&8 34 12.2&66 11 18.5&17&3&179&10&145&0.3&1.3&0.4&9&0.7&0.5\\
\enddata
\label{tbl-DDO53}
\end{deluxetable}
\clearpage

\begin{deluxetable}{cccccccccccccc}
\tabletypesize{\scriptsize}
\rotate
\tablecaption{NGC\,2841 hole properties}
\tablewidth{0pt}
\tablehead{
\colhead{\textit{No}} & \colhead{\textit{$\alpha$}} & \colhead{\textit{$\delta$}} & \colhead{$V_\mathrm{hel}$} & \colhead{\textit{Type}} &
\colhead{$d$} & \colhead{$V_\mathrm{exp}$} &\colhead{\textit{P.A.}} & \colhead{\textit{Axial Ratio}} & 
\colhead{R} & \colhead{$n_\mathrm{HI}$} & \colhead{$t_\mathrm{kin}$} & \colhead{$log(E_\mathrm{E})$} & 
 \colhead{$log(M_\mathrm{HI})$} \\
\colhead{} & \colhead{($h\ \ \ m\ \ \ s$)} & \colhead{($\arcdeg \ \ \  \arcmin \ \ \  \arcsec$)} & \colhead{(km\,s$^{-1}$)} & \colhead{($1-3$)} &
\colhead{(pc)} &\colhead{(km\,s$^{-1}$)} & \colhead{($^{\circ}$)} & \colhead{} & 
\colhead{(kpc)} & \colhead{(cm$^{-3}$)} & \colhead{(Myr)} & \colhead{($10^{50}\, \mathrm{erg}$)} & 
\colhead{($10^{4}\, \mathrm{M_{\odot}}$)} }
\startdata
1&9 21 36.4&51 06 55.8&379&1&1333&9&161&1.0&38.4&0.0&72&1.8&1.7\\
2&9 21 36.9&51 02 25.8&539&1&1136&9&163&0.5&34.3&0.0&62&1.9&1.7\\
3&9 21 40.3&51 07 58.9&363&1&1597&9&162&0.8&44.6&0.0&87&2.0&1.8\\
4&9 21 44.5&51 01 40.9&384&1&1264&9&0&0.9&23.0&0.1&69&2.5&2.6\\
5&9 21 45.0&51 01 10.9&405&1&1253&9&145&0.8&23.5&0.1&68&2.3&2.1\\
6&9 21 45.6&51 01 57.4&374&1&886&9&31&0.9&21.0&0.1&48&2.0&1.7\\
7&9 21 47.1&51 02 21.4&363&1&1474&9&108&0.5&19.4&0.1&80&2.7&2.6\\
8&9 21 52.1&50 59 50.0&441&1&1421&9&75&0.8&15.0&0.1&77&2.7&2.6\\
\enddata
\label{tbl-NGC2841}
\end{deluxetable}
\clearpage

\begin{deluxetable}{cccccccccccccc}
\tabletypesize{\scriptsize}
\rotate
\tablecaption{Holmberg\,I hole properties}
\tablewidth{0pt}
\tablehead{
\colhead{\textit{No}} & \colhead{\textit{$\alpha$}} & \colhead{\textit{$\delta$}} & \colhead{$V_\mathrm{hel}$} & \colhead{\textit{Type}} &
\colhead{$d$} & \colhead{$V_\mathrm{exp}$} &\colhead{\textit{P.A.}} & \colhead{\textit{Axial Ratio}} & 
\colhead{R} & \colhead{$n_\mathrm{HI}$} & \colhead{$t_\mathrm{kin}$} & \colhead{$log(E_\mathrm{E})$} & 
 \colhead{$log(M_\mathrm{HI})$} \\
\colhead{} & \colhead{($h\ \ \ m\ \ \ s$)} & \colhead{($\arcdeg \ \ \  \arcmin \ \ \  \arcsec$)} & \colhead{(km\,s$^{-1}$)} & \colhead{($1-3$)} &
\colhead{(pc)} &\colhead{(km\,s$^{-1}$)} & \colhead{($^{\circ}$)} & \colhead{} & 
\colhead{(kpc)} & \colhead{(cm$^{-3}$)} & \colhead{(Myr)} & \colhead{($10^{50}\, \mathrm{erg}$)} & 
\colhead{($10^{4}\, \mathrm{M_{\odot}}$)} }
\startdata
1&9 40 13.6&71 11 29.3&130&1&491&6&123&0.8&1.8&0.2&40&1.6&1.6\\
2&9 40 22.1&71 12 03.5&146&2&195&16&13&0.9&1.6&0.5&6&0.7&1.3\\
3&9 40 28.6&71 12 07.7&138&3&363&12&45&1.0&1.4&0.5&15&1.5&1.9\\
4&9 40 30.0&71 11 03.2&135&1&744&6&4&1.0&0.2&0.4&61&2.8&2.4\\
5&9 40 33.6&71 12 49.5&153&3&418&12&45&0.6&2.1&0.3&17&1.4&1.9\\
6&9 40 59.1&71 12 36.3&138&1&668&6&45&0.8&3.1&0.0&54&1.2&1.1\\
\enddata
\label{tbl-HoI}
\end{deluxetable}
\clearpage

\begin{deluxetable}{cccccccccccccc}
\tabletypesize{\scriptsize}
\rotate
\tablecaption{NGC\,2976 hole properties}
\tablewidth{0pt}
\tablehead{
\colhead{\textit{No}} & \colhead{\textit{$\alpha$}} & \colhead{\textit{$\delta$}} & \colhead{$V_\mathrm{hel}$} & \colhead{\textit{Type}} &
\colhead{$d$} & \colhead{$V_\mathrm{exp}$} &\colhead{\textit{P.A.}} & \colhead{\textit{Axial Ratio}} & 
\colhead{R} & \colhead{$n_\mathrm{HI}$} & \colhead{$t_\mathrm{kin}$} & \colhead{$log(E_\mathrm{E})$} & 
 \colhead{$log(M_\mathrm{HI})$} \\
\colhead{} & \colhead{($h\ \ \ m\ \ \ s$)} & \colhead{($\arcdeg \ \ \  \arcmin \ \ \  \arcsec$)} & \colhead{(km\,s$^{-1}$)} & \colhead{($1-3$)} &
\colhead{(pc)} &\colhead{(km\,s$^{-1}$)} & \colhead{($^{\circ}$)} & \colhead{} & 
\colhead{(kpc)} & \colhead{(cm$^{-3}$)} & \colhead{(Myr)} & \colhead{($10^{50}\, \mathrm{erg}$)} & 
\colhead{($10^{4}\, \mathrm{M_{\odot}}$)} }
\startdata
1&9 47 06.0&67 54 13.5&-7&1&158&9&0&0.6&2.7&0.1&9&-0.2&-0.3\\
2&9 47 07.8&67 56 55.5&59&1&140&9&104&0.5&2.2&0.2&8&0.0&-0.2\\
3&9 47 11.0&67 55 58.5&59&1&169&12&44&0.7&1.1&1.2&7&1.4&1.3\\
4&9 47 11.6&67 54 15.0&-17&2&179&15&62&0.8&1.6&0.4&6&1.0&0.5\\
5&9 47 13.7&67 54 48.0&-2&2&155&15&0&1.0&0.5&1.5&5&1.5&1.2\\
6&9 47 13.7&67 55 46.5&28&1&187&9&155&0.7&0.9&1.3&10&1.3&1.4\\
7&9 47 14.5&67 55 55.5&39&3&127&23&150&0.7&1.2&1.0&3&1.3&0.4\\
8&9 47 14.8&67 54 57.0&-2&1&158&9&78&0.4&0.1&2.3&9&1.4&1.2\\
9&9 47 19.6&67 53 42.0&-63&2&134&15&135&0.6&1.5&1.2&4&1.1&0.6\\
10&9 47 21.4&67 54 24.0&-33&3&164&15&17&0.9&1.0&1.8&5&1.6&1.5\\
11&9 47 23.8&67 53 03.0&-63&1&239&9&81&0.8&2.2&0.1&13&0.5&0.3\\
12&9 47 27.5&67 54 43.5&-33&1&190&9&0&1.0&2.4&0.2&10&0.5&0.3\\
13&9 47 28.9&67 55 18.0&-7&1&206&9&93&0.8&3.1&0.1&11&0.0&-0.1\\
\enddata
\label{tbl-NGC2976}
\end{deluxetable}
\clearpage

\begin{deluxetable}{cccccccccccccc}
\tabletypesize{\scriptsize}
\rotate
\tablecaption{NGC\,3031 hole properties}
\tablewidth{0pt}
\tablehead{
\colhead{\textit{No}} & \colhead{\textit{$\alpha$}} & \colhead{\textit{$\delta$}} & \colhead{$V_\mathrm{hel}$} & \colhead{\textit{Type}} &
\colhead{$d$} & \colhead{$V_\mathrm{exp}$} &\colhead{\textit{P.A.}} & \colhead{\textit{Axial Ratio}} & 
\colhead{R} & \colhead{$n_\mathrm{HI}$} & \colhead{$t_\mathrm{kin}$} & \colhead{$log(E_\mathrm{E})$} & 
 \colhead{$log(M_\mathrm{HI})$} \\
\colhead{} & \colhead{($h\ \ \ m\ \ \ s$)} & \colhead{($\arcdeg \ \ \  \arcmin \ \ \  \arcsec$)} & \colhead{(km\,s$^{-1}$)} & \colhead{($1-3$)} &
\colhead{(pc)} &\colhead{(km\,s$^{-1}$)} & \colhead{($^{\circ}$)} & \colhead{} & 
\colhead{(kpc)} & \colhead{(cm$^{-3}$)} & \colhead{(Myr)} & \colhead{($10^{50}\, \mathrm{erg}$)} & 
\colhead{($10^{4}\, \mathrm{M_{\odot}}$)} }
\startdata
1&9 53 56.2&69 12 36.5&119&2&131&12&79&0.5&14.2&0.2&5&0.1&-0.3\\
2&9 53 58.8&69 13 44.0&132&2&140&8&26&0.8&14.3&0.2&9&0.0&-0.1\\
3&9 53 59.4&69 13 26.0&119&1&155&8&35&0.5&14.1&0.2&10&0.0&-0.1\\
4&9 54 00.8&69 13 50.0&127&1&86&8&164&0.7&14.2&0.2&5&-0.7&-0.8\\
5&9 54 02.7&69 13 08.0&114&1&179&8&16&0.6&13.7&0.1&11&0.1&0.0\\
6&9 54 04.5&69 13 12.4&122&1&140&8&170&0.8&13.5&0.2&9&-0.2&-0.3\\
7&9 54 05.1&69 14 04.9&135&1&147&8&98&0.7&13.9&0.2&9&0.1&-0.1\\
8&9 54 11.3&69 13 46.8&132&2&108&8&86&0.7&13.3&0.2&7&-0.4&-0.5\\
\enddata
\label{tbl-NGC3031}
\end{deluxetable}
\clearpage

\begin{deluxetable}{cccccccccccccc}
\tabletypesize{\scriptsize}
\rotate
\tablecaption{NGC\,3184 hole properties}
\tablewidth{0pt}
\tablehead{
\colhead{\textit{No}} & \colhead{\textit{$\alpha$}} & \colhead{\textit{$\delta$}} & \colhead{$V_\mathrm{hel}$} & \colhead{\textit{Type}} &
\colhead{$d$} & \colhead{$V_\mathrm{exp}$} &\colhead{\textit{P.A.}} & \colhead{\textit{Axial Ratio}} & 
\colhead{R} & \colhead{$n_\mathrm{HI}$} & \colhead{$t_\mathrm{kin}$} & \colhead{$log(E_\mathrm{E})$} & 
 \colhead{$log(M_\mathrm{HI})$} \\
\colhead{} & \colhead{($h\ \ \ m\ \ \ s$)} & \colhead{($\arcdeg \ \ \  \arcmin \ \ \  \arcsec$)} & \colhead{(km\,s$^{-1}$)} & \colhead{($1-3$)} &
\colhead{(pc)} &\colhead{(km\,s$^{-1}$)} & \colhead{($^{\circ}$)} & \colhead{} & 
\colhead{(kpc)} & \colhead{(cm$^{-3}$)} & \colhead{(Myr)} & \colhead{($10^{50}\, \mathrm{erg}$)} & 
\colhead{($10^{4}\, \mathrm{M_{\odot}}$)} }
\startdata
1&10 18 04.9&41 25 32.5&586&1&823&7&152&0.7&7.6&0.7&58&2.9&2.8\\
2&10 18 05.3&41 26 41.5&566&1&944&7&12&0.6&8.3&0.6&66&3.0&2.9\\
3&10 18 05.4&41 23 37.0&625&1&856&7&0&0.5&9.4&0.4&60&2.7&2.7\\
4&10 18 05.7&41 25 01.0&594&2&798&13&5&0.6&7.3&0.7&30&3.2&2.7\\
5&10 18 06.4&41 25 59.5&576&1&798&7&162&0.6&6.9&0.9&56&3.0&2.8\\
6&10 18 06.5&41 27 25.0&550&1&944&7&12&0.6&9.2&0.4&66&2.7&2.7\\
7&10 18 07.6&41 26 05.5&571&1&742&7&104&0.5&6.2&1.0&52&2.9&2.7\\
8&10 18 08.5&41 22 26.5&646&1&923&7&117&0.7&11.2&0.2&64&2.4&2.6\\
\enddata
\label{tbl-NGC3184}
\end{deluxetable}
\clearpage

\begin{deluxetable}{cccccccccccccc}
\tabletypesize{\scriptsize}
\rotate
\tablecaption{IC\,2574 hole properties}
\tablewidth{0pt}
\tablehead{
\colhead{\textit{No}} & \colhead{\textit{$\alpha$}} & \colhead{\textit{$\delta$}} & \colhead{$V_\mathrm{hel}$} & \colhead{\textit{Type}} &
\colhead{$d$} & \colhead{$V_\mathrm{exp}$} &\colhead{\textit{P.A.}} & \colhead{\textit{Axial Ratio}} & 
\colhead{R} & \colhead{$n_\mathrm{HI}$} & \colhead{$t_\mathrm{kin}$} & \colhead{$log(E_\mathrm{E})$} & 
 \colhead{$log(M_\mathrm{HI})$} \\
\colhead{} & \colhead{($h\ \ \ m\ \ \ s$)} & \colhead{($\arcdeg \ \ \  \arcmin \ \ \  \arcsec$)} & \colhead{(km\,s$^{-1}$)} & \colhead{($1-3$)} &
\colhead{(pc)} &\colhead{(km\,s$^{-1}$)} & \colhead{($^{\circ}$)} & \colhead{} & 
\colhead{(kpc)} & \colhead{(cm$^{-3}$)} & \colhead{(Myr)} & \colhead{($10^{50}\, \mathrm{erg}$)} & 
\colhead{($10^{4}\, \mathrm{M_{\odot}}$)} }
\startdata
1&10 27 30.6&68 21 29.2&4&1&1078&7&27&0.9&7.4&0.3&75&2.8&3.0\\
2&10 27 34.1&68 20 32.2&0&3&429&10&0&1.0&7.8&0.2&21&1.7&1.4\\
3&10 27 45.8&68 24 36.5&19&3&351&17&0&1.0&4.8&0.3&10&1.9&1.2\\
4&10 27 47.2&68 22 47.0&19&2&541&10&12&0.6&5.1&0.4&26&2.2&1.9\\
5&10 27 47.4&68 21 20.0&9&1&468&7&0&1.0&6.1&0.3&33&1.8&1.7\\
6&10 27 54.0&68 24 16.9&25&3&509&10&143&0.7&3.9&0.3&25&2.1&1.7\\
7&10 27 56.1&68 23 49.9&30&1&2021&7&30&0.7&3.7&0.4&141&3.8&3.5\\
8&10 27 59.4&68 24 30.3&25&1&565&7&123&0.7&3.2&0.3&39&1.9&2.2\\
\enddata
\label{tbl-IC2574}
\end{deluxetable}
\clearpage

\begin{deluxetable}{cccccccccccccc}
\tabletypesize{\scriptsize}
\rotate
\tablecaption{NGC\,3521 hole properties}
\tablewidth{0pt}
\tablehead{
\colhead{\textit{No}} & \colhead{\textit{$\alpha$}} & \colhead{\textit{$\delta$}} & \colhead{$V_\mathrm{hel}$} & \colhead{\textit{Type}} &
\colhead{$d$} & \colhead{$V_\mathrm{exp}$} &\colhead{\textit{P.A.}} & \colhead{\textit{Axial Ratio}} & 
\colhead{R} & \colhead{$n_\mathrm{HI}$} & \colhead{$t_\mathrm{kin}$} & \colhead{$log(E_\mathrm{E})$} & 
 \colhead{$log(M_\mathrm{HI})$} \\
\colhead{} & \colhead{($h\ \ \ m\ \ \ s$)} & \colhead{($\arcdeg \ \ \  \arcmin \ \ \  \arcsec$)} & \colhead{(km\,s$^{-1}$)} & \colhead{($1-3$)} &
\colhead{(pc)} &\colhead{(km\,s$^{-1}$)} & \colhead{($^{\circ}$)} & \colhead{} & 
\colhead{(kpc)} & \colhead{(cm$^{-3}$)} & \colhead{(Myr)} & \colhead{($10^{50}\, \mathrm{erg}$)} & 
\colhead{($10^{4}\, \mathrm{M_{\odot}}$)} }
\startdata
1&11 05 42.2&0 00 15.0&944&1&828&10&161&0.9&11.5&0.3&41&2.8&2.7\\
2&11 05 42.7&0 00 10.5&970&1&794&10&131&0.7&10.3&0.3&39&2.7&2.6\\
3&11 05 43.0&0 01 55.5&996&1&660&10&126&0.5&13.5&0.2&32&2.1&1.8\\
4&11 05 45.8&0 01 28.5&939&1&672&10&146&0.9&5.1&1.2&33&3.1&2.6\\
5&11 05 47.8&0 02 46.5&727&1&603&10&129&0.6&4.6&1.3&29&3.0&2.5\\
6&11 05 48.0&0 01 07.5&981&1&676&10&139&0.6&3.9&1.4&33&3.2&2.6\\
7&11 05 48.1&0 02 54.0&696&1&934&10&0&1.0&4.7&1.3&46&3.6&2.9\\
8&11 05 50.0&0 04 16.5&597&2&928&16&131&0.8&7.8&0.7&28&3.5&2.9\\
9&11 05 52.1&0 04 31.5&597&1&1011&10&95&1.0&7.8&0.6&49&3.3&2.9\\
10&11 05 53.4&0 03 19.5&701&1&1205&10&126&0.6&9.0&0.5&59&3.5&3.1\\
11&11 05 54.1&0 05 42.0&597&1&1122&10&145&0.7&11.9&0.3&55&3.1&2.9\\
12&11 05 57.2&0 03 55.5&385&1&1137&10&88&0.8&16.6&0.1&56&2.7&2.8\\
13&11 05 59.0&0 06 13.5&628&1&921&10&146&0.7&18.6&0.1&45&2.3&2.0\\
\enddata
\label{tbl-NGC3521}
\end{deluxetable}
\clearpage

\begin{deluxetable}{cccccccccccccc}
\tabletypesize{\scriptsize}
\rotate
\tablecaption{NGC\,3627 hole properties}
\tablewidth{0pt}
\tablehead{
\colhead{\textit{No}} & \colhead{\textit{$\alpha$}} & \colhead{\textit{$\delta$}} & \colhead{$V_\mathrm{hel}$} & \colhead{\textit{Type}} &
\colhead{$d$} & \colhead{$V_\mathrm{exp}$} &\colhead{\textit{P.A.}} & \colhead{\textit{Axial Ratio}} & 
\colhead{R} & \colhead{$n_\mathrm{HI}$} & \colhead{$t_\mathrm{kin}$} & \colhead{$log(E_\mathrm{E})$} & 
 \colhead{$log(M_\mathrm{HI})$} \\
\colhead{} & \colhead{($h\ \ \ m\ \ \ s$)} & \colhead{($\arcdeg \ \ \  \arcmin \ \ \  \arcsec$)} & \colhead{(km\,s$^{-1}$)} & \colhead{($1-3$)} &
\colhead{(pc)} &\colhead{(km\,s$^{-1}$)} & \colhead{($^{\circ}$)} & \colhead{} & 
\colhead{(kpc)} & \colhead{(cm$^{-3}$)} & \colhead{(Myr)} & \colhead{($10^{50}\, \mathrm{erg}$)} & 
\colhead{($10^{4}\, \mathrm{M_{\odot}}$)} }
\startdata
1&11 20 13.1&12 57 01.0&849&1&466&13&160&0.8&7.8&0.2&18&1.9&1.4\\
2&11 20 13.2&13 00 01.0&616&1&425&13&142&0.9&2.6&1.0&16&2.6&2.0\\
3&11 20 14.3&13 01 13.0&554&1&359&13&0&1.0&4.7&0.2&14&1.6&1.5\\
4&11 20 14.4&13 00 44.5&554&2&538&30&0&1.0&3.4&1.3&9&3.5&2.4\\
5&11 20 15.0&13 00 17.5&590&2&626&15&148&0.9&2.2&1.1&20&3.2&2.4\\
6&11 20 15.4&12 58 07.0&859&1&538&13&0&1.0&3.7&0.8&20&2.7&2.2\\
7&11 20 15.5&12 58 41.5&844&2&987&65&0&1.0&2.2&0.4&7&4.2&2.8\\
8&11 20 15.6&13 00 56.5&548&3&401&13&175&0.8&4.3&0.7&15&2.3&2.0\\
\enddata
\label{tbl-NGC3627}
\end{deluxetable}
\clearpage

\begin{deluxetable}{cccccccccccccc}
\tabletypesize{\scriptsize}
\rotate
\tablecaption{NGC\,4214 hole properties}
\tablewidth{0pt}
\tablehead{
\colhead{\textit{No}} & \colhead{\textit{$\alpha$}} & \colhead{\textit{$\delta$}} & \colhead{$V_\mathrm{hel}$} & \colhead{\textit{Type}} &
\colhead{$d$} & \colhead{$V_\mathrm{exp}$} &\colhead{\textit{P.A.}} & \colhead{\textit{Axial Ratio}} & 
\colhead{R} & \colhead{$n_\mathrm{HI}$} & \colhead{$t_\mathrm{kin}$} & \colhead{$log(E_\mathrm{E})$} & 
 \colhead{$log(M_\mathrm{HI})$} \\
\colhead{} & \colhead{($h\ \ \ m\ \ \ s$)} & \colhead{($\arcdeg \ \ \  \arcmin \ \ \  \arcsec$)} & \colhead{(km\,s$^{-1}$)} & \colhead{($1-3$)} &
\colhead{(pc)} &\colhead{(km\,s$^{-1}$)} & \colhead{($^{\circ}$)} & \colhead{} & 
\colhead{(kpc)} & \colhead{(cm$^{-3}$)} & \colhead{(Myr)} & \colhead{($10^{50}\, \mathrm{erg}$)} & 
\colhead{($10^{4}\, \mathrm{M_{\odot}}$)} }
\startdata
1&12 15 15.4&36 19 49.4&266&2&368&13&55&0.8&4.5&0.3&14&1.7&1.2\\
2&12 15 16.2&36 19 08.9&266&1&342&6&14&0.6&4.2&0.3&28&1.2&1.2\\
3&12 15 16.2&36 20 08.9&261&2&498&10&11&0.6&4.6&0.3&24&1.9&1.6\\
4&12 15 19.3&36 19 43.4&258&3&257&10&0&1.0&3.7&0.4&13&1.3&1.0\\
5&12 15 22.5&36 18 56.9&258&1&382&6&51&0.8&3.0&0.2&31&1.2&1.7\\
6&12 15 26.5&36 25 06.0&285&1&298&6&60&0.8&6.7&0.0&24&0.0&0.1\\
7&12 15 28.3&36 17 21.0&271&1&433&6&46&0.5&2.8&0.4&35&1.6&1.9\\
8&12 15 28.6&36 18 27.0&262&1&812&6&52&0.9&2.1&0.4&66&2.5&2.4\\
\enddata
\label{tbl-NGC4214}
\end{deluxetable}
\clearpage

\begin{deluxetable}{cccccccccccccc}
\tabletypesize{\scriptsize}
\rotate
\tablecaption{NGC\,4449 hole properties}
\tablewidth{0pt}
\tablehead{
\colhead{\textit{No}} & \colhead{\textit{$\alpha$}} & \colhead{\textit{$\delta$}} & \colhead{$V_\mathrm{hel}$} & \colhead{\textit{Type}} &
\colhead{$d$} & \colhead{$V_\mathrm{exp}$} &\colhead{\textit{P.A.}} & \colhead{\textit{Axial Ratio}} & 
\colhead{R} & \colhead{$n_\mathrm{HI}$} & \colhead{$t_\mathrm{kin}$} & \colhead{$log(E_\mathrm{E})$} & 
 \colhead{$log(M_\mathrm{HI})$} \\
\colhead{} & \colhead{($h\ \ \ m\ \ \ s$)} & \colhead{($\arcdeg \ \ \  \arcmin \ \ \  \arcsec$)} & \colhead{(km\,s$^{-1}$)} & \colhead{($1-3$)} &
\colhead{(pc)} &\colhead{(km\,s$^{-1}$)} & \colhead{($^{\circ}$)} & \colhead{} & 
\colhead{(kpc)} & \colhead{(cm$^{-3}$)} & \colhead{(Myr)} & \colhead{($10^{50}\, \mathrm{erg}$)} & 
\colhead{($10^{4}\, \mathrm{M_{\odot}}$)} }
\startdata
1&12 27 44.9&44 04 53.1&267&1&550&10&7&0.9&8.1&0.0&27&0.5&0.4\\
2&12 27 50.3&44 02 50.2&247&1&600&10&0&0.8&5.9&0.0&29&0.7&0.5\\
3&12 27 51.0&44 02 50.3&236&1&716&10&122&0.7&5.9&0.0&35&0.9&0.7\\
4&12 28 04.1&44 04 58.0&180&2&784&26&124&0.9&2.0&0.7&15&3.6&2.6\\
5&12 28 07.9&44 03 22.0&195&3&458&20&38&0.9&4.0&0.2&11&2.2&1.4\\
6&12 28 08.3&44 04 43.0&190&1&798&10&44&0.8&1.5&1.1&39&3.3&2.7\\
7&12 28 09.5&44 05 58.0&206&1&735&10&0&1.0&1.3&1.2&36&3.2&2.6\\
8&12 28 10.1&44 06 04.0&211&1&901&10&25&0.6&1.5&1.2&44&3.5&2.8\\
\enddata
\label{tbl-NGC4449}
\end{deluxetable}
\clearpage

\begin{deluxetable}{cccccccccccccc}
\tabletypesize{\scriptsize}
\rotate
\tablecaption{NGC\,4736 hole properties}
\tablewidth{0pt}
\tablehead{
\colhead{\textit{No}} & \colhead{\textit{$\alpha$}} & \colhead{\textit{$\delta$}} & \colhead{$V_\mathrm{hel}$} & \colhead{\textit{Type}} &
\colhead{$d$} & \colhead{$V_\mathrm{exp}$} &\colhead{\textit{P.A.}} & \colhead{\textit{Axial Ratio}} & 
\colhead{R} & \colhead{$n_\mathrm{HI}$} & \colhead{$t_\mathrm{kin}$} & \colhead{$log(E_\mathrm{E})$} & 
 \colhead{$log(M_\mathrm{HI})$} \\
\colhead{} & \colhead{($h\ \ \ m\ \ \ s$)} & \colhead{($\arcdeg \ \ \  \arcmin \ \ \  \arcsec$)} & \colhead{(km\,s$^{-1}$)} & \colhead{($1-3$)} &
\colhead{(pc)} &\colhead{(km\,s$^{-1}$)} & \colhead{($^{\circ}$)} & \colhead{} & 
\colhead{(kpc)} & \colhead{(cm$^{-3}$)} & \colhead{(Myr)} & \colhead{($10^{50}\, \mathrm{erg}$)} & 
\colhead{($10^{4}\, \mathrm{M_{\odot}}$)} }
\startdata
1&12 50 30.9&41 08 57.0&407&1&371&9&2&0.8&6.1&0.1&20&0.7&0.6\\
2&12 50 37.2&41 11 13.6&370&1&262&9&176&0.6&7.3&0.0&14&0.0&-0.1\\
3&12 50 39.1&41 04 15.1&339&1&359&9&134&0.8&7.0&0.0&19&0.3&0.2\\
4&12 50 39.3&41 04 46.6&339&1&381&9&163&0.8&6.2&0.0&21&0.5&0.4\\
5&12 50 41.1&41 07 36.1&401&1&364&9&151&0.7&3.1&0.4&20&1.6&1.6\\
6&12 50 44.6&41 07 27.1&401&2&414&20&42&0.8&2.3&1.1&10&2.8&1.9\\
7&12 50 45.0&41 06 57.1&386&2&361&26&0&1.0&2.3&1.2&7&2.8&1.8\\
8&12 50 45.2&41 07 46.6&407&3&303&15&23&0.8&2.3&1.2&10&2.2&1.6\\
\enddata
\label{tbl-NGC4736}
\end{deluxetable}
\clearpage

\begin{deluxetable}{cccccccccccccc}
\tabletypesize{\scriptsize}
\rotate
\tablecaption{DDO\,154 hole properties}
\tablewidth{0pt}
\tablehead{
\colhead{\textit{No}} & \colhead{\textit{$\alpha$}} & \colhead{\textit{$\delta$}} & \colhead{$V_\mathrm{hel}$} & \colhead{\textit{Type}} &
\colhead{$d$} & \colhead{$V_\mathrm{exp}$} &\colhead{\textit{P.A.}} & \colhead{\textit{Axial Ratio}} & 
\colhead{R} & \colhead{$n_\mathrm{HI}$} & \colhead{$t_\mathrm{kin}$} & \colhead{$log(E_\mathrm{E})$} & 
 \colhead{$log(M_\mathrm{HI})$} \\
\colhead{} & \colhead{($h\ \ \ m\ \ \ s$)} & \colhead{($\arcdeg \ \ \  \arcmin \ \ \  \arcsec$)} & \colhead{(km\,s$^{-1}$)} & \colhead{($1-3$)} &
\colhead{(pc)} &\colhead{(km\,s$^{-1}$)} & \colhead{($^{\circ}$)} & \colhead{} & 
\colhead{(kpc)} & \colhead{(cm$^{-3}$)} & \colhead{(Myr)} & \colhead{($10^{50}\, \mathrm{erg}$)} & 
\colhead{($10^{4}\, \mathrm{M_{\odot}}$)} }
\startdata
1&12 53 56.7&27 07 53.0&411&1&472&8&53&0.7&3.2&0.1&29&1.3&1.2\\
2&12 53 58.6&27 07 53.0&400&1&685&8&110&0.8&2.6&0.1&42&1.8&1.7\\
3&12 54 03.0&27 07 54.5&400&1&460&8&118&0.7&2.3&0.2&28&1.5&1.3\\
4&12 54 03.1&27 09 17.0&380&1&361&8&162&0.8&1.6&0.2&22&1.2&1.1\\
5&12 54 06.3&27 08 21.5&382&1&604&8&139&0.9&2.2&0.2&37&1.9&1.7\\
6&12 54 07.6&27 09 53.0&362&3&396&6&0&1.0&1.3&0.2&32&1.1&1.2\\
7&12 54 08.4&27 09 14.0&372&2&454&13&163&0.8&1.1&0.2&17&1.9&1.5\\
8&12 54 15.6&27 11 02.0&328&2&271&13&0&1.0&3.6&0.1&10&0.7&0.3\\
9&12 54 15.9&27 09 41.0&354&1&416&8&173&0.9&4.1&0.1&25&1.0&0.8\\
\enddata
\label{tbl-DDO154}
\end{deluxetable}
\clearpage

\begin{deluxetable}{cccccccccccccc}
\tabletypesize{\scriptsize}
\rotate
\tablecaption{NGC\,5194 hole properties}
\tablewidth{0pt}
\tablehead{
\colhead{\textit{No}} & \colhead{\textit{$\alpha$}} & \colhead{\textit{$\delta$}} & \colhead{$V_\mathrm{hel}$} & \colhead{\textit{Type}} &
\colhead{$d$} & \colhead{$V_\mathrm{exp}$} &\colhead{\textit{P.A.}} & \colhead{\textit{Axial Ratio}} & 
\colhead{R} & \colhead{$n_\mathrm{HI}$} & \colhead{$t_\mathrm{kin}$} & \colhead{$log(E_\mathrm{E})$} & 
 \colhead{$log(M_\mathrm{HI})$}  \\
\colhead{} & \colhead{($h\ \ \ m\ \ \ s$)} & \colhead{($\arcdeg \ \ \  \arcmin \ \ \  \arcsec$)} & \colhead{(km\,s$^{-1}$)} & \colhead{($1-3$)} &
\colhead{(pc)} &\colhead{(km\,s$^{-1}$)} & \colhead{($^{\circ}$)} & \colhead{} & 
\colhead{(kpc)} & \colhead{(cm$^{-3}$)} & \colhead{(Myr)} & \colhead{($10^{50}\, \mathrm{erg}$)} & 
\colhead{($10^{4}\, \mathrm{M_{\odot}}$)} }
\startdata
1&13 29 33.2&47 11 43.4&473&1&1344&8&161&0.8&8.2&0.2&47&3.3&2.8\\
2&13 29 35.7&47 09 59.9&483&1&368&8&135&0.9&8.2&0.4&12.9&2.0&1.4\\
3&13 29 35.7&47 10 32.9&462&1&349&8&0&1.0&7.7&0.4&12.2&1.9&1.4\\
4&13 29 39.8&47 09 31.5&499&1&828&32&93&0.8&7.5&0.4&12.7&3.5&2.7\\
5&13 29 40.4&47 10 24.0&488&1&1903&8&23&0.8&6.0&0.8&116&4.2&3.9\\
6&13 29 41.7&47 08 36.0&509&1&658&20&103&0.5&8.6&0.4&16.1&2.9&2.5\\
7&13 29 41.7&47 12 48.0&400&1&349&8&0&1.0&5.2&0.7&12.2&2.2&2.0\\
8&13 29 42.7&47 12 19.5&416&1&717&8&80&0.9&4.4&0.7&25.1&3.1&2.5\\
\enddata
\label{tbl-NGC5194}
\end{deluxetable}
\clearpage

\begin{deluxetable}{cccccccccccccc}
\tabletypesize{\scriptsize}
\rotate
\tablecaption{NGC\,6946 hole properties}
\tablewidth{0pt}
\tablehead{
\colhead{\textit{No}} & \colhead{\textit{$\alpha$}} & \colhead{\textit{$\delta$}} & \colhead{$V_\mathrm{hel}$} & \colhead{\textit{Type}} &
\colhead{$d$} & \colhead{$V_\mathrm{exp}$} &\colhead{\textit{P.A.}} & \colhead{\textit{Axial Ratio}} & 
\colhead{R} & \colhead{$n_\mathrm{HI}$} & \colhead{$t_\mathrm{kin}$} & \colhead{$log(E_\mathrm{E})$} & 
 \colhead{$log(M_\mathrm{HI})$}  \\
\colhead{} & \colhead{($h\ \ \ m\ \ \ s$)} & \colhead{($\arcdeg \ \ \  \arcmin \ \ \  \arcsec$)} & \colhead{(km\,s$^{-1}$)} & \colhead{($1-3$)} &
\colhead{(pc)} &\colhead{(km\,s$^{-1}$)} & \colhead{($^{\circ}$)} & \colhead{} & 
\colhead{(kpc)} & \colhead{(cm$^{-3}$)} & \colhead{(Myr)} & \colhead{($10^{50}\, \mathrm{erg}$)} & 
\colhead{($10^{4}\, \mathrm{M_{\odot}}$)} }
\startdata
1&20 33 46.1&60 09 27.5&145&1&970&8&166&0.5&14.9&0.1&59&2.2&2.0\\
2&20 34 07.2&60 09 40.0&132&1&892&8&156&0.8&10.3&0.3&55&2.6&2.6\\
3&20 34 11.5&60 06 29.6&158&1&1695&8&117&0.6&9.9&0.3&104&3.5&3.1\\
4&20 34 13.2&60 08 52.2&145&2&543&15&177&0.9&8.6&0.6&18&2.7&2.3\\
5&20 34 18.5&60 08 47.7&142&3&198&10&93&0.8&7.5&1.0&10&1.3&1.0\\
6&20 34 19.5&60 07 37.3&153&1&1138&8&148&0.7&7.5&0.8&70&3.5&3.0\\
7&20 34 19.6&60 09 31.3&137&1&1222&8&123&0.8&7.4&1.0&75&3.7&3.2\\
8&20 34 20.6&60 10 16.3&122&1&485&8&130&0.9&7.7&1.3&30&2.6&2.5\\
\enddata
\label{tbl-NGC6946}
\end{deluxetable}
\clearpage

\begin{deluxetable}{cccccccccccccc}
\tabletypesize{\scriptsize}
\rotate
\tablecaption{NGC\,7793 hole properties}
\tablewidth{0pt}
\tablehead{
\colhead{\textit{No}} & \colhead{\textit{$\alpha$}} & \colhead{\textit{$\delta$}} & \colhead{$V_\mathrm{hel}$} & \colhead{\textit{Type}} &
\colhead{$d$} & \colhead{$V_\mathrm{exp}$} &\colhead{\textit{P.A.}} & \colhead{\textit{Axial Ratio}} & 
\colhead{R} & \colhead{$n_\mathrm{HI}$} & \colhead{$t_\mathrm{kin}$} & \colhead{$log(E_\mathrm{E})$} & 
 \colhead{$log(M_\mathrm{HI})$}  \\
\colhead{} & \colhead{($h\ \ \ m\ \ \ s$)} & \colhead{($\arcdeg \ \ \  \arcmin \ \ \  \arcsec$)} & \colhead{(km\,s$^{-1}$)} & \colhead{($1-3$)} &
\colhead{(pc)} &\colhead{(km\,s$^{-1}$)} & \colhead{($^{\circ}$)} & \colhead{} & 
\colhead{(kpc)} & \colhead{(cm$^{-3}$)} & \colhead{(Myr)} & \colhead{($10^{50}\, \mathrm{erg}$)} & 
\colhead{($10^{4}\, \mathrm{M_{\odot}}$)} }
\startdata
1&23 57 29.1&-32 35 04.4&310&1&443&8&22&0.8&5.2&0.3&27&1.7&1.5\\
2&23 57 35.5&-32 35 57.0&286&1&754&8&10&0.7&3.9&0.7&46&2.8&2.7\\
3&23 57 35.7&-32 34 58.5&302&1&313&8&174&0.6&3.5&0.9&19&1.8&2.0\\
4&23 57 36.7&-32 35 46.5&286&1&852&8&64&0.9&3.5&0.8&52&3.1&2.8\\
5&23 57 37.5&-32 33 07.5&291&2&336&16&13&0.7&4.3&0.4&10&2.0&1.4\\
6&23 57 41.1&-32 33 60.0&299&2&401&13&45&0.6&2.7&1.1&15&2.5&2.2\\
7&23 57 41.4&-32 37 07.5&253&1&189&8&0&0.8&3.9&0.8&12&1.0&0.8\\
8&23 57 43.3&-32 36 33.0&248&1&434&8&175&0.4&2.7&1.0&27&2.3&2.2\\
\enddata
\label{tbl-NGC7793}
\end{deluxetable}
\clearpage

\section{Figures}
\label{figures}

\setcounter{figure}{0} 
\renewcommand{\thefigure}{C\arabic{figure}}

\begin{figure}
\begin{center}
\includegraphics[angle=0,scale=.75]{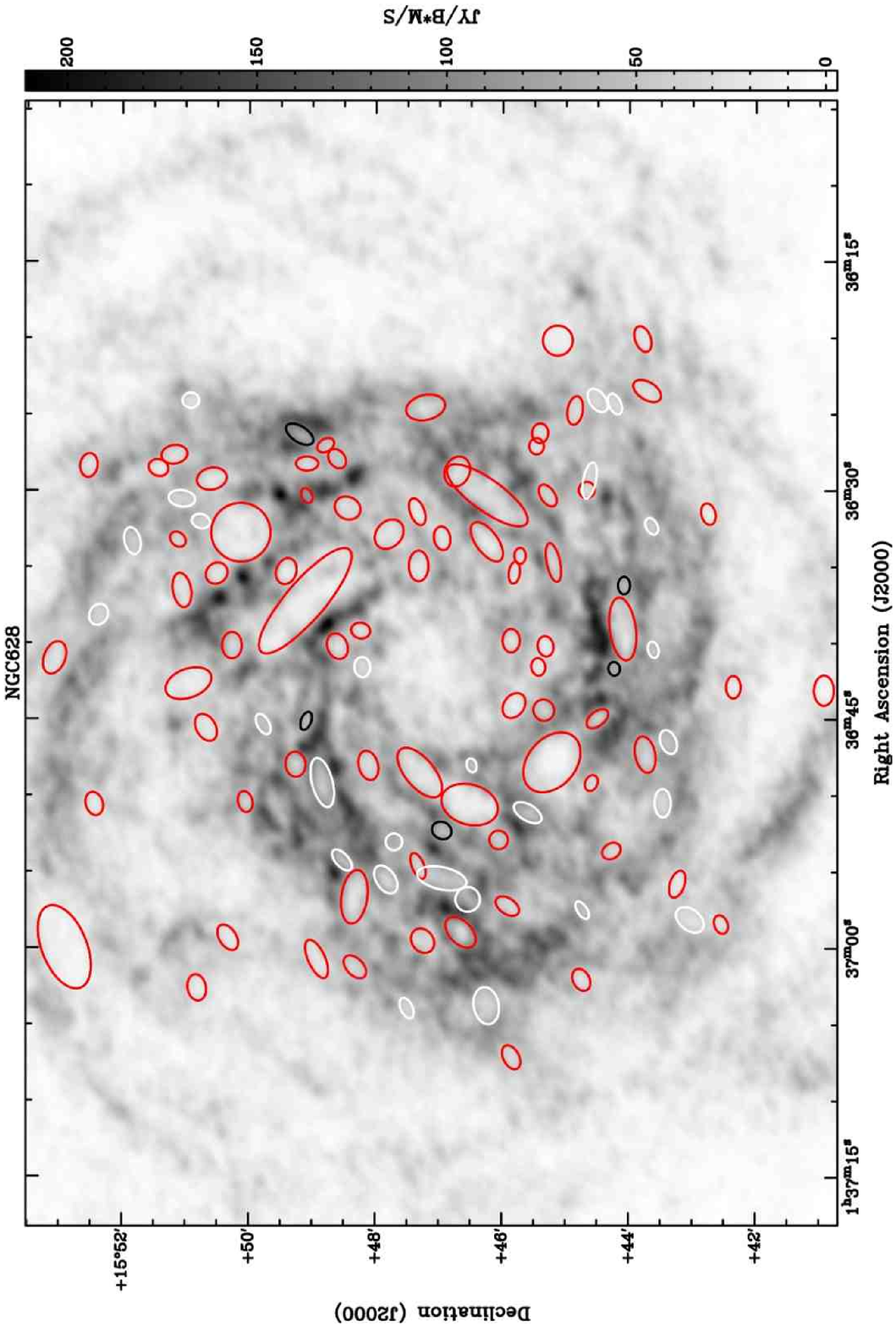}
\end{center}
\caption{Position of the \ion{H}{1} holes in NGC\,628. 
The colours illustrate the different type of the holes: red - type 1, white - type 2 and black - type 3.
\label{NGC628-map}}
\end{figure}
\clearpage

\begin{figure}
\begin{center}
\includegraphics[angle=0,scale=.75]{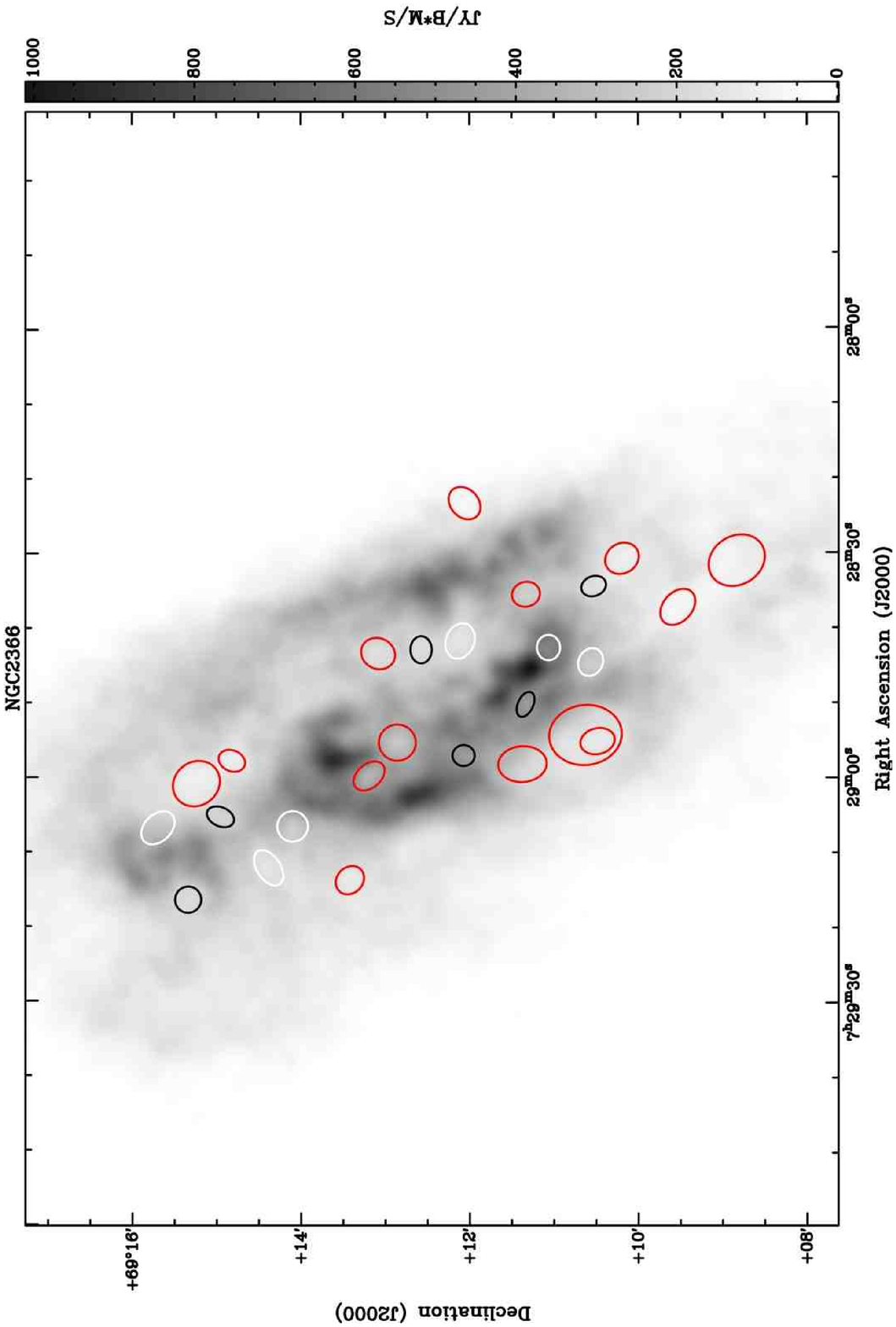}
\end{center}
\caption{Position of the \ion{H}{1} holes in NGC\,2366. 
The colours illustrate the different type of the holes: red - type 1, white - type 2 and black - type 3.
\label{NGC2366-map}}
\end{figure}
\clearpage

\begin{figure}
\begin{center}
\includegraphics[angle=0,scale=.75]{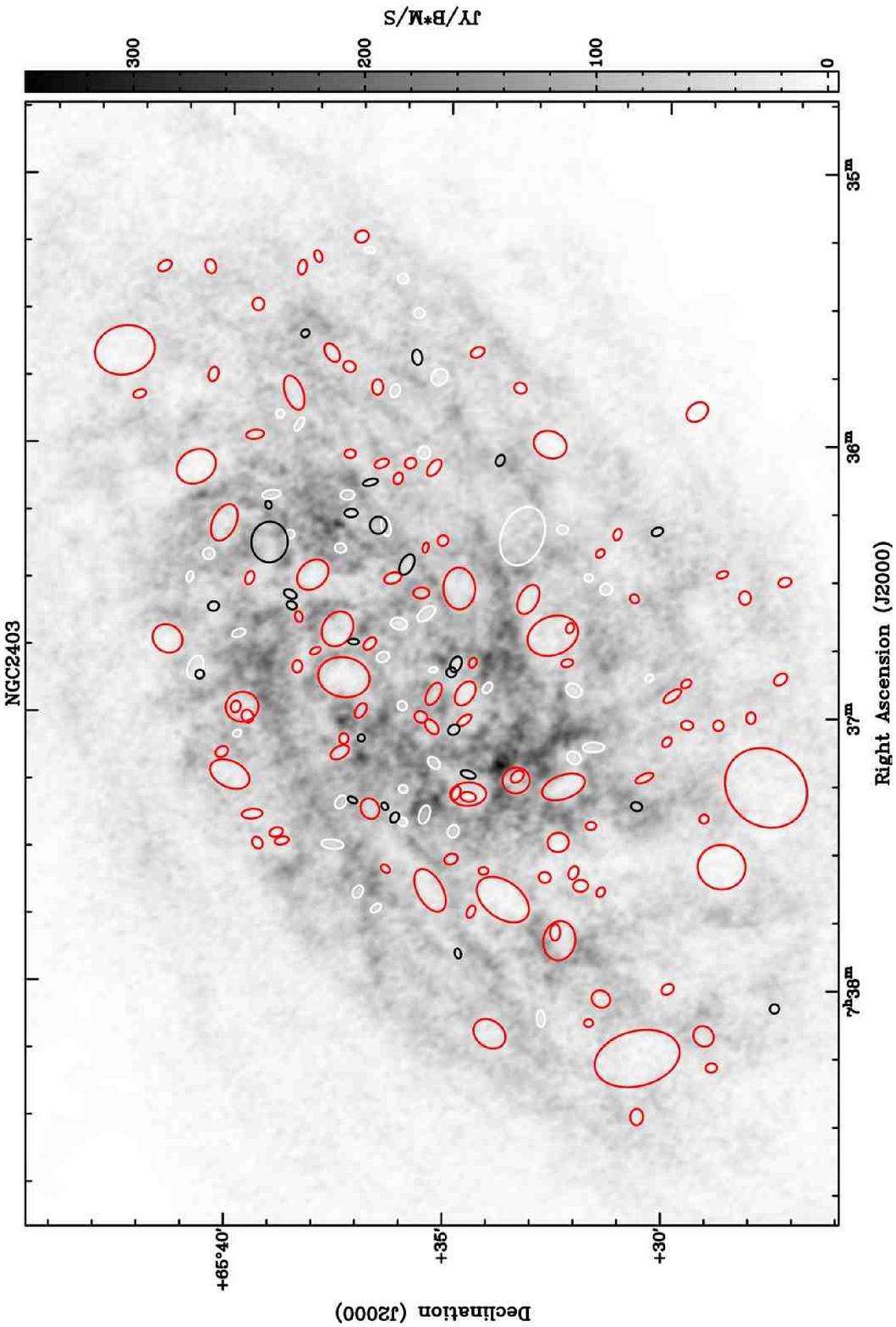}
\end{center}
\caption{Position of the \ion{H}{1} holes in NGC\,2403. 
The colours illustrate the different type of the holes: red - type 1, white - type 2 and black - type 3.
\label{NGC2403-map}}
\end{figure}
\clearpage

\begin{figure}
\begin{center}
\includegraphics[angle=0,scale=.75]{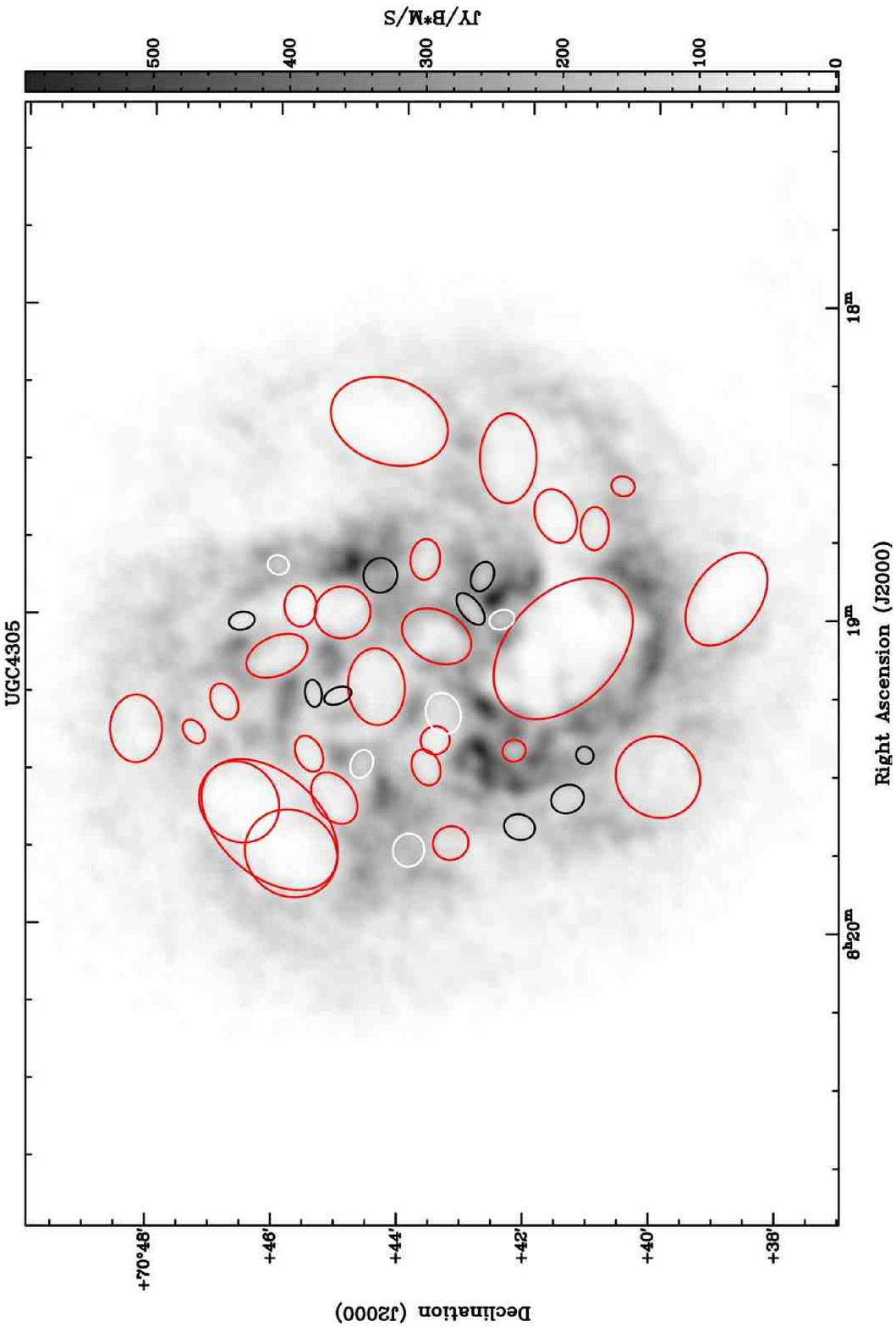}
\end{center}
\caption{Position of the \ion{H}{1} holes in Holmberg\,II.
The colours illustrate the different type of the holes: red - type 1, white - type 2 and black - type 3.
\label{HoII-map}}
\end{figure}
\clearpage

\begin{figure}
\begin{center}
\includegraphics[angle=0,scale=.60]{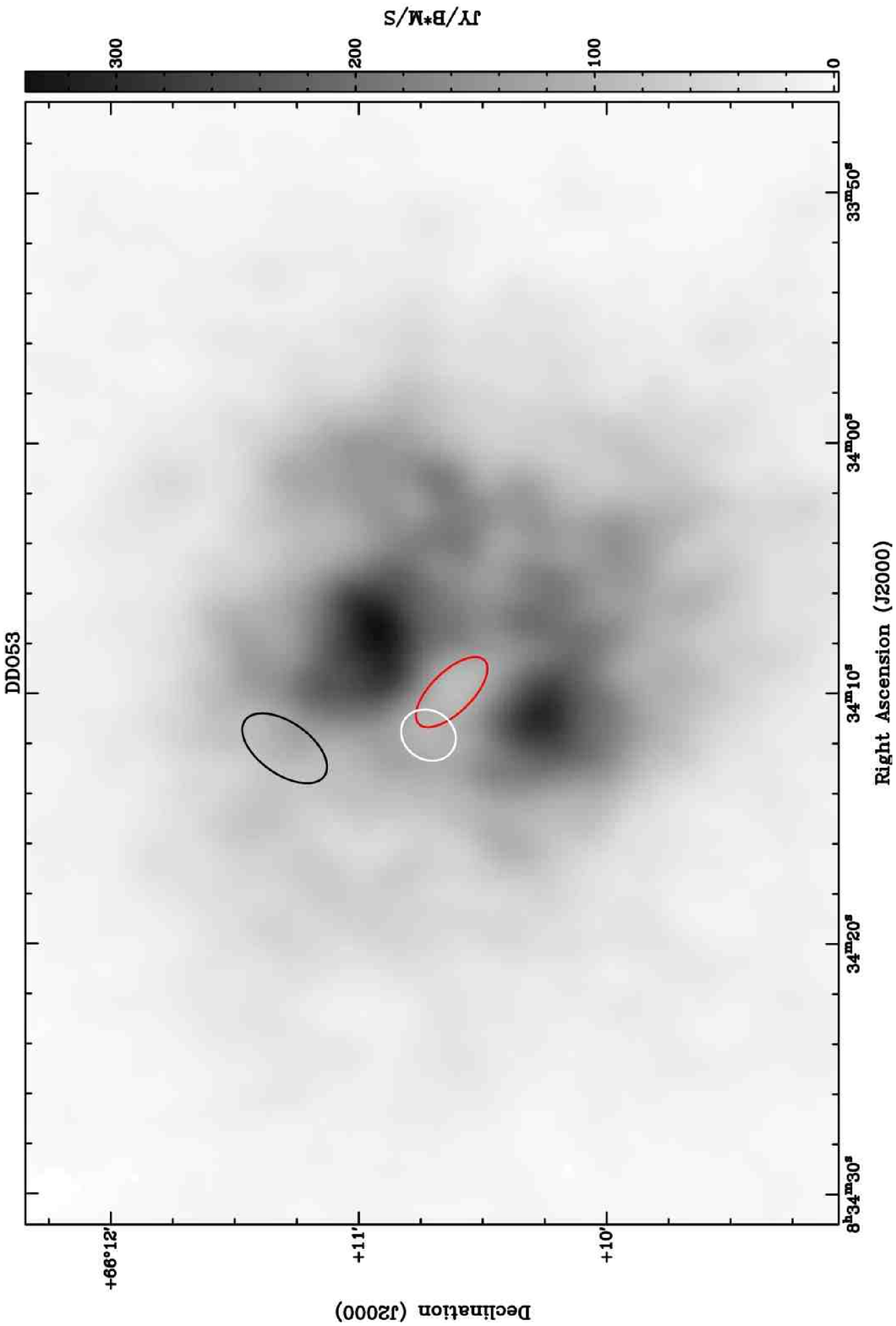}
\end{center}
\caption{Position of the \ion{H}{1} holes in DDO\,53.
The colours illustrate the different type of the holes: red - type 1, white - type 2 and black - type 3.
\label{DDO53-map}}
\end{figure}
\clearpage

\begin{figure}
\begin{center}
\includegraphics[angle=0,scale=.75]{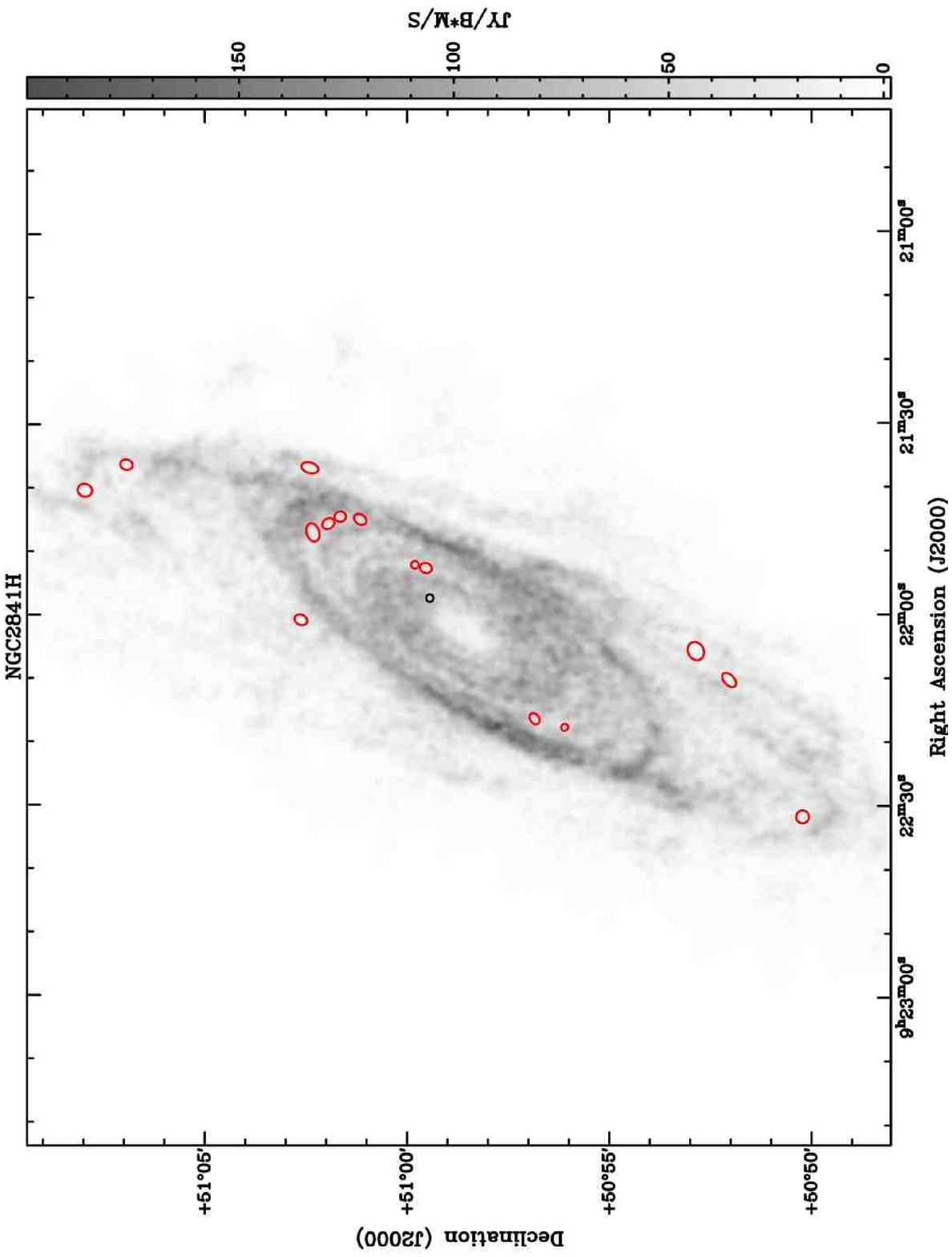}
\end{center}
\caption{Position of the \ion{H}{1} holes in NGC\,2841. 
The colours illustrate the different type of the holes: red - type 1, white - type 2 and black - type 3.
\label{NGC2841-map}}
\end{figure}
\clearpage

\begin{figure}
\begin{center}
\includegraphics[angle=0,scale=.75]{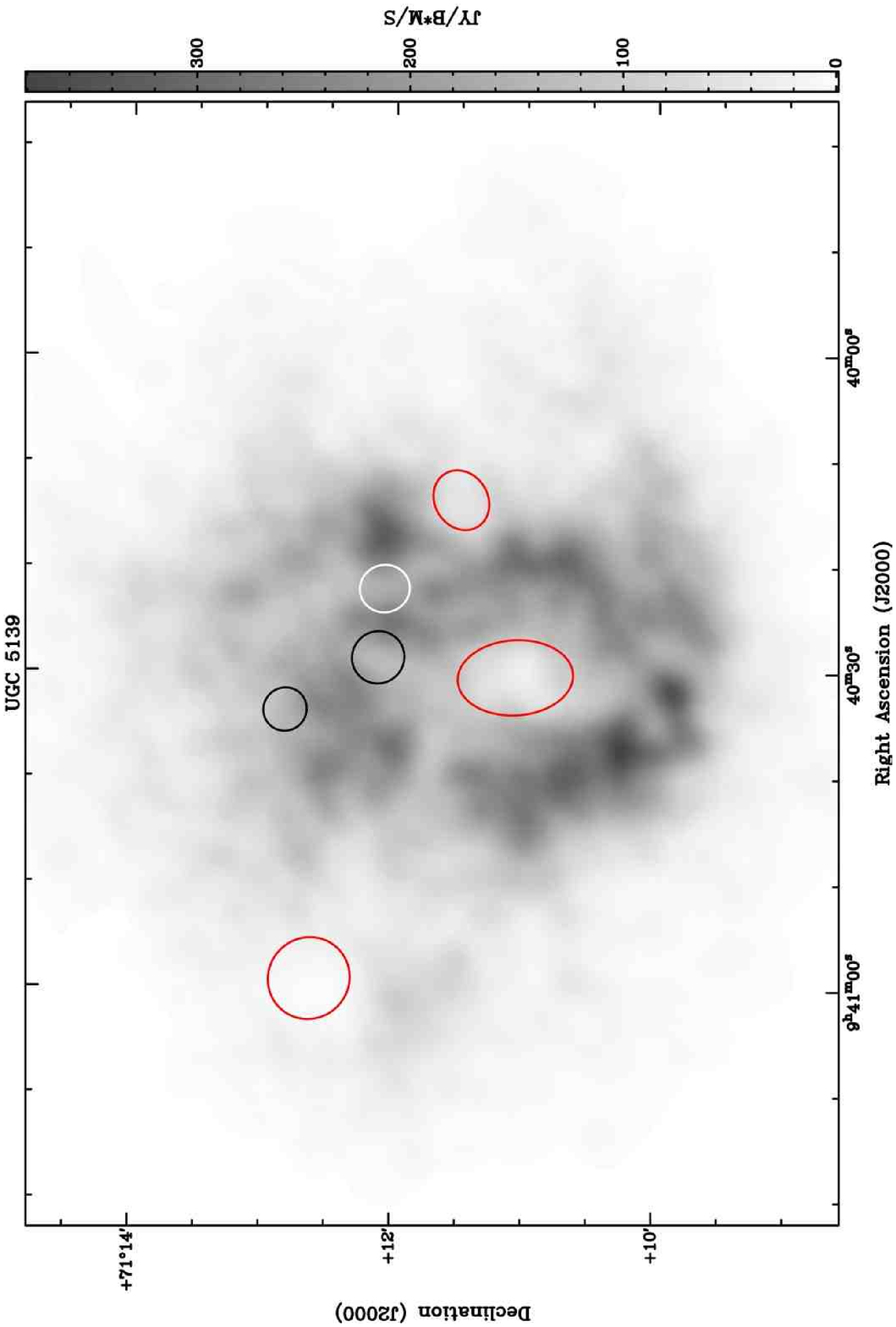}
\end{center}
\caption{Position of the \ion{H}{1} holes in Holmberg\,I. 
The colours illustrate the different type of the holes: red - type 1, white - type 2 and black - type 3.
\label{HoI-map}}
\end{figure}
\clearpage

\begin{figure}
\begin{center}
\includegraphics[angle=0,scale=.75]{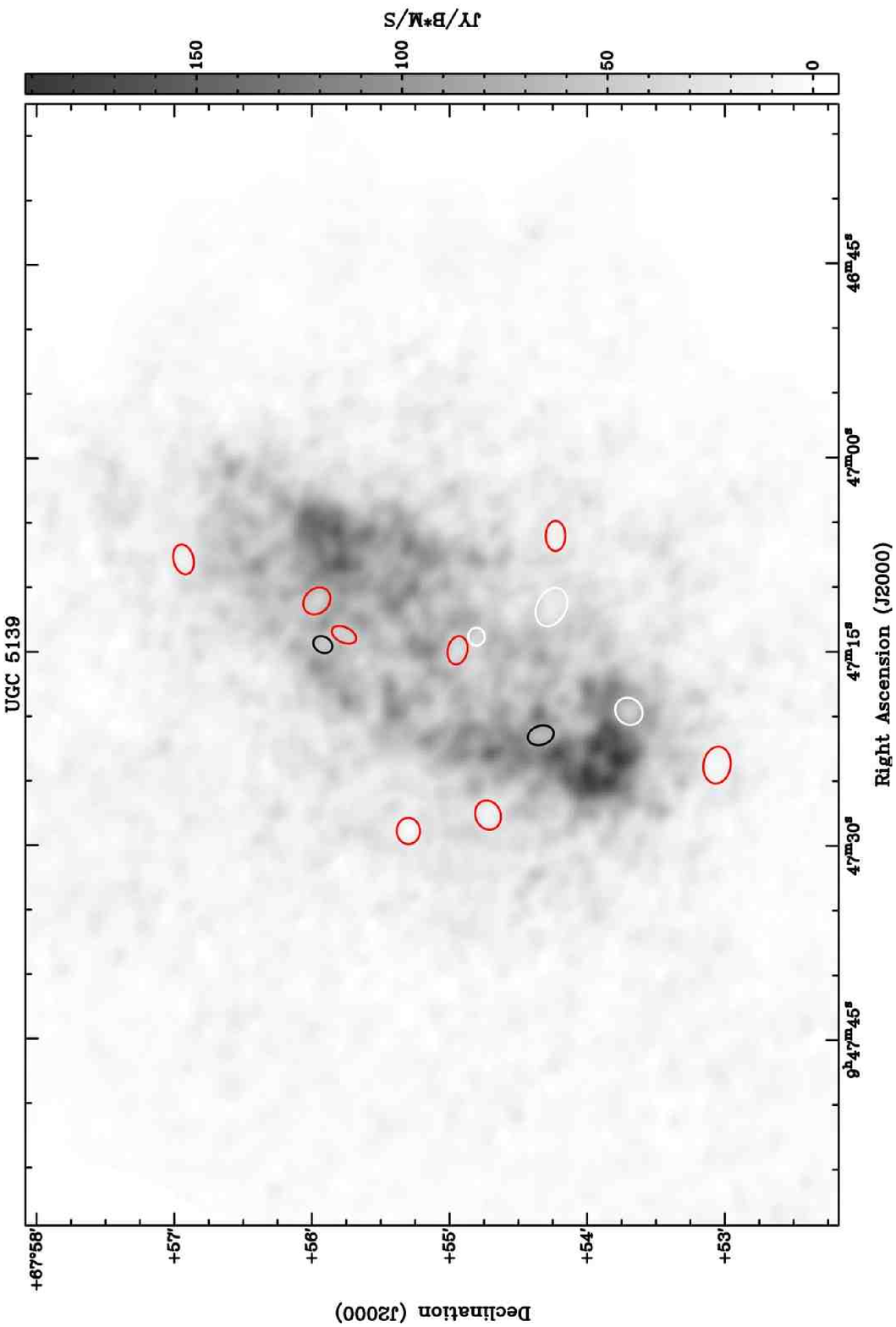}
\end{center}
\caption{Position of the \ion{H}{1} holes in NGC\,2976. 
The colours illustrate the different type of the holes: red - type 1, white - type 2 and black - type 3.
\label{NGC2976-map}}
\end{figure}
\clearpage

\begin{figure}
\begin{center}
\includegraphics[angle=0,scale=.75]{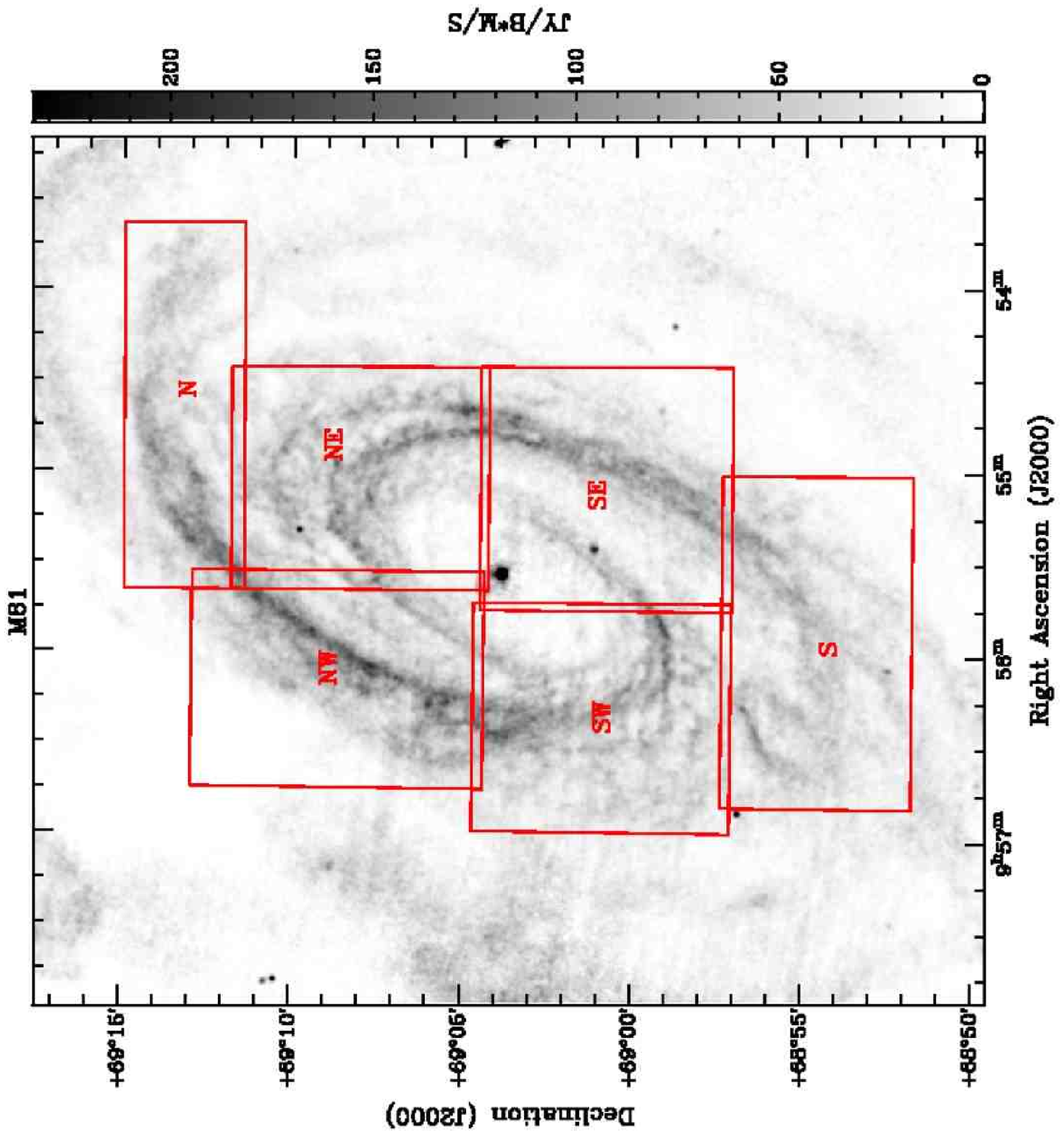}
\end{center}
\caption{NGC\,3031 divided into 6 fields. \label{M81-areas}}
\end{figure}
\clearpage

\begin{figure}
\begin{center}
\includegraphics[angle=0,scale=.59]{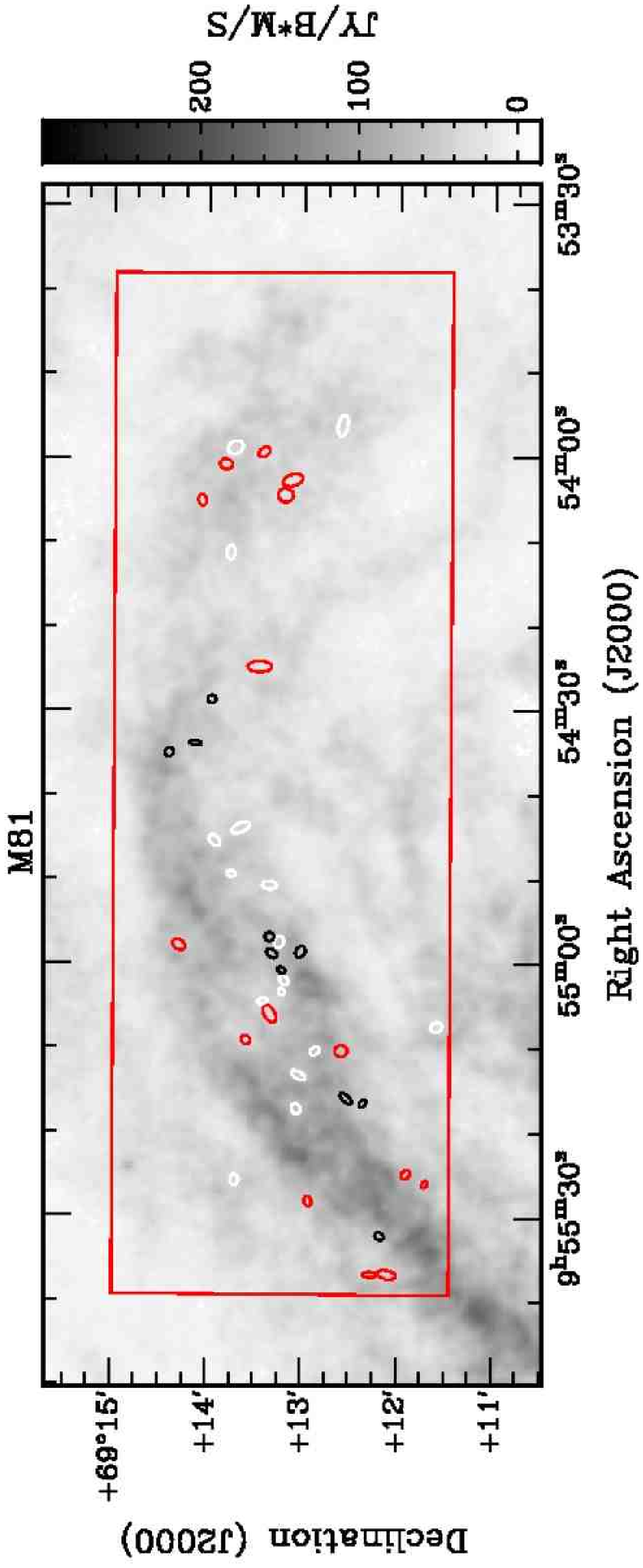}
\end{center}
\caption{Position of the \ion{H}{1} holes in the north field of NGC\,3031. 
The colours illustrate the different type of the holes: red - type 1, white - type 2 and black - type 3. \label{M81-N}}
\end{figure}

\begin{figure}
\begin{center}
\includegraphics[angle=0,scale=.71]{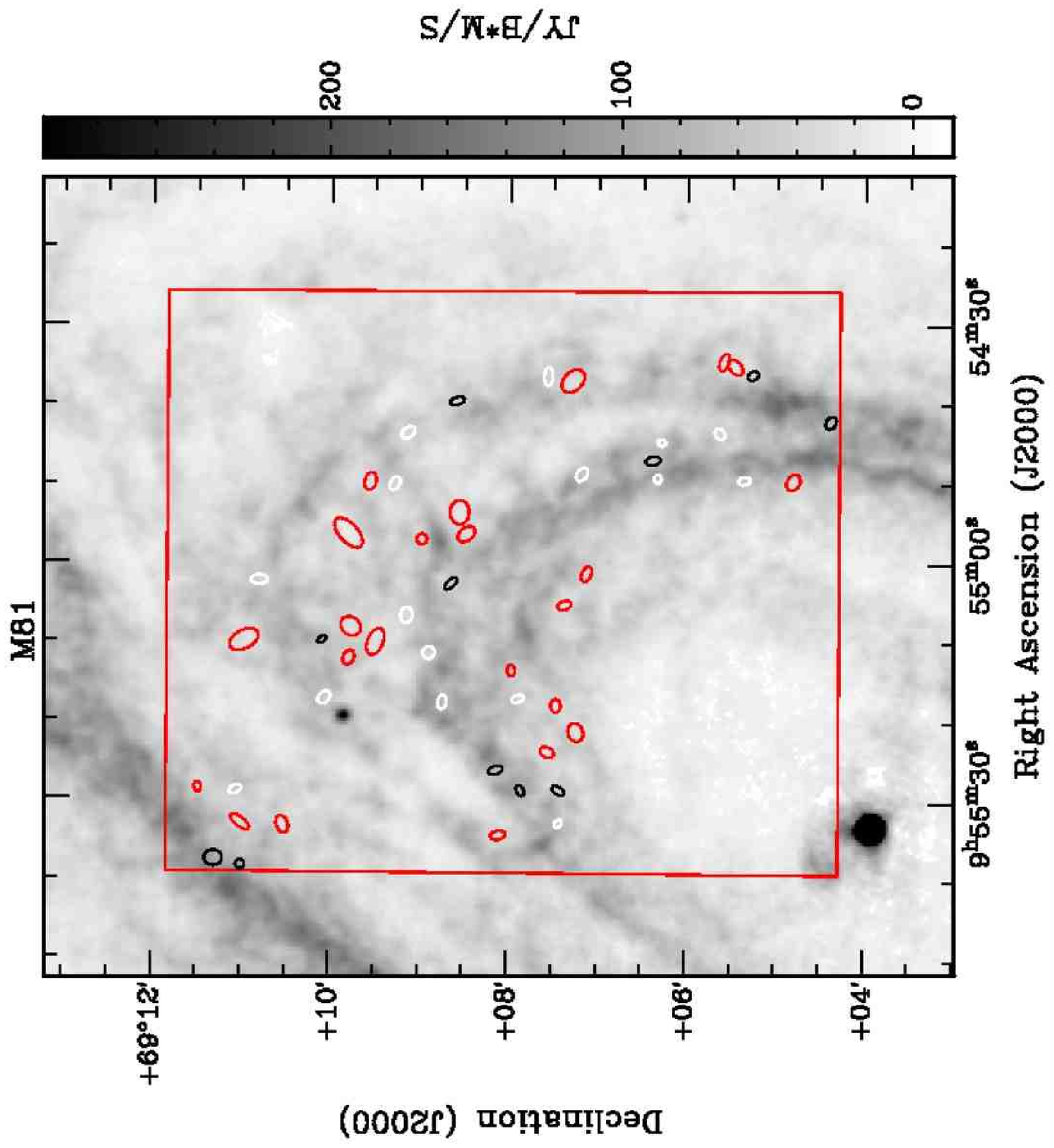}
\end{center}
\caption{Position of the \ion{H}{1} holes in the north--east field of NGC\,3031. 
The colours illustrate the different type of the holes: red - type 1, white - type 2 and black - type 3. \label{M81-NE}}
\end{figure}

\begin{figure}
\begin{center}
\includegraphics[angle=0,scale=.74]{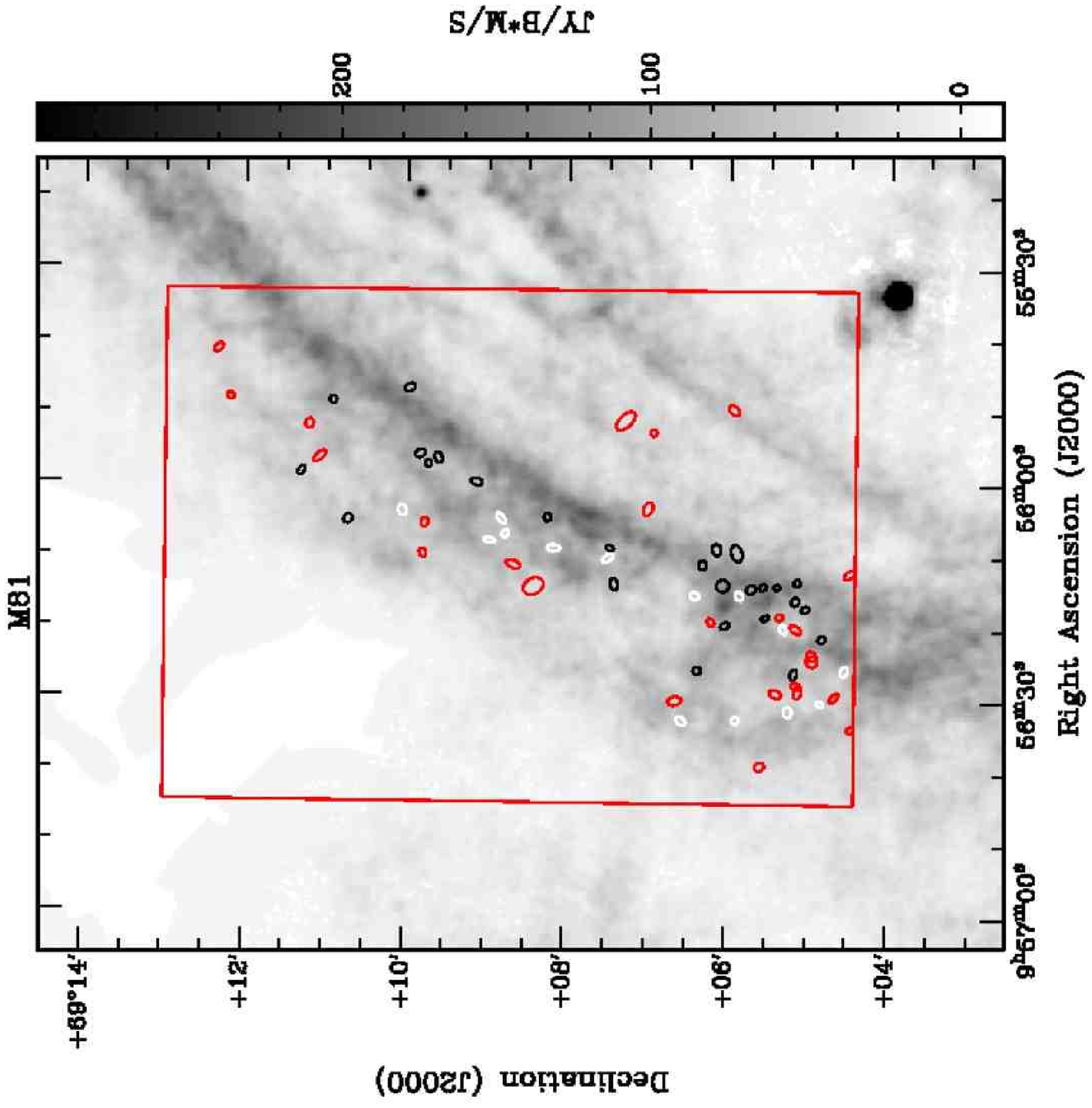}
\end{center}
\caption{Position of the \ion{H}{1} holes in the north--west field of NGC\,3031. 
The colours illustrate the different type of the holes: red - type 1, white - type 2 and black - type 3. \label{M81-NW}}
\end{figure}

\begin{figure}
\begin{center}
\includegraphics[angle=0,scale=.60]{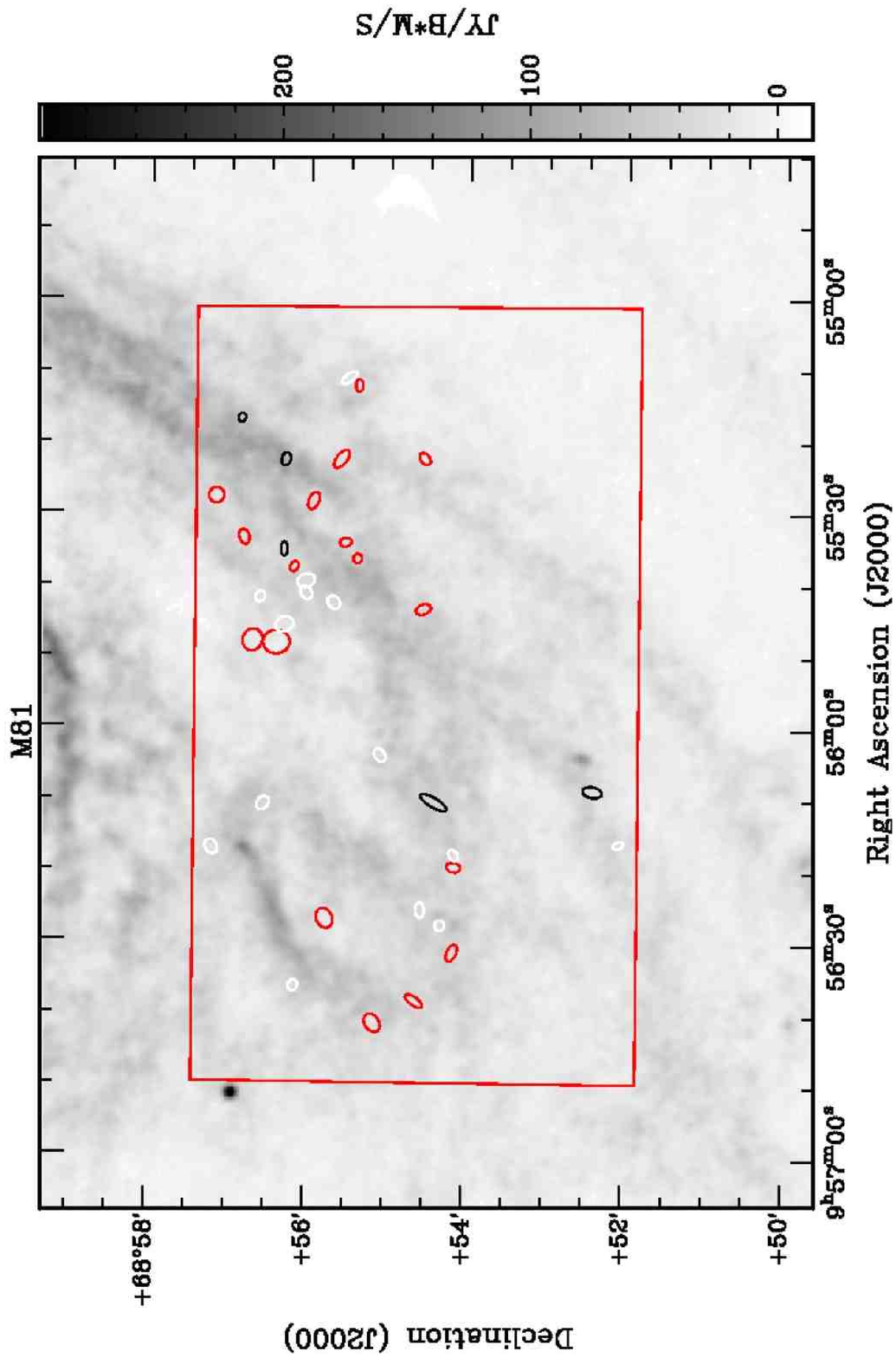}
\end{center}
\caption{Position of the \ion{H}{1} holes in the south field of NGC\,3031. 
The colours illustrate the different type of the holes: red - type 1, white - type 2 and black - type 3. \label{M81-S}}
\end{figure}

\begin{figure}
\begin{center}
\includegraphics[angle=0,scale=.65]{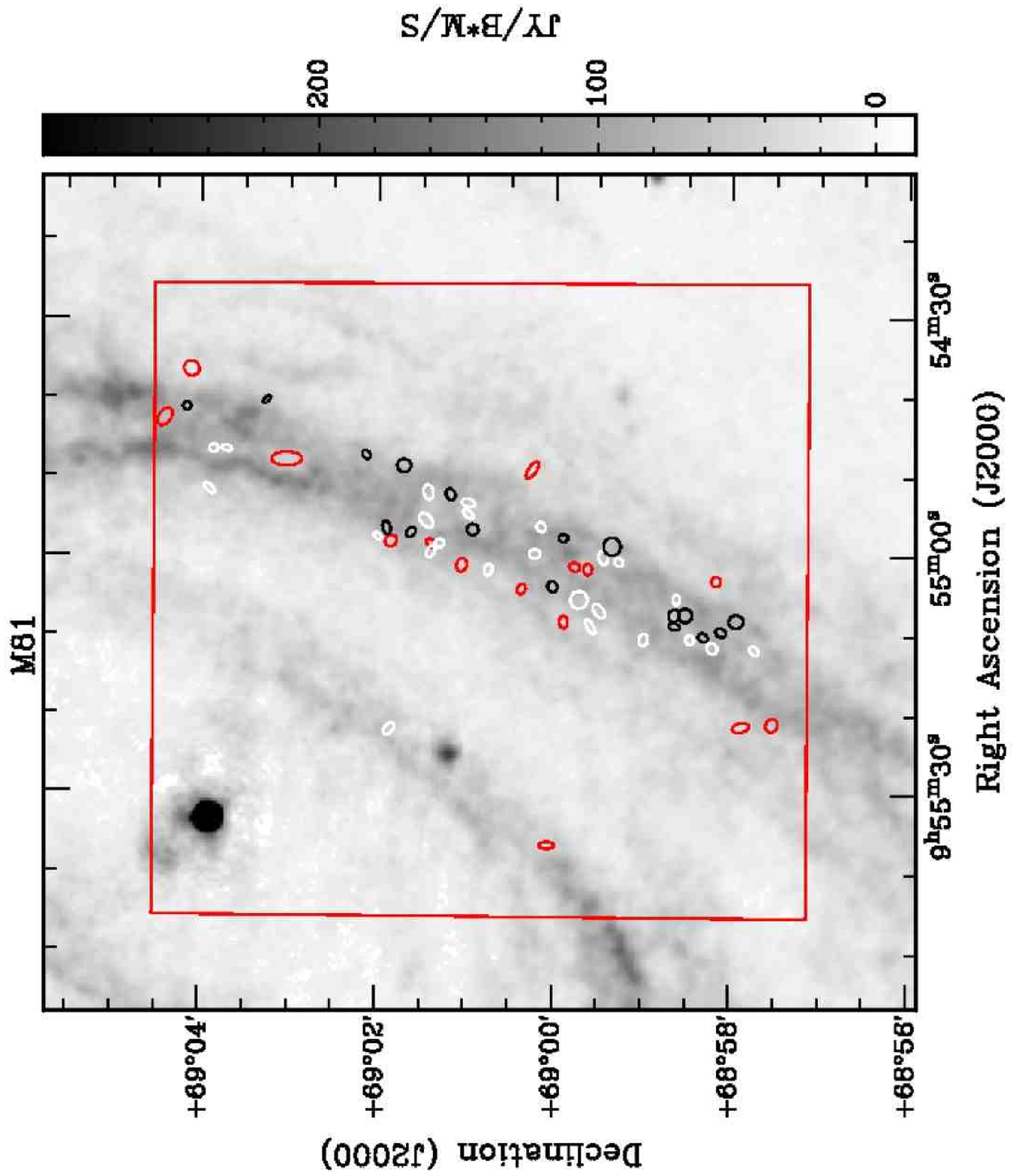}
\end{center}
\caption{Position of the \ion{H}{1} holes in the south--east field of NGC\,3031. 
The colours illustrate the different type of the holes: red - type 1, white - type 2 and black - type 3. \label{M81-SE}}
\end{figure}

\begin{figure}
\begin{center}
\includegraphics[angle=0,scale=.65]{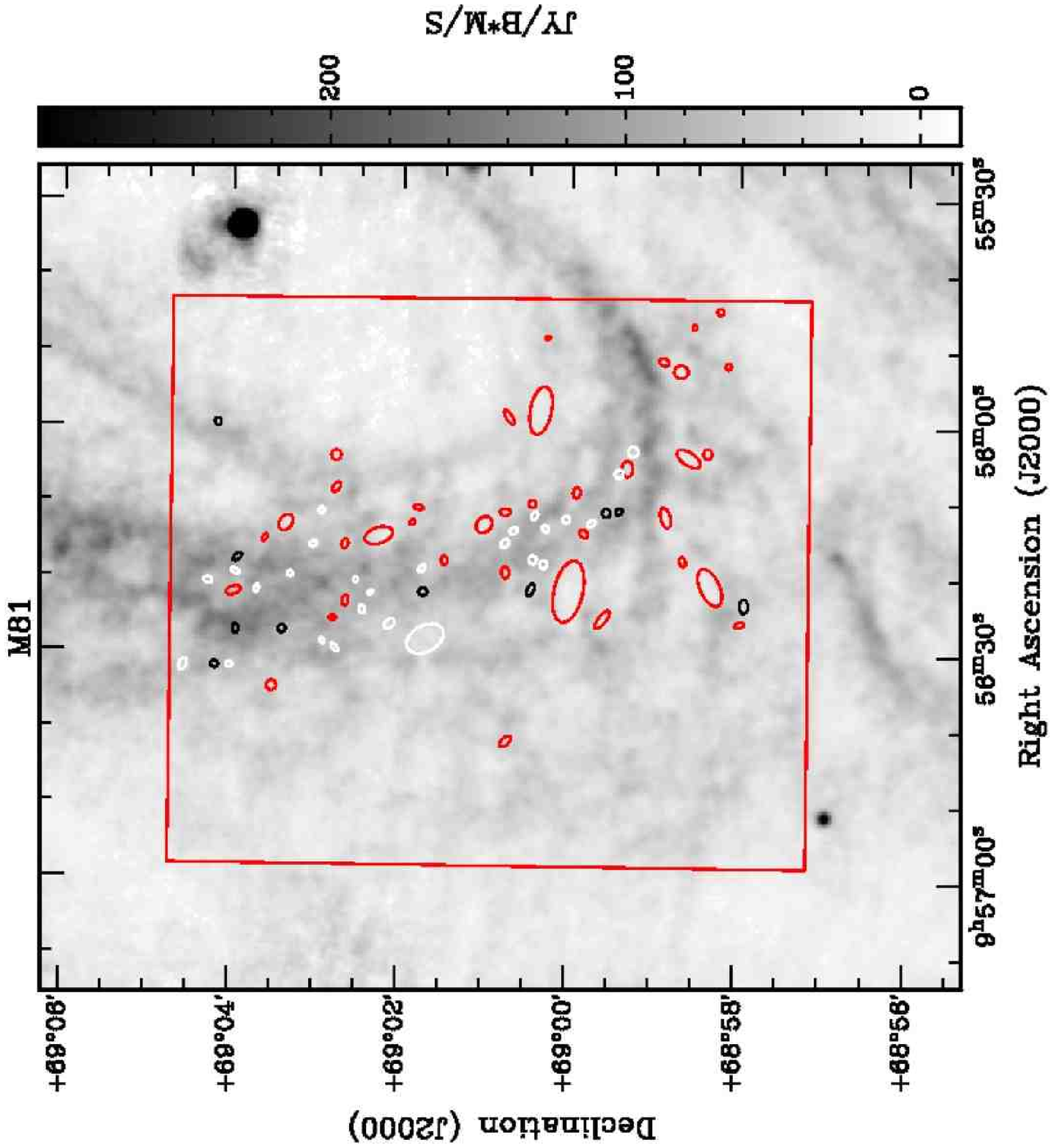}
\end{center}
\caption{Position of the \ion{H}{1} holes in the south--west field of NGC\,3031. 
The colours illustrate the different type of the holes: red - type 1, white - type 2 and black - type 3. \label{M81-SW}}
\end{figure}
\clearpage

\begin{figure}
\begin{center}
\includegraphics[angle=0,scale=.75]{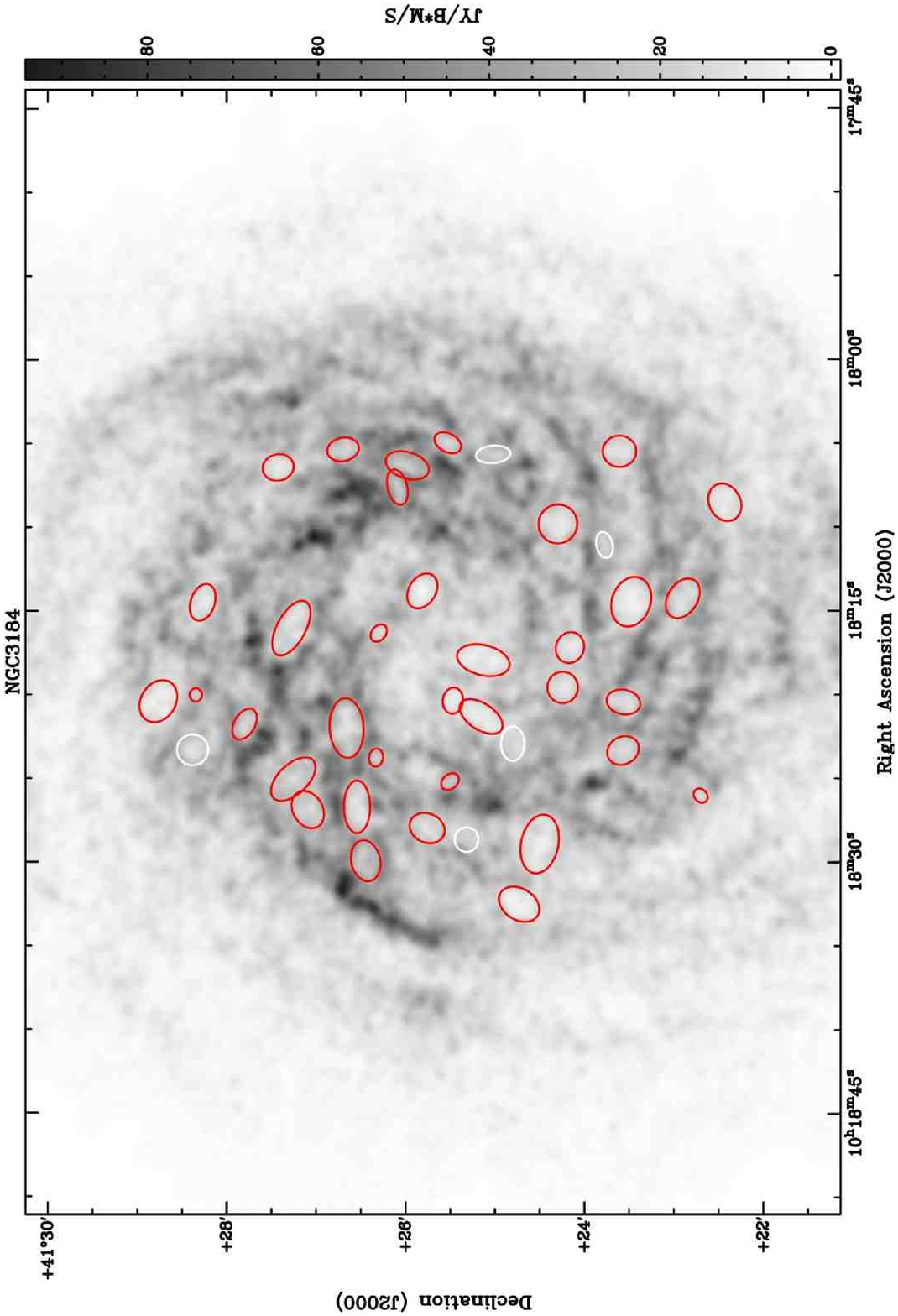}
\end{center}
\caption{Position of the \ion{H}{1} holes in NGC\,3184. 
The colours illustrate the different type of the holes: red - type 1, white - type 2 and black - type 3.
\label{NGC3184}}
\end{figure}
\clearpage

\begin{figure}
\begin{center}
\includegraphics[angle=0,scale=.75]{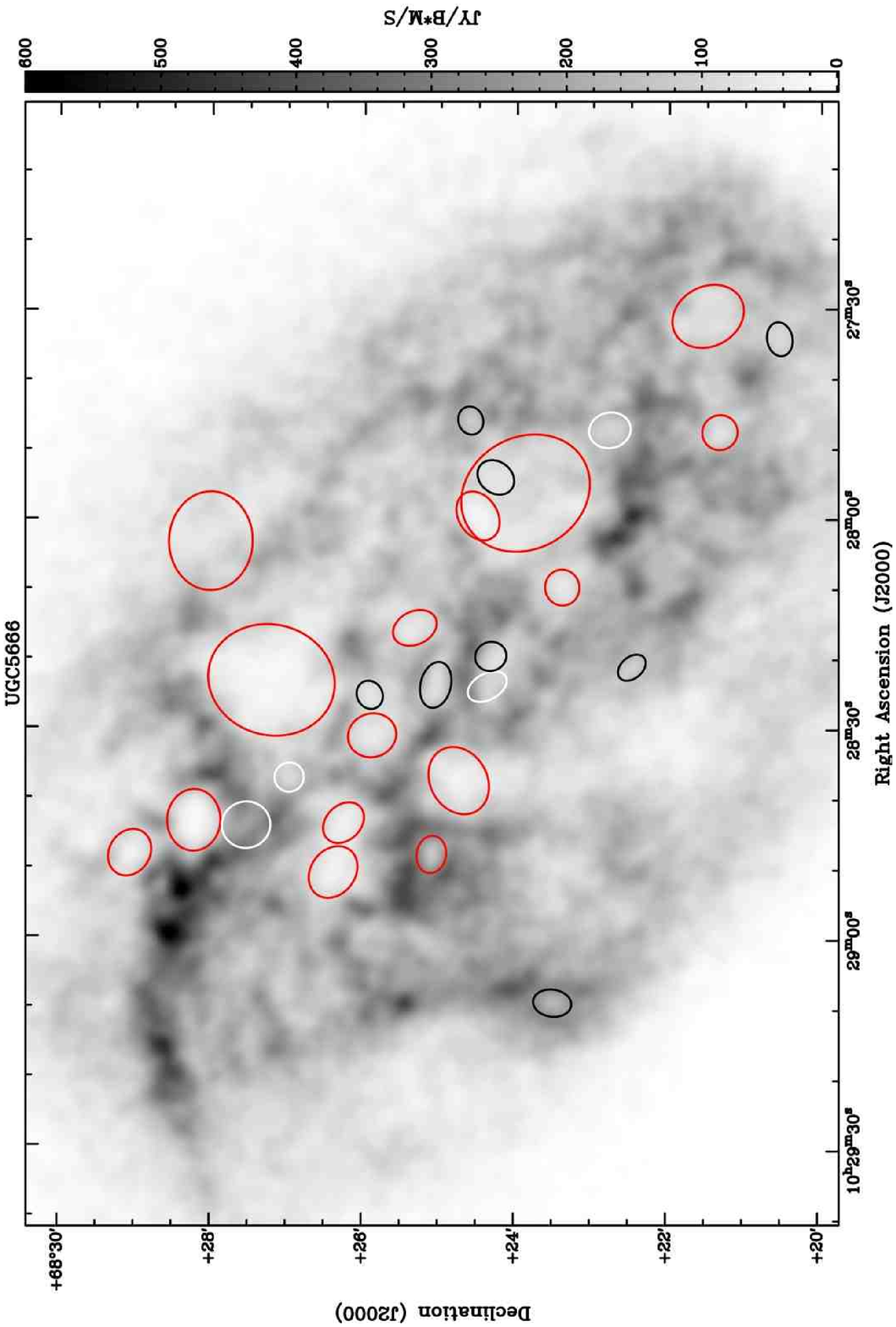}
\end{center}
\caption{Position of the \ion{H}{1} holes in IC\,2574. 
The colours illustrate the different type of the holes: red - type 1, white - type 2 and black - type 3.
\label{IC2574-map}}
\end{figure}
\clearpage

\begin{figure}
\begin{center}
\includegraphics[angle=0,scale=.75]{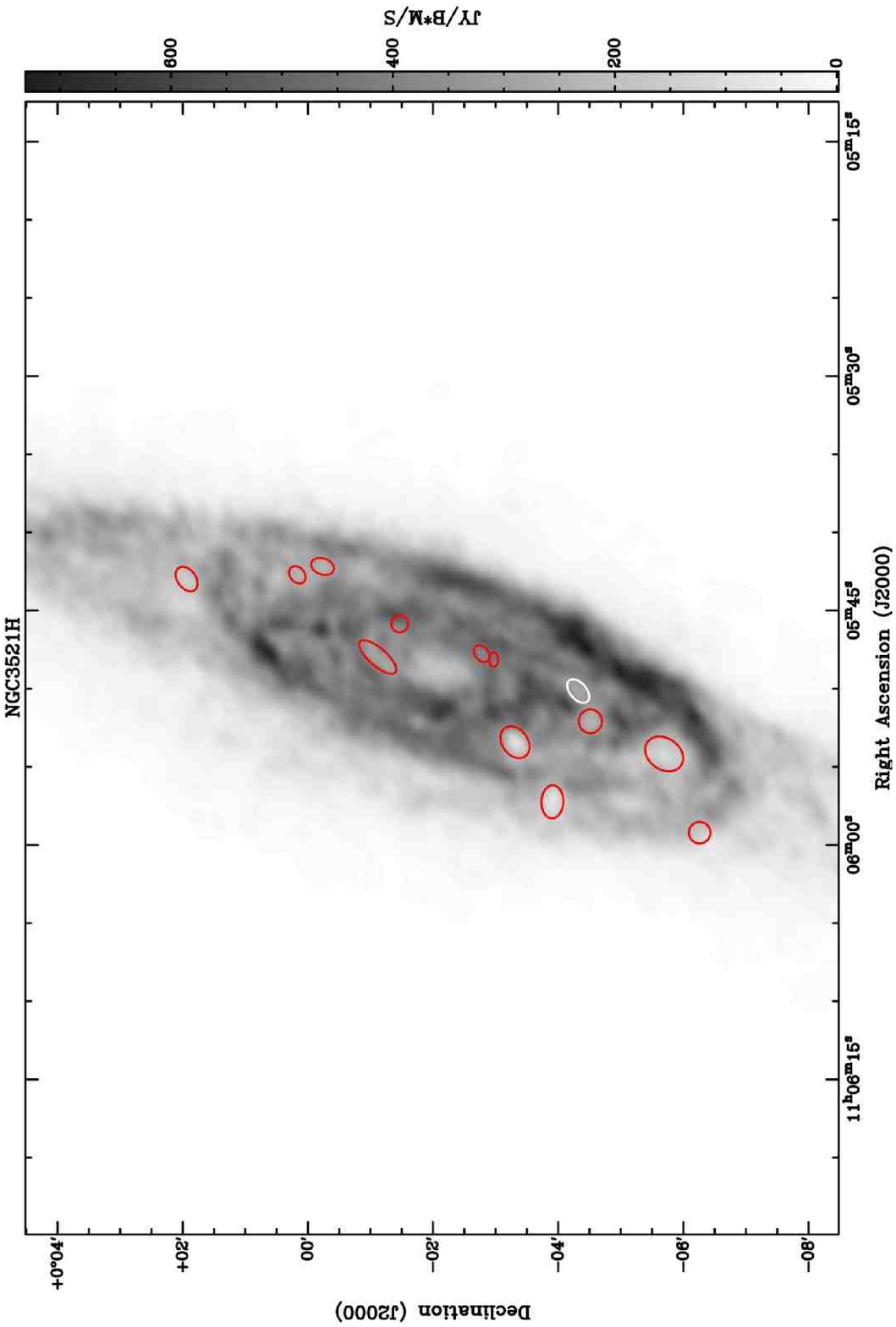}
\end{center}
\caption{Position of the \ion{H}{1} holes in NGC\,3521. 
The colours illustrate the different type of the holes: red - type 1, white - type 2 and black - type 3.
\label{NGC3521-map}}
\end{figure}
\clearpage

\begin{figure}
\begin{center}
\includegraphics[angle=0,scale=.75]{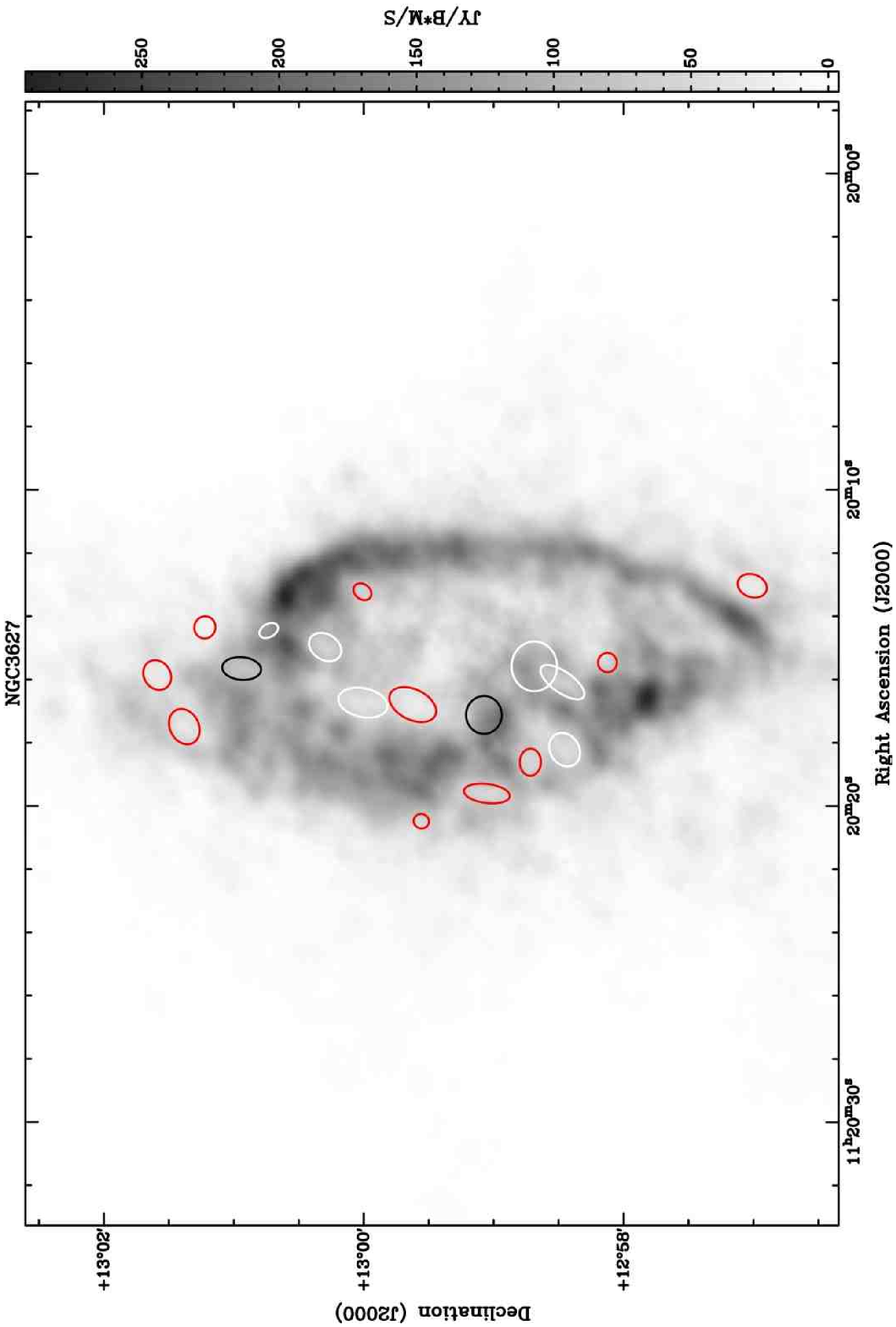}
\end{center}
\caption{Position of the \ion{H}{1} holes in NGC\,3627. 
The colours illustrate the different type of the holes: red - type 1, white - type 2 and black - type 3.
\label{NGC3627-map}}
\end{figure}
\clearpage

\begin{figure}
\begin{center}
\includegraphics[angle=0,scale=.75]{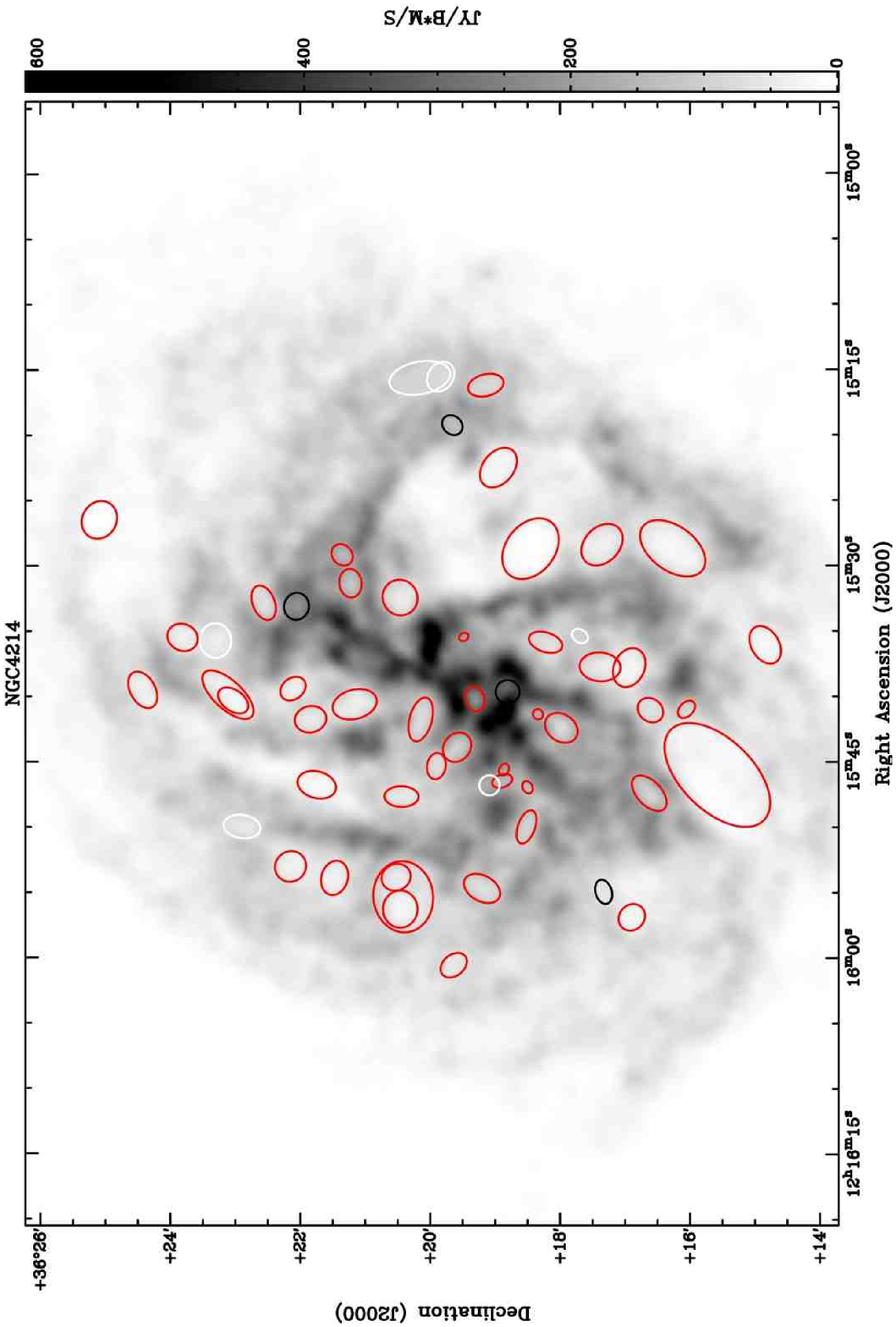}
\end{center}
\caption{Position of the \ion{H}{1} holes in NGC\,4214. 
The colours illustrate the different type of the holes: red - type 1, white - type 2 and black - type 3.
\label{NGC4214-map}}
\end{figure}
\clearpage

\begin{figure}
\begin{center}
\includegraphics[angle=0,scale=.75]{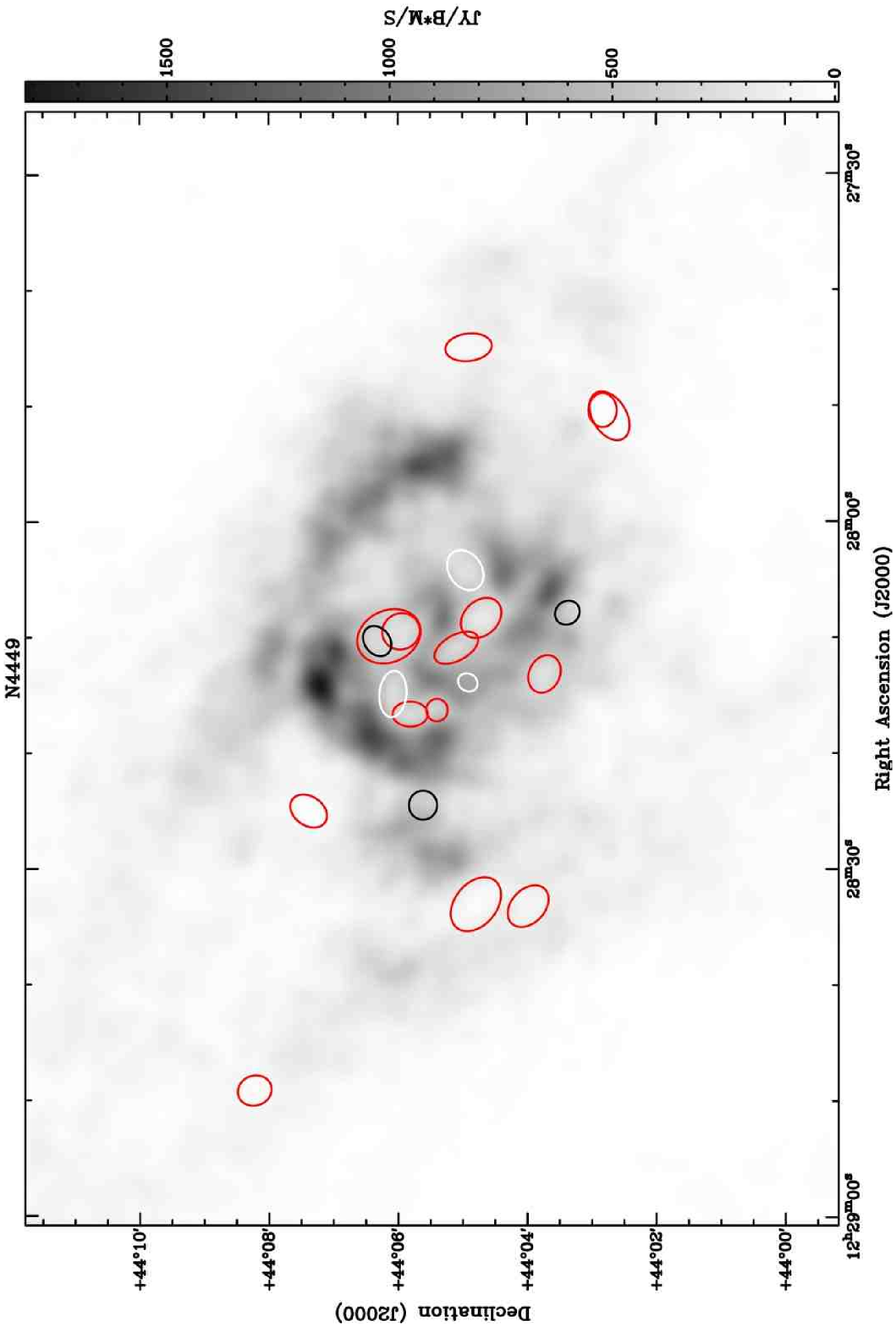}
\end{center}
\caption{Position of the \ion{H}{1} holes in NGC\,4449. 
The colours illustrate the different type of the holes: red - type 1, white - type 2 and black - type 3.
\label{NGC4449-map}}
\end{figure}
\clearpage

\begin{figure}
\begin{center}
\includegraphics[angle=0,scale=.75]{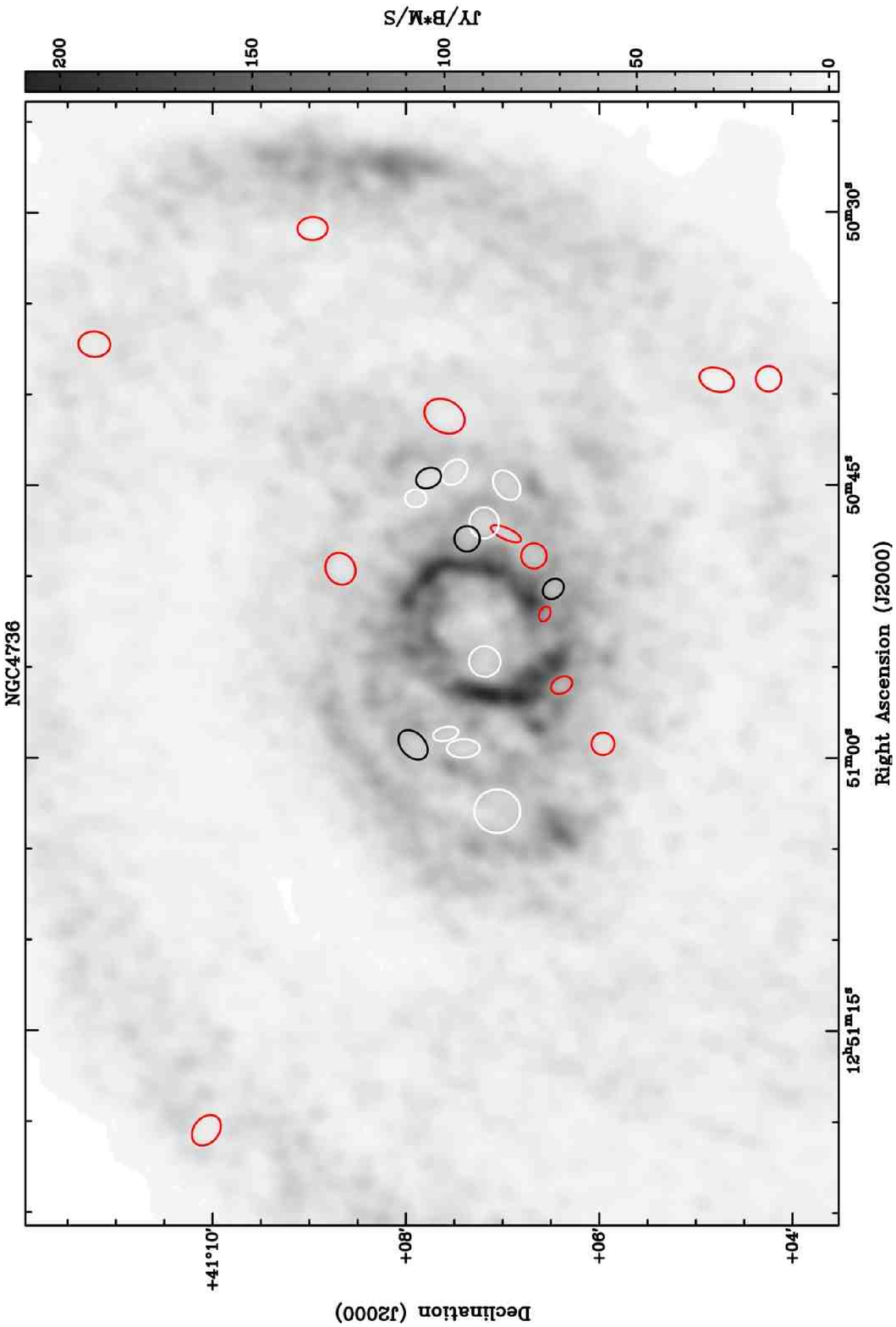}
\end{center}
\caption{Position of the \ion{H}{1} holes in NGC\,4736. 
The colours illustrate the different type of the holes: red - type 1, white - type 2 and black - type 3.
\label{NGC4736-map}}
\end{figure}
\clearpage

\begin{figure}
\begin{center}
\includegraphics[angle=0,scale=.75]{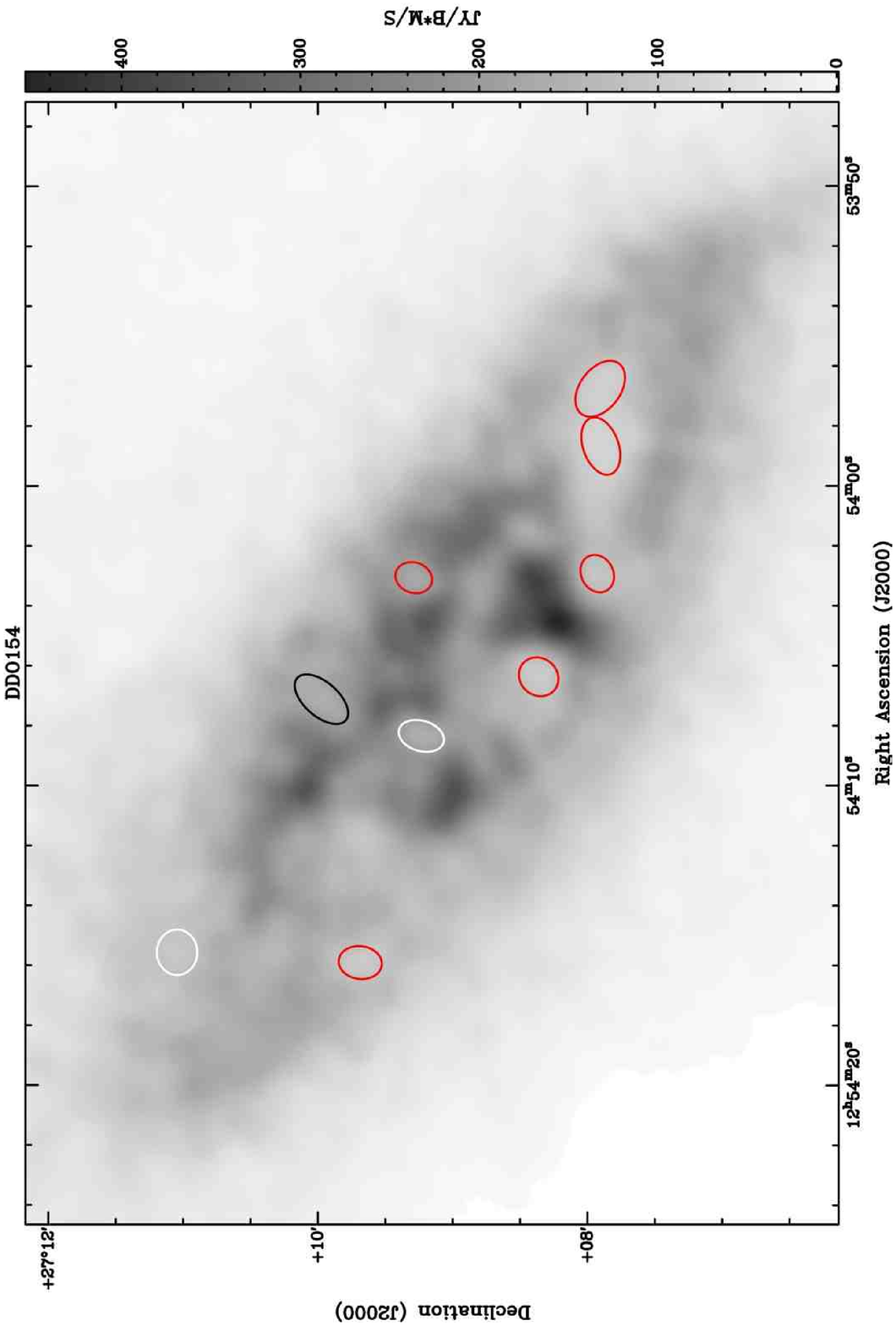}
\end{center}
\caption{Position of the \ion{H}{1} holes in DDO\,154. 
The colours illustrate the different type of the holes: red - type 1, white - type 2 and black - type 3.
\label{DDO154-map}}
\end{figure}
\clearpage

\begin{figure}
\begin{center}
\includegraphics[angle=0,scale=.75]{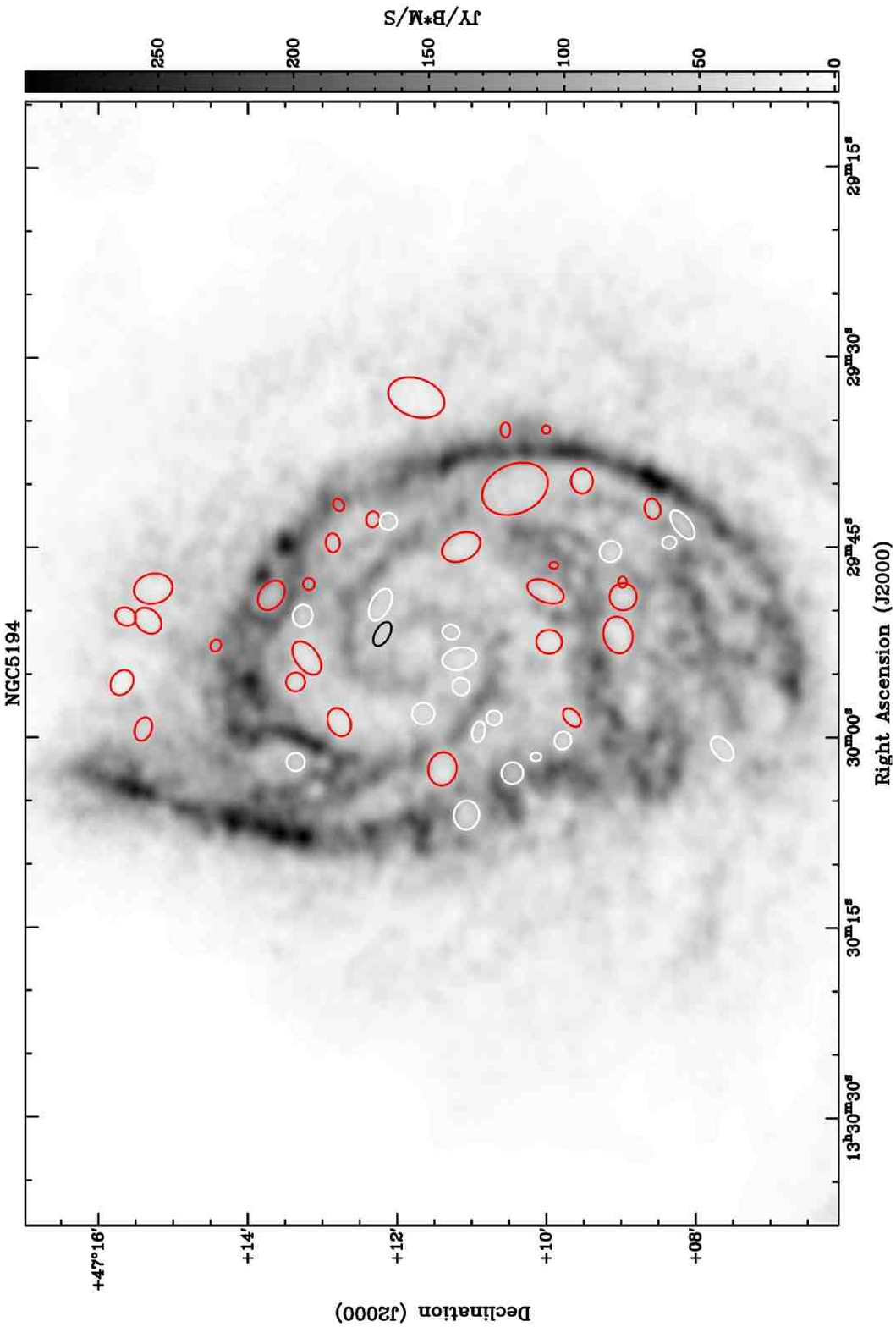}
\end{center}
\caption{Position of the \ion{H}{1} holes in NGC\,5194. 
The colours illustrate the different type of the holes: red - type 1, white - type 2 and black - type 3.
\label{NGC5194-map}}
\end{figure}
\clearpage

\begin{figure}
\begin{center}
\includegraphics[angle=0,scale=.75]{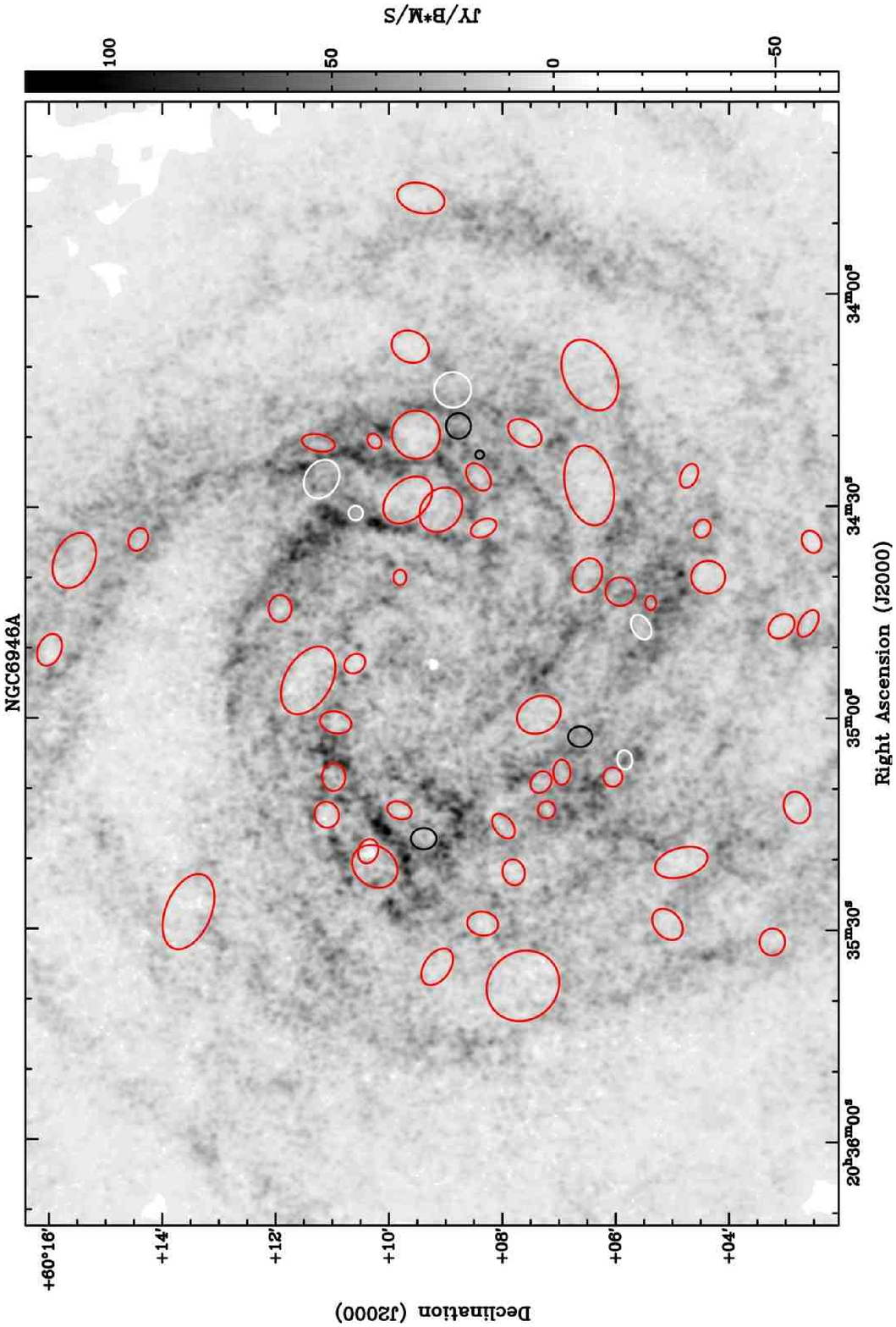}
\end{center}
\caption{Position of the \ion{H}{1} holes in NGC\,6946. 
The colours illustrate the different type of the holes: red - type 1, white - type 2 and black - type 3.
\label{NGC6946-map}}
\end{figure}
\clearpage

\begin{figure}
\begin{center}
\includegraphics[angle=0,scale=.75]{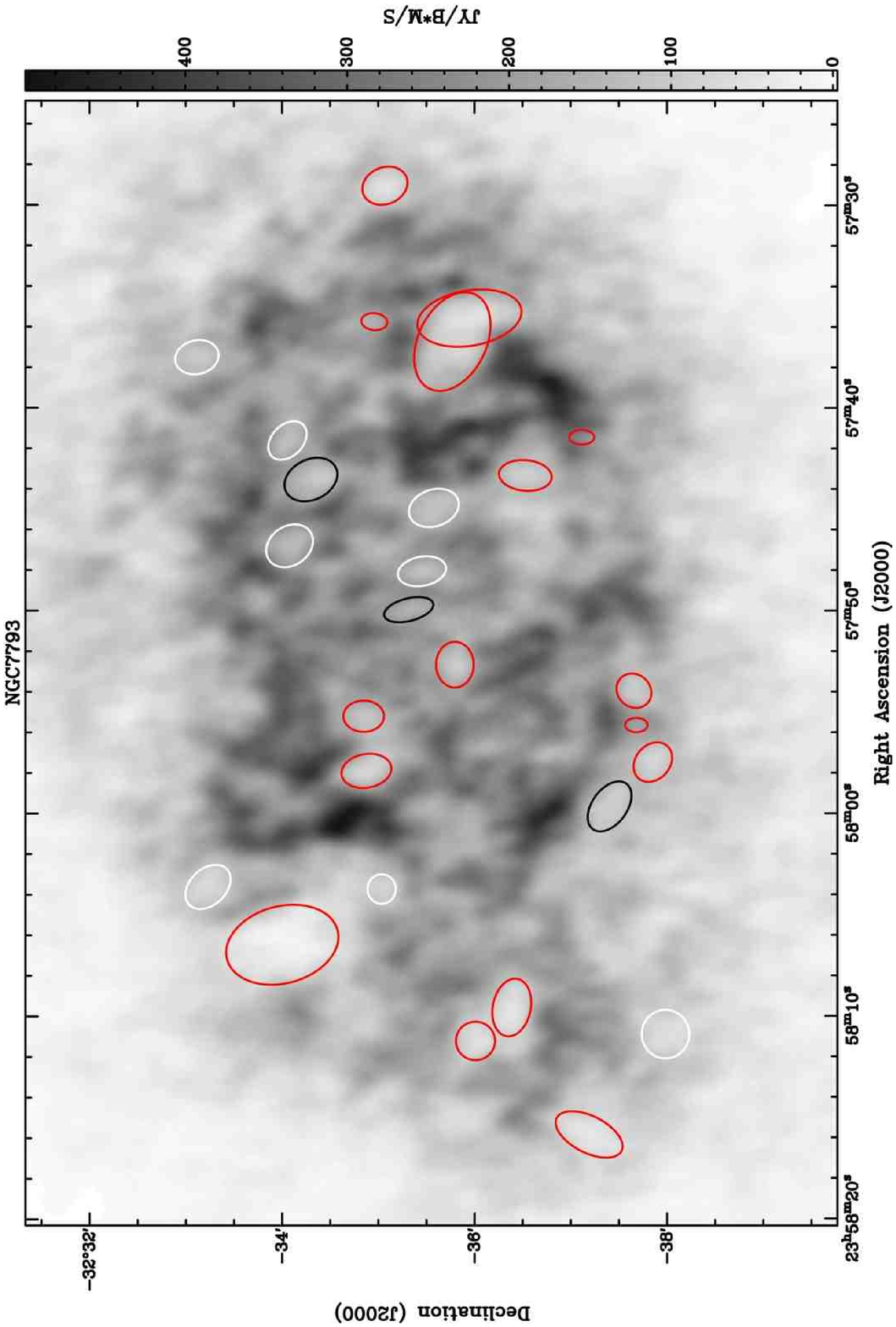}
\end{center}
\caption{Position of the \ion{H}{1} holes in NGC\,7793. 
The colours illustrate the different type of the holes: red - type 1, white - type 2 and black - type 3.
\label{NGC7793-map}}
\end{figure}

\end{document}